\documentclass[journal]{IEEEtran}
\usepackage{setspace}
\usepackage{graphicx}
\usepackage{amstext}
\usepackage{algorithm}
\usepackage{algorithmic}
\usepackage{mathrsfs}
\usepackage{amssymb}
\usepackage{amsmath}
\usepackage{bm}
\allowdisplaybreaks[4]
\usepackage{epstopdf}
\usepackage{multicol}
\usepackage{stfloats}
\usepackage{url}
\usepackage{color}
\usepackage{enumerate}
\usepackage[colorlinks,
            linkcolor=black,
            anchorcolor=blue,
            citecolor=black
            ]{hyperref}
\usepackage{breqn}
\usepackage{bbm}
\usepackage{cite}
\usepackage{subfig}
\usepackage{caption}
\usepackage{enumitem}
\usepackage{mathrsfs}
\usepackage{multirow}

\DeclareMathAlphabet{\mathscrbf}{OMS}{mdugm}{b}{n}

\usepackage[T1]{fontenc}

\DeclareFontFamily{OML}{zavm}{\skewchar\font=127 }
\DeclareFontShape{OML}{zavm}{m}{it}{<-> s*[.78] zavmri7m}{}
\DeclareFontShape{OML}{zavm}{b}{it}{<-> s*[.78] zavmbi7m}{}
\DeclareFontShape{OML}{zavm}{m}{sl}{<->ssub * zavm/m/it}{}
\DeclareFontShape{OML}{zavm}{bx}{it}{<->ssub * zavm/b/it}{}
\DeclareFontShape{OML}{zavm}{b}{sl}{<->ssub * zavm/b/it}{}
\DeclareFontShape{OML}{zavm}{bx}{sl}{<->ssub * zavm/b/sl}{}
\DeclareMathAlphabet{\mathsf}{OML}{zavm}{m}{it} 

\newtheorem{Lemma}{Lemma}
\newtheorem{Remark}{Remark}

\newtheorem{Problem}{Problem}
\newtheorem{Definition}{Definition}

\begin{document}
\title{  Joint Optimal Software Caching, Computation Offloading and Communications Resource Allocation  for Mobile Edge Computing}
\author{Wanli~Wen,~\IEEEmembership{Member,~IEEE,}
        Ying~Cui,~\IEEEmembership{Member,~IEEE,}
        Tony~Q.~S.~Quek,~\IEEEmembership{Fellow,~IEEE,}
        Fu-Chun~Zheng,~\IEEEmembership{Senior Member,~IEEE,}
        and~Shi~Jin,~\IEEEmembership{Senior Member,~IEEE}

\thanks{Copyright (c) 2015 IEEE. Personal use of this material is permitted. However, permission to use this material for any other purposes must be obtained from the IEEE by sending a request to pubs-permissions@ieee.org. Manuscript received November 18, 2019; revised February 24, 2020; accepted May~5, 2020.(\textit{Corresponding author: Fu-Chun~Zheng})

W. Wen and T. Q. S. Quek are with the Information Systems Technology and Design Pillar, Singapore University of Technology
and Design, Singapore 487372 (e-mail: wanli_wen@sutd.edu.sg; tonyquek@sutd.edu.sg).

Y. Cui is with the Department of Electronic Engineering, Shanghai Jiao Tong University, Shanghai 200240, China (e-mail: cuiying@sjtu.edu.cn).

F.-C. Zheng is with the School of Electronic and Information Engineering, Harbin Institute of Technology, Shenzhen 518055, China (e-mail: fzheng@ieee.org).

S. Jin is with the National Mobile Communications Research Laboratory, Southeast University, Nanjing 210096, China (e-mail: jinshi@seu.edu.cn).
}}

\maketitle

\begin{abstract}
As software may be used by multiple users, caching popular software at the wireless edge has been considered to save computation and communications resources for mobile edge computing (MEC). However, fetching uncached software from the core network and multicasting popular software to users have so far been ignored. Thus, existing design is incomplete and less practical.
In this paper, we propose a joint caching, computation and communications mechanism which involves software fetching, caching and multicasting, as well as task input data uploading, task executing (with non-negligible time duration) and computation result downloading, and mathematically characterize it. Then, we optimize the joint caching, offloading and time allocation policy to minimize the weighted sum energy consumption subject to the caching and deadline constraints. The problem is a challenging two-timescale mixed integer nonlinear programming (MINLP) problem, and is NP-hard in general. We convert it into an equivalent convex MINLP problem by using some appropriate transformations and propose two low-complexity algorithms to obtain suboptimal solutions of the original non-convex MINLP problem.  Specifically, the first suboptimal solution is obtained by solving a relaxed convex problem using the consensus alternating direction method of multipliers (ADMM), and then rounding its optimal solution properly. The second suboptimal solution is proposed by obtaining a stationary point of an equivalent difference of convex (DC) problem using the penalty convex-concave procedure (Penalty-CCP) and ADMM. Finally, by numerical results, we show that the proposed solutions outperform existing schemes and reveal their advantages in efficiently utilizing storage, computation and communications resources.
\end{abstract}

\IEEEpeerreviewmaketitle
\section{Introduction}

With rapid development of mobile devices such as smart phones and tablet computers, a wide-range of mobile application services with advanced features such as augmented reality (AR), mobile online gaming and ultra-high-definition video streaming, are constantly emerging.  These listed application services are usually both latency-sensitive and computation-intensive. However, mobile devices are often constrained with limited battery capacity and computation capability.  Mobile edge computing (MEC) is one promising technology which provides the computing capability to support these application services at the wireless edge {\cite{MECsurveyMao, Rodrigues2017TC, Rodrigues2018TC, MECCachingWang2017}}. {Note that, in the context of MEC, we refer to application services as computation services.} In an MEC system, each computation service is processed by a particular software, and a user demanding a computation service will generate a specific computation task. A user's computation task can be executed remotely at the MEC server attached to a serving node (e.g., base station or access point) via computation offloading or {local computation}.

In order to design an energy-efficient MEC system, it is required to jointly optimize the  computation and communications resources among distributed users and MEC servers. Emerging research toward this direction considers optimal resource allocation for various types of multi-user MEC systems \cite{DinhMEC2017TCOM, MEC2016ISITLiu, JSAC2016YouMec, MECWangTVT2017, MECChen2016CNconf, MECMchenICC2016, JSAC2017MECsun, MECChenJSAC2018,  MECShan2018Conf, MECGuo2017Conf, MECWangTWC, MECYouTWC, JRAODinMEC2019CL, 2019CLMECXiong, Ding2019TVT}. Specifically, the authors in \cite{DinhMEC2017TCOM} and \cite{MEC2016ISITLiu} study single-user MEC systems with elastic tasks, and minimize the weighted sum of the system latency and energy consumption; the authors in \cite{JSAC2016YouMec} investigate a single-user MEC system with inelastic tasks, and minimize the total energy consumption under the deadline constraint; the authors in \cite{MECWangTVT2017, MECChen2016CNconf, MECMchenICC2016, JSAC2017MECsun, MECChenJSAC2018,  MECShan2018Conf} examine multi-user MEC systems with elastic tasks, and minimize the weighted sum of the system delay and energy consumption \cite{MECWangTVT2017, MECChen2016CNconf, MECMchenICC2016} or minimize the system delay under the transmit power constraint \cite{JSAC2017MECsun, MECChenJSAC2018, MECShan2018Conf}; the authors in \cite{MECGuo2017Conf, MECWangTWC, MECYouTWC, JRAODinMEC2019CL, 2019CLMECXiong, Ding2019TVT} study multi-user MEC systems with inelastic tasks, and minimize the total energy consumption under the deadline constraint.  Note that the works \cite{DinhMEC2017TCOM, MEC2016ISITLiu, JSAC2016YouMec, MECWangTVT2017, MECChen2016CNconf, MECMchenICC2016, JSAC2017MECsun, MECChenJSAC2018,  MECShan2018Conf, MECGuo2017Conf, MECWangTWC, MECYouTWC, JRAODinMEC2019CL, 2019CLMECXiong} do not consider caching at serving nodes.

In an MEC system, the computation task input data, computation results and software may be reused by multiple users and thus can be stored in advance at the wireless edge to save computation and {communications} resources of an MEC system \cite{MECsurveyMao}. Therefore, it is increasingly important to jointly optimize caching, computation and communications resources for design energy-efficient cache-assisted MEC systems.  Recently, the works in \cite{MeccachingAccess2018Yang, MECcachingICC2018Sun, MECcachingBitrateTMC2018, 2019arXivBiCachMec, MECAccessXu2017, MeccachingTVT2017Zhou,  4CNdikumanarxiv2018, MeccachingIWCMC2018Zhao,  MeccachingTVT2018Liu, EECachMEC2018Hao, 2018arXiv181007797T, MECCachingCui2017, MECcachingXuJie2018arxiv, XuJie2018MECserviceCach}  consider optimal caching, computation offloading, and transmission power and time allocation to reduce costs (e.g., energy and latency) of MEC systems.  Specifically, in \cite{MeccachingAccess2018Yang, MECcachingICC2018Sun, MECcachingBitrateTMC2018, 2019arXivBiCachMec}, the authors study single-user cache-assisted MEC systems with elastic tasks, and minimize the communications resource consumption \cite{MeccachingAccess2018Yang, MECcachingICC2018Sun}, the system latency \cite{MECcachingBitrateTMC2018} or the weighted sum of the system latency and energy consumption \cite{2019arXivBiCachMec}. {The proposed solutions in \cite{MeccachingAccess2018Yang, MECcachingICC2018Sun, MECcachingBitrateTMC2018, 2019arXivBiCachMec} may not be suitable for multi-user MEC systems.} In \cite{MECAccessXu2017, 4CNdikumanarxiv2018, MeccachingTVT2017Zhou,  MeccachingIWCMC2018Zhao,  MECcachingXuJie2018arxiv, XuJie2018MECserviceCach}, the authors investigate multi-user cache-assisted MEC systems with elastic tasks, and maximize the total network revenue \cite{MECAccessXu2017, MeccachingTVT2017Zhou, XuJie2018MECserviceCach},  or minimize the system latency \cite{4CNdikumanarxiv2018, MeccachingIWCMC2018Zhao,  MECcachingXuJie2018arxiv}. In \cite{MeccachingTVT2018Liu, EECachMEC2018Hao, 2018arXiv181007797T, MECCachingCui2017}, the authors study multi-user cache-assisted MEC systems with inelastic tasks and maximize the total network revenue \cite{MeccachingTVT2018Liu}, or minimize the total energy consumption \cite{EECachMEC2018Hao, 2018arXiv181007797T, MECCachingCui2017} under the deadline constraint.

Note that in \cite{MECAccessXu2017, MeccachingTVT2017Zhou,  4CNdikumanarxiv2018, MeccachingIWCMC2018Zhao,  MeccachingTVT2018Liu, EECachMEC2018Hao, 2018arXiv181007797T, MECCachingCui2017, MECcachingXuJie2018arxiv, XuJie2018MECserviceCach}, it is assumed that the size of the task input data or computation result of each task is negligible. Hence, the resource consumption for transmitting task input data from users to a serving node or computation results from a serving node to users is not considered. This assumption may not hold for many application services with large sizes of task input data or computation results, such as AR and multi-media transformation.  In \cite{MECcachingBitrateTMC2018, MECAccessXu2017, MeccachingTVT2017Zhou,  4CNdikumanarxiv2018, MECcachingXuJie2018arxiv,XuJie2018MECserviceCach, 2018arXiv181007797T, MECCachingCui2017}, on the other hand, it is assumed that task execution durations at the serving node or users are negligible. This assumption may not be suitable for some applications with computation tasks of high workloads, such as online video gaming and 3D modeling/rendering.

Furthermore, the authors in \cite{MECAccessXu2017, MeccachingTVT2018Liu, EECachMEC2018Hao, 4CNdikumanarxiv2018, 2018arXiv181007797T, MeccachingTVT2017Zhou} consider caching task input data at a serving node to save communications resource, and the authors in \cite{2018arXiv181007797T, MeccachingTVT2017Zhou,  MECCachingCui2017, MeccachingIWCMC2018Zhao} consider caching computation results at a serving node to save computation {resource consumption}. Since task input data or computation results of different tasks are usually not the same, even if they correspond to the same service, the resulting reduction of communications or computation resource may not be significant. In addition, an implicit assumption in \cite{MECAccessXu2017, MeccachingTVT2017Zhou,  4CNdikumanarxiv2018, MeccachingIWCMC2018Zhao,  MeccachingTVT2018Liu, EECachMEC2018Hao, 2018arXiv181007797T, MECCachingCui2017} is that an MEC server or users  can process whatever types of computation tasks that are offloaded from users without considering the availability of the software at the wireless edge. This may not hold in practice, since {only a limited number of software can be stored at the wireless edge due to} limited storage resource. Recently, the authors in \cite{MECcachingXuJie2018arxiv, XuJie2018MECserviceCach} consider {limited storage resource and study optimal caching of software} at serving nodes. However, in \cite{MECcachingXuJie2018arxiv, XuJie2018MECserviceCach}, fetching {uncached software} via the backhaul link is not considered, and the tasks corresponding to the service processed by an uncached software have to be transmitted to a remote cloud for executing. In addition, in \cite{XuJie2018MECserviceCach}, only one software can be stored at the serving node. Thus, the model regarding software in \cite{MECcachingXuJie2018arxiv, XuJie2018MECserviceCach} {may not} fully exploit the potential advantage of MEC {systems}.

\begin{figure*}[!t]
    \centering
         {\includegraphics[width= 1.0\textwidth]{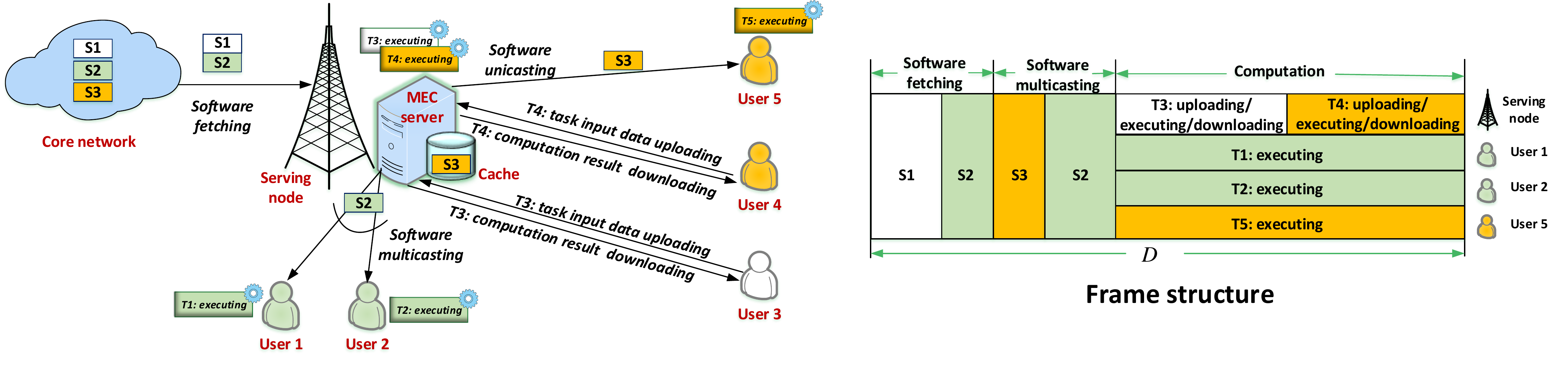}}
        \caption{The system model at $N=3$ and $K=5$. Here, S1, S2 and S3 represent the software {corresponding to} services $1$, $2$ and~$3$ in the set $\mathcal{X}$. Only S3 is cached at the serving node. T1, $...$, T5 represent the tasks of users $1,\cdots,5$ in the set $\mathcal{K}$. The color of each user corresponds to the service it demands. {T1, T2 and T5 are executed locally, T3 is executed at the serving node, and T4 is executed at the serving node.}}\label{figsystemodel}
\end{figure*}

In summary, {further studies are required} to design {more practical energy-}efficient cache-assisted MEC systems by jointly optimizing caching, computation and {communications} resource {allocation}.   In this paper, we {consider joint optimization of software} caching, computation {offloading} and {communications} {resource allocation} in a multi-user cache-assisted MEC system, consisting of one serving node with caching and computing capabilities and multiple users {each with} computing capability.
The main contributions are summarized below.
\begin{itemize}
  \item  First, we propose a joint caching, computation and  communications mechanism which involves software fetching, caching and multicasting, as well as task input data uploading, task \-executing (with non-negligible time duration) and computation result downloading and mathematically characterize it. Note that the proposed mechanism {is more practical and} can exploit more advantages of MEC systems than the existing ones \cite{DinhMEC2017TCOM, MEC2016ISITLiu, JSAC2016YouMec, MECWangTVT2017, MECChen2016CNconf, MECMchenICC2016, JSAC2017MECsun, MECChenJSAC2018,  MECShan2018Conf, MECGuo2017Conf, MECWangTWC, MECYouTWC, JRAODinMEC2019CL, 2019CLMECXiong, MeccachingAccess2018Yang, MECcachingICC2018Sun, MECcachingBitrateTMC2018, MECAccessXu2017, MeccachingTVT2017Zhou,  4CNdikumanarxiv2018, MeccachingIWCMC2018Zhao, MECcachingXuJie2018arxiv,XuJie2018MECserviceCach, MeccachingTVT2018Liu, EECachMEC2018Hao, 2018arXiv181007797T, MECCachingCui2017, 2019arXivBiCachMec}.  {To the best of our knowledge, this is the first work considering software fetching and multicasting besides caching.}
  \item  Then, we optimize the joint caching, offloading and time allocation policy to minimize the weighted sum energy consumption  subject to the caching and deadline constraints. The problem is a challenging two-timescale mixed integer nonlinear programming (MINLP) problem and is NP-hard in general. We convert it into an equivalent convex MINLP problem by using some appropriate transformations and propose two low-complexity algorithms to obtain suboptimal solutions of the original non-convex MINLP problem. Specifically, the first suboptimal solution is obtained by {solving a relaxed convex problem using} alternating direction method of multipliers (ADMM), and rounding its optimal solution {properly}. The second suboptimal solution is obtained by {obtaining} a stationary point {of an equivalent difference of convex (DC) problem} using the penalty convex-concave procedure (Penalty-CCP) and ADMM.
  \item Finally, {by numerical results, we show} that the proposed solutions outperform existing schemes and reveal their advantages in efficiently utilizing storage, computation and communications resources.  {In addition, we show that the suboptimal solution based on penalty CCP outperforms the one based on continuous relaxation at the cost of computational complexity~increase.}
\end{itemize}

\section{System Model}

As shown in Fig.~\ref{figsystemodel}, we consider a multi-user cache-assisted MEC system with one  single-antenna serving node and $K$ single-antenna users, denoted by set $\mathcal{K}\triangleq\{1,2,\cdots,K\}$. The serving node has caching and computing {capabilities}. Each user also has {its own} computing {capability}. Key notations used in the rest of this paper are
summarized in Table~\ref{tabNotations}.
\subsection{Service, Task, Channel and System State}
Consider $N$ computation-intensive and latency-sensitive computation {services}, denoted by set $\mathcal{X}\triangleq\{1,2,\cdots,N\}$.  {Each user $k$ randomly requests a computation service}, denoted by $X_k\in\mathcal{X}$.  Let $\mathbf{X}\triangleq(X_k)_{k\in\mathcal{K}} \in\mathcal{X}^K$ denote the random system service state. Each computation service may be demanded by multiple users.   Let $\mathcal{K}_n(\mathbf{X})\triangleq \{k\in\mathcal{K}:X_k=n\}$ and $K_n(\mathbf{X})\triangleq {|\mathcal{K}_n(\mathbf{X})| =} \sum_{k\in\mathcal{K}}\mathbbm{1}\left(X_k=n\right)$ denote the set and number of users who requests service $n$ at the random system service state $\mathbf{X}$, where $\mathbbm{1}\left(\cdot\right)$ denotes the indicator function.

\begin{table*}
\centering\small
\caption{Notations.}\label{tabNotations}
\begin{tabular}{p{1.2cm} p{7.8cm} | p{1.4cm} p{5.8cm}}
  \hline\hline
  Notation  & Definition                                      & Notation  & Definition \\ \hline
  $\mathcal{K}$ &   Set of users (tasks)                              & $K$ & Number of users. \\
  $\mathcal{X}$ &   Set of services.             & $N$ & Number of services.\\
  ${\mathcal{L}_{\texttt{u}}^{\mathrm{dat}}}$ & Finite space of the size of input data.  & $B$ & System Bandwidth. \\
  ${\mathcal{L}_{\texttt{e}}^{\mathrm{dat}}}$ & Finite space of computation load.   & $D$ & System Deadline. \\
  ${\mathcal{L}_{\texttt{d}}^{\mathrm{dat}}}$ & Finite space of the size of computation result. & $L_{\texttt{u},k}^{\mathrm{dat}}$ & Size of input data of task $k$. \\
  $\mathcal{H}$                               & Finite channel state space.  & $L_{\texttt{e},k}^{\mathrm{dat}}$ & Computation load of task $k$.\\
  $\mathcal{Q}$                             & System state space. & $L_{\texttt{d},k}^{\mathrm{dat}}$ & Size of computation result of task $k$. \\
  $\mathbf{X}$  &  Random system service state.               &  $X_k$ & Service requested by user $k$. \\
  ${\mathbf{L}_{\texttt{u}}^{\mathrm{dat}}}$  & Random system task input data state.         & $H_k$ & Channel state of user $k$.\\
  ${\mathbf{L}_{\texttt{e}}^{\mathrm{dat}}}$  & Random system task computation load state.   & $H_{\texttt{d},n}(\mathbf{X}, \mathbf{H})$ & The smallest  value  among the channel states of all the $K_n(X)$ users in $\mathcal{K}_n(\mathbf{X})$.\\
  ${\mathbf{L}_{\texttt{d}}^{\mathrm{dat}}}$  & Random system task computation result state. & $R$ &  Transmission rate  via backhaul link.\\
  $\mathbf{H}$                                & Random system channel state.                 & $C$ & Cache size. \\
  $\mathbf{Q}$                                & Random system state.                         & $F_{\mathrm{sn}}$ & CPU frequency of serving node. \\
  $\mathcal{K}_n(\mathbf{X})$ & Set of users who requests service $n$ at the random system service state $\mathbf{X}$.       & $K_n(\mathbf{X})$ & Number of users who requests service $n$ at the random system service state $\mathbf{X}$.\\
  $\mathbf{c}$                                & Caching action.                              & $F_k$ &  CPU frequency of user $k$.\\
  $\mathbf{o}$                                & Offloading action.                           & $\mu_{\mathrm{sn}}$ & Constant factor at serving node.\\
  $\mathbf{t}_{\texttt{u}}^{\mathrm{dat}}$           & Time allocation action for uploading the task input data of all users. & $\mu_k$ &  Constant factor at user $k$.\\
  $\mathbf{t}_{\texttt{d}}^{\mathrm{dat}}$           & Time allocation action for downloading  the computation result of all users.  & $\omega_k$ & Weight factor for user $k$.\\
  $\mathbf{t}_{\texttt{d}}^{\mathrm{sfw}}$           & Time allocation action for multicasting all software.  & $l_{\texttt{d},n}^{\mathrm{sfw}}$ & Size of software $n$.   \\
  $\mathbf{O}$                           & A mapping from system state $\mathbf{Q}$ to offloading action $\mathbf{o}$. & $n_0$ & Power of the complex additive white Gaussian noise over  entire bandwidth $B$.\\
  $\mathbf{T}_{\texttt{u}}^{\mathrm{dat}}$ & A mapping from system state $\mathbf{Q}$ to time allocation action $\mathbf{t}_{\texttt{u}}^{\mathrm{dat}}$ for uploading the task input data of all users.  &  $\epsilon$ & A positive constant.\\
  $\mathbf{T}_{\texttt{d}}^{\mathrm{dat}}$ & A mapping from system state $\mathbf{Q}$ to time allocation action $\mathbf{t}_{\texttt{d}}^{\mathrm{dat}}$ for downloading the computation result of all users.  & $\rho$ &  A penalty parameter for ADMM algorithm.\\
  $\mathbf{T}_{\texttt{d}}^{\mathrm{sfw}}$ & A mapping from system state $\mathbf{Q}$ to time allocation action $\mathbf{t}_{\texttt{d}}^{\mathrm{sfw}}$ for multicasting all software.  & \\
  \hline\hline
\end{tabular}
\end{table*}

Each user $k$ has a specific task for the requested service $X_k$. {The task of user $k$ is also referred to as task $k$. Task $k$} is specified by {three {random} task parameters, i.e.,} {the size of the task input data} $L_{\texttt{u},k}^{\mathrm{dat}}\in {\mathcal{L}_{\texttt{u}}^{\mathrm{dat}}}$ (in bits), the computation load $L_{\texttt{e},k}^{\mathrm{dat}}\in {\mathcal{L}_{\texttt{e}}^{\mathrm{dat}}}$ (in number of CPU-cycles) and {the size of the computation result} $L_{\texttt{d},k}^{\mathrm{dat}}\in {\mathcal{L}_{\texttt{d}}^{\mathrm{dat}}}$ (in bits).  Here, ${\mathcal{L}_{\texttt{u}}^{\mathrm{dat}}}$, ${\mathcal{L}_{\texttt{e}}^{\mathrm{dat}}}$ and ${\mathcal{L}_{\texttt{d}}^{\mathrm{dat}}}$ denote the {corresponding} finite spaces. {Let $\mathbf{L}_{\texttt{u}}^{\mathrm{dat}} \triangleq (L_{\texttt{u},k}^{\mathrm{dat}})_{k\in\mathcal{K}}\in ({\mathcal{L}_{\texttt{u}}^{\mathrm{dat}}})^K$, $\mathbf{L}_{\texttt{e}}^{\mathrm{dat}} \triangleq (L_{\texttt{e},k}^{\mathrm{dat}})_{k\in\mathcal{K}}\in ({\mathcal{L}_{\texttt{e}}^{\mathrm{dat}}})^K$ and $\mathbf{L}_{\texttt{d}}^{\mathrm{dat}} \triangleq (L_{\texttt{d},k}^{\mathrm{dat}})_{k\in\mathcal{K}}\in ({\mathcal{L}_{\texttt{d}}^{\mathrm{dat}}})^K$ denote the random system task input data {size} state, random system computation load state and random system computation result {size} state, respectively.}  The {task of each user is generated at the time $0$ and has to be completed by the deadline $D$ (in seconds), where $D$ {reflects} the user latency requirement.\footnote{We assume that all tasks have the same deadline. The optimization results obtained in this paper can be extended to study a more general task scenario, where tasks may have different deadlines. The {extension} will be considered in our future work.}}  We consider the operation of the MEC system in {the} time interval $[0,D]$.

{We study Time-Division Duplexing (TDD) mode and consider Time Division Multiple Access (TDMA) system.} The entire bandwidth {is $B$ (Hz)}. Assume {that} channel {conditions do not change} during {$[0,D]$}. {Let $H_k\in\mathcal{H}$  denote the random channel state of user $k$, representing the power gain of the channel (over the entire bandwidth $B$) between user $k$ and the {serving node}, where $\mathcal{H}$ denotes the finite channel state space.}
Let $\mathbf{H}\triangleq(H_k)_{k\in\mathcal{K}} \in\mathcal{H}^K$ denote the random system channel state. {Let {$H_{\texttt{d},n}(\mathbf{X}, \mathbf{H})\triangleq \min_{k\in\mathcal{K}_n(\mathbf{X})}H_k$}  denote the smallest  value  among the channel states of all the {$K_n(\mathbf{X})$} users in $\mathcal{K}_n(\mathbf{X})$}. For notation simplicity, if $\mathcal{K}_n(\mathbf{X})=\emptyset$, we define $H_{\texttt{d},n}(\mathbf{X}, \mathbf{H}){=}\infty$.

The random system state consists of the random system service state $\mathbf{X}$, {the random system task input data {size} state ${\mathbf{L}_{\texttt{u}}^{\mathrm{dat}}}$, the random system computation load~state ${\mathbf{L}_{\texttt{e}}^{\mathrm{dat}}}$, the random system computation result {size} state ${\mathbf{L}_{\texttt{d}}^{\mathrm{dat}}}$} and the random system channel state $\mathbf{H}$, {and is} denoted by $\mathbf{Q}\triangleq (\mathbf{X}, { {\mathbf{L}_{\texttt{u}}^{\mathrm{dat}}},{\mathbf{L}_{\texttt{e}}^{\mathrm{dat}}},{\mathbf{L}_{\texttt{d}}^{\mathrm{dat}}},} \mathbf{H})\in \mathcal{Q} \triangleq \mathcal{X}^K \times ({\mathcal{L}_{\texttt{u}}^{\mathrm{dat}}})^K\times ({\mathcal{L}_{\texttt{e}}^{\mathrm{dat}}})^K\times ({\mathcal{L}_{\texttt{e}}^{\mathrm{dat}}})^K \times \mathcal{H}^K$. Let $p_{\mathbf{Q}}(\mathbf{q})\triangleq\Pr[\mathbf{Q}=\mathbf{q}]$ denote the probability of the random system state $\mathbf{Q}$ being $\mathbf{q}\triangleq (\mathbf{x}, { {\mathbf{l}_{\texttt{u}}^{\mathrm{dat}}},{\mathbf{l}_{\texttt{e}}^{\mathrm{dat}}},{\mathbf{l}_{\texttt{d}}^{\mathrm{dat}}},} \mathbf{h}) \in\mathcal{Q}$, where $\mathbf{x}\triangleq (x_k)_{k\in\mathcal{K}} \in\mathcal{X}^K$, {$\mathbf{l}_{\texttt{u}}^{\mathrm{dat}} \triangleq (l_{\texttt{u},k}^{\mathrm{dat}})_{k\in\mathcal{K}}\in ({\mathcal{L}_{\texttt{u}}^{\mathrm{dat}}})^K$, $\mathbf{l}_{\texttt{e}}^{\mathrm{dat}} \triangleq (l_{\texttt{e},k}^{\mathrm{dat}})_{k\in\mathcal{K}}\in ({\mathcal{L}_{\texttt{e}}^{\mathrm{dat}}})^K$, $\mathbf{l}_{\texttt{d}}^{\mathrm{dat}} \triangleq (l_{\texttt{d},k}^{\mathrm{dat}})_{k\in\mathcal{K}}\in ({\mathcal{L}_{\texttt{e}}^{\mathrm{dat}}})^K$ and}  $\mathbf{h}\triangleq (h_k)_{k\in\mathcal{K}} \in\mathcal{H}^K$. Note that $p_{\mathbf{Q}}(\mathbf{q})\ge 0$ for all $\mathbf{q}\in \mathcal{Q}$ and $\sum_{\mathbf{q}\in \mathcal{Q}}p_{\mathbf{Q}}(\mathbf{q})=~1$.  Each user $k$ can inform the serving node of its service index $X_k$ and three task parameters $(L_{\texttt{u},k}^{\mathrm{dat}},L_{\texttt{e},k}^{\mathrm{dat}},L_{\texttt{d},k}^{\mathrm{dat}})$, e.g., via some feedback mechanism. In addition, the {serving node} can easily obtain the channel {state} of each user $k$, i.e., $H_k$, e.g., by channel sounding. Thus, we assume that the {serving node} is aware of the system state~$\mathbf{Q}$.

\subsection{Software Fetching and Caching}
We consider that each service $n\in\mathcal{X}$ is {processed} by a particular software, also {indexed} by $n\in\mathcal{X}$.  Let $l_{\texttt{d},n}^{\mathrm{sfw}}$ denote the size of software $n$ (in bits).   Suppose all software in $\mathcal{X}$ are available at the core network. The serving node is connected to the core network via a backhaul link of {transmission rate} $R$ (in bits per second, bps). {In the MEC system,} the serving node needs to get access to the software {that can process} the {services requested} by the users.  The time duration {for the serving node to fetch software $n$} over the backhaul link is given by $T_{\mathrm{B,\texttt{d} },n}^{\mathrm{sfw}}=  {l_{\texttt{d},n}^{\mathrm{sfw}}}/{R}$, for all $n\in\mathcal{X}$.\footnote{In this paper, we focus on energy consumption at the wireless edge. Thus, we do not consider the energy consumption for fetching a software over the backhaul link.}

The {serving node} is equipped with a cache of size $C$ (in bits). Recall that each service may be demanded by multiple users{. Thus,} the software for each service can be reusable. {To save fetching cost}, we consider caching {some} software in $\mathcal{X}$ at the {serving node}.\footnote{Caching some popular software at wireless edge is highly desirable in some practical scenarios \cite{MECsurveyMao}. For example, users in a museum can use the AR/VR services to get a better immersive experience, and thus, it is desirable to cache those software related to some popular AR/VR services in advance at the serving node in this area to provide higher quality services. Another example is that users are likely to play some online video games at certain times (e.g., after dinner or before sleep), and thus, it is wise for the game providers to cache some popular video game software at the serving node to reduce the huge computation loads to the users.}  {Let $c_n$ denote the caching action for software $n$ at the {serving node}, where
{\setlength{\arraycolsep}{0.0em}
\begin{eqnarray}
c_n\in\{0,1\}, \; n\in\mathcal{X}.\label{eqconstrcn1}
\end{eqnarray}\setlength{\arraycolsep}{5pt}}Here, $c_n = 1$ means that software $n$ is cached, and $c_n =~0$ otherwise.  Note that if software $n$ has been cached at the serving node, i.e., $c_n=1$, fetching software $n$ from the core network is no longer necessary; otherwise, the server needs to fetch software $n$ when there is some user demanding service $n$.}  Under the cache size constraint at the {serving node}, we~have
{\setlength{\arraycolsep}{0.0em}
\begin{eqnarray}
\sum _{n\in\mathcal{X}}c_nl_{\texttt{d},n}^{\mathrm{sfw}}\le C.\label{eqconstrcn2}
\end{eqnarray}\setlength{\arraycolsep}{5pt}}Let $\mathbf{c}\triangleq(c_n)_{n\in\mathcal{X}}$ denote  the system caching action.

\begin{Remark}[Software Fetching and Caching]
The authors in \cite{MECAccessXu2017, MeccachingTVT2017Zhou,  4CNdikumanarxiv2018, MeccachingIWCMC2018Zhao,  MeccachingTVT2018Liu, EECachMEC2018Hao, 2018arXiv181007797T, MECCachingCui2017} implicitly assume that an MEC server  can process whatever types of tasks without considering the availability of software at the wireless edge. The authors in \cite{MECcachingXuJie2018arxiv, XuJie2018MECserviceCach} consider caching the software at the serving node{, but do not consider} fetching {uncached software} via the backhaul link. {Thus,} the tasks corresponding to the service processed by an uncached software have to be transmitted {via the backhaul link} to a remote cloud for executing{, failing to exploit potential advantages} of MEC systems.
\end{Remark}

\subsection{Computation}

The serving node has computing capability by running a server with fixed CPU frequency $F_{\mathrm{{sn}}}$ (in number of CPU-cycles per second). Each user $k$ also has computing capability with fixed CPU frequency $F_k$ (in number of CPU-cycles per second). Usually, $F_{\mathrm{{sn}}}$ is much larger than $F_k$, $k\in\mathcal{K}$. Task $k$ can be executed either remotely at the serving node via computation offloading or locally at {user $k$.} {Different from \cite{4CNdikumanarxiv2018, MECAccessXu2017, MECcachingBitrateTMC2018, MeccachingTVT2017Zhou, MECcachingXuJie2018arxiv, 2018arXiv181007797T, MECCachingCui2017}, we will consider time durations for executing tasks at both the serving node and the users, so as to properly model computation-intensive tasks in practice.} Let $o_k$ denote the computation offloading action of user $k$, where
{\setlength{\arraycolsep}{0.0em}
\begin{eqnarray}
o_k\in\{0,1\},\; k\in\mathcal{K}.\label{eqconstrok}
\end{eqnarray}\setlength{\arraycolsep}{5pt}}Here, $o_k=1$ means that task $k$ is executed at the {serving node} and $o_k=0$ means that task $k$ is executed at user $k$. Let $\mathbf{o}\triangleq \left(o_k\right)_{k\in\mathcal{K}}$ denote the  computation offloading action.

\subsubsection{Computing at Serving Node}

In the case that task $k$ is executed at the serving node, {i.e., $o_k=1$}, there are three stages: i) user $k$ uploads the task input data of $L_{\texttt{u},k}^{\mathrm{dat}}$ bits to the {serving node}, ii) the {serving node} executes task $k$ {(which requires $L_{\texttt{e},k}^{\mathrm{dat}}$ CPU-cycles)}, and iii) {the serving node transmits} the computation result of $L_{\texttt{d},k}^{\mathrm{dat}}$ bits {to user $k$}. {The time duration and} energy consumption for executing task $k$ at the serving node {are given by ${L_{\texttt{e},k}^{\mathrm{dat}}} / {F_{\mathrm{{sn}}}}$ (in seconds) and} $\mu_{\mathrm{sn}} L_{\texttt{e},k}^{\mathrm{dat}} F_{\mathrm{sn}}^2$ {(in Joule)}, respectively, where $\mu_{\mathrm{sn}}$ is a constant factor  determined by the switched capacitance of the MEC server \cite{MECsurveyMao}.

Let $t_{\texttt{u},k}^{\mathrm{dat}}$ ($t_{\texttt{d},k}^{\mathrm{dat}}$) denote the time duration for transmitting the task input data (computation result) of user $k$ and
$p_{\texttt{u},k}^{\mathrm{dat}}$ {(in Watt)} the transmission power of user $k$. Then, the {maximum} achievable transmission rate {(in bps)} is $r_{\texttt{u},k}=B\log_2\left(1+ {p_{\texttt{u},k}^{\mathrm{dat}} H_k}/{n_0}\right)$, where $n_0$ is the power of the complex additive white Gaussian noise {over the entire bandwidth $B$}. On the other hand, the transmission rate should be fixed as $r_{\texttt{u},k}= {L_{\texttt{u},k}^{\mathrm{dat}}} /{t_{\texttt{u},k}^{\mathrm{dat}}}$, since this is the most energy-efficient transmission method for transmitting  ${L_{\texttt{u},k}^{\mathrm{dat}}}$ bits in ${t_{\texttt{u},k}^{\mathrm{dat}}}$ seconds \cite{MECYouTWC}. Define $g(x)\triangleq n_0(2^{ {x}/{B}}-1)$. Then, we have $
p_{\texttt{u},k}^{\mathrm{dat}}=\frac{1}{H_k}g\left({ L_{\texttt{u},k}^{\mathrm{dat}}}/{t_{\texttt{u},k}^{\mathrm{dat}}}\right)$. Thus, the energy consumption at user $k$ for uploading the {task input data} {of $L_{\texttt{u},k}^{\mathrm{dat}}$ bits} to the {serving node} is given by{
\begin{equation*}
E_{\texttt{u},k}^{\mathrm{dat}}(t_{\texttt{u},k}^{\mathrm{dat}}, L_{\texttt{u},k}^{\mathrm{dat}}, H_k ) =
\begin{cases}
\frac{t_{\texttt{u},k}^{\mathrm{dat}}}{{H_k}}g\left( {\frac{{ L_{\texttt{u},k}^{\mathrm{dat}}}}{{ t_{\texttt{u},k}^{\mathrm{dat}}}}}\right), & \mbox{if } t_{\texttt{u},k}^{\mathrm{dat}}>0, \\
0, & \mbox{otherwise}.
\end{cases}
\end{equation*}}Similarly, the energy consumption at the serving node for transmitting the computation result {of $L_{\texttt{d},k}^{\mathrm{dat}}$ bits} to user $k$ is {
\begin{equation*}
E_{\texttt{d},k}^{\mathrm{dat}}(t_{\texttt{d},k}^{\mathrm{dat}}, L_{\texttt{d},k}^{\mathrm{dat}}, H_k ) =
\begin{cases} \frac{t_{\texttt{d},k}^{\mathrm{dat}}}{{H_k}}g\left( \frac{L_{\texttt{d},k}^{\mathrm{dat}}}{ t_{\texttt{d},k}^{\mathrm{dat}}} \right), & \mbox{if } t_{\texttt{d},k}^{\mathrm{dat}}>0, \\
0, & \mbox{otherwise}.
\end{cases}
\end{equation*}}

\subsubsection{Computing at Users}
{In the case that task $k$ is executed at user $k$,  i.e., $o_k=0$, there are two stages: i) the serving node transmits the corresponding software $X_k$ to user $k$, ii) user $k$ executes task $k$ locally.  To avoid  transmitting the same software to multiple users {over multiple times} for energy saving, we consider {software} multicasting.} {In particular, we multicast software $n$ to the users in $\{k\in\mathcal{K}_n(\mathbf{X}):o_k=0\}$.} As a matter of fact, both multicast and unicast may happen (with different probabilities). Without loss of generality, we {still} refer to this transmission as multicast.   The time duration and  energy consumption for executing task $k$ at user $k$ are given by ${L_{\texttt{e},k} ^{\mathrm{dat}}}/{F_k}$ and $\mu_k L_{\texttt{e},k}^{\mathrm{dat}} F_k^2$, respectively, where $\mu_k$ is a constant factor  determined by the switched capacitance of the device of user $k$ \cite{MECsurveyMao}.

Let $t_{\texttt{d},n}^{\mathrm{sfw}}$ denote the time duration for multicasting software $n$. Then, the energy consumption at the serving node for {multicasting} software $n$ of $l_{\texttt{d},n}^{\mathrm{sfw}}$ bits  is given by{
\begin{equation*}
E_{\texttt{d},n}^{\mathrm{sfw}} (t_{\texttt{d},n}^{\mathrm{sfw}},\mathbf{X,H} ) =
\begin{cases} \frac{t_{\texttt{d},n}^{\mathrm{sfw}}}{{H_{\texttt{d},n}(\mathbf{X},\mathbf{H})}}g\left( \frac{l_{\texttt{d},n}^{\mathrm{sfw}}}{ t_{\texttt{d},n}^{\mathrm{sfw}}} \right), & \mbox{if } {t_{\texttt{d},n}^{\mathrm{sfw}}}>0, \\
0, & \mbox{otherwise}.
\end{cases}
\end{equation*}}

\begin{Remark}[Software Multicasting]
To the best of our knowledge, software multicasting in MEC systems has {\-never} been considered in existing literature. It enjoys more multicasting opportunities than multicasting of computation results proposed in our previous work \cite{MECCachingCui2017}, as the computation results of different tasks are usually not the same, even if they correspond to the same service.
\end{Remark}

\subsection{Deadline Constraints and Total Energy Consumption}
{For ease of implementation, we consider three successive phases, for fetching software, multicasting software, and computation, respectively, as illustrated in Fig. \ref{figsystemodel}.} {Under the considered frame structure in Fig. \ref{figsystemodel}}, we have two deadline constraints, as shown in (\ref{eqconstrSumTimeTDMA}) and (\ref{eqconstrSumTimeTDMAlocaluser}), where
\begin{figure*}[!t]
{\begingroup\makeatletter\def\f@size{9.5}\check@mathfonts
\def\maketag@@@#1{\hbox{\m@th\normalsize\normalfont#1}}\setlength{\arraycolsep}{0.0em}
\begin{eqnarray}
&&\sum\limits_{n\in\mathcal{X}}   \left( \mathbbm{1}\left(K_n(\mathbf{X})\ge 1\right)(1-c_n)T_{\mathrm{B,\texttt{d} },n}^{\mathrm{sfw}} + \left(\max_{k\in\mathcal{K}_n(\mathbf{X})} (1-o_k)\right) t_{\texttt{d},n}^{\mathrm{sfw}} \right)  + \sum\limits_{k\in\mathcal{K}}o_k \left(t_{\texttt{u},k}^{\mathrm{dat}} +  {L_{\texttt{e},k}^{\mathrm{dat}}} / {F_{\mathrm{{sn}}}} + t_{\texttt{d},k}^{\mathrm{dat}}  \right) \le D,\label{eqconstrSumTimeTDMA}\\
&&\sum\limits_{n\in\mathcal{X}}  \left( \mathbbm{1}\left(K_n(\mathbf{X})\ge 1\right)(1-c_n)T_{\mathrm{B,\texttt{d} },n}^{\mathrm{sfw}} + \left(\max_{k\in\mathcal{K}_n(\mathbf{X})} (1-o_k)\right) t_{\texttt{d},n}^{\mathrm{sfw}} \right)  + (1-o_k)  {L_{\texttt{e},k} ^{\mathrm{dat}}} / {F_k} \le D,\;   k\in\mathcal{K}.\label{eqconstrSumTimeTDMAlocaluser}
\end{eqnarray}\setlength{\arraycolsep}{5pt}\endgroup}
\noindent\rule[0.15\baselineskip]{\textwidth}{0.1pt}
\end{figure*}{\setlength{\arraycolsep}{0.0em}
\begin{eqnarray}
&&0\le t_{\texttt{u},k}^{\mathrm{dat}} \le o_kD,   \quad k\in\mathcal{K},  \label{eqconstrtuk}\\
&&0\le t_{\texttt{d},k}^{\mathrm{dat}} \le o_kD,   \quad k\in\mathcal{K},  \label{eqconstrtdk}\\
&&0\le t_{\texttt{d},n}^{\mathrm{sfw}} \le  \max_{k\in\mathcal{K}_n(\mathbf{X})}(1-o_k) D, \quad n\in\mathcal{X}.\label{eqconstrtdksfw}
\end{eqnarray}\setlength{\arraycolsep}{5pt}}{Here, $\max_{k\in\mathcal{K}_n(\mathbf{X})} (1-o_k)$ indicates whether software $n$ is requested by some users, i.e., $\max_{k\in\mathcal{K}_n(\mathbf{X})} (1-o_k)=1$ means that software $n$ is requested by some users, and $\max_{k\in\mathcal{K}_n(\mathbf{X})} (1-o_k)=0$, otherwise. To facilitate understanding of the derivation of the deadline constraints in (\ref{eqconstrSumTimeTDMA}) and (\ref{eqconstrSumTimeTDMAlocaluser}), we point that} the first terms in (\ref{eqconstrSumTimeTDMA}) and (\ref{eqconstrSumTimeTDMAlocaluser}) are the same, representing the total time duration for software fetching and multicasting. The second term in (\ref{eqconstrSumTimeTDMA}) {represents the total time duration for uploading, executing and downloading {the} tasks being offloaded. The second term in (\ref{eqconstrSumTimeTDMAlocaluser}) represents the time duration for executing task $k$ at user $k$, which is zero if task  $k$ is offloaded.} {Please also note that, due to the consideration of software fetching, caching and multicasting, as well as task input data uploading, task executing (with non-negligible time duration) and computation result downloading, the deadline constraints in (\ref{eqconstrSumTimeTDMA}) and (\ref{eqconstrSumTimeTDMAlocaluser}) are more {comprehensive} than those in \cite{MeccachingAccess2018Yang, MECcachingICC2018Sun, MECcachingBitrateTMC2018, MECAccessXu2017, MeccachingTVT2017Zhou,  4CNdikumanarxiv2018, MeccachingIWCMC2018Zhao,  MeccachingTVT2018Liu, EECachMEC2018Hao, 2018arXiv181007797T, MECCachingCui2017, MECcachingXuJie2018arxiv, XuJie2018MECserviceCach}.}

Finally, the weighted sum energy consumption {(at the wireless edge)} is given~by (\ref{eqTDMAtotalEngery}),
{\begingroup\makeatletter\def\f@size{9.5}\check@mathfonts
\def\maketag@@@#1{\hbox{\m@th\normalsize\normalfont#1}}
\begin{figure*}[t]
\begin{align}\label{eqTDMAtotalEngery}
E( {{\mathbf{o}},{\mathbf{t}}_{\texttt{u}}^{\mathrm{dat}}, {\mathbf{t}}_{\texttt{d}}^{\mathrm{dat}}, {\mathbf{t}}_{\texttt{d}}^{\mathrm{sfw}},  \mathbf{Q}} ) &=  \sum _{n \in {\cal X}} \left(\max_{k\in\mathcal{K}_n(\mathbf{X})} (1-o_k)\right) E_{\texttt{d},n}^{\mathrm{sfw}}( t_{\texttt{d},n}^{\mathrm{sfw}},\mathbf{X,H}) + \sum _{k \in {\cal K}}  {o_k}\Big( {\omega_k}E_{\texttt{u},k}^{\mathrm{dat}} ( {t_{\texttt{u},k}^{\mathrm{dat}}, L_{\texttt{u},k}^{\mathrm{dat}}, {H_k}}  )\nonumber\\
 & + \mu_{\mathrm{sn}} L_{\texttt{e},k}^{\mathrm{dat}} F_{\mathrm{sn}}^2 +  E_{\texttt{d},k}^{\mathrm{dat}} ( t_{\texttt{d},k}^{\mathrm{dat}}, L_{\texttt{d},k}^{\mathrm{dat}}, H_k  ) \Big) + \sum _{k \in {\cal K}} (1 - {o_k}){\omega_k}\mu_k L_{\texttt{e},k}^{\mathrm{dat}} F_k^2.
\end{align}
\noindent\rule[0\baselineskip]{\textwidth}{0.1pt}
\end{figure*}\endgroup}
where  $\mathbf{t}_{\texttt{u}}^{\mathrm{dat}}\triangleq (t_{\texttt{u},k}^{\mathrm{dat}})_{k\in\mathcal{K}}$, $\mathbf{t}_{\texttt{d}}^{\mathrm{dat}}\triangleq (t_{\texttt{d},k}^{\mathrm{dat}})_{k\in\mathcal{K}}$ and $\mathbf{t}_{\texttt{d}}^{\mathrm{sfw}}\triangleq (t_{\texttt{d},n}^{\mathrm{sfw}})_{n\in\mathcal{X}}$. In addition, in (\ref{eqTDMAtotalEngery}), $\omega_k>0$  {represents} the weight factor for user $k$. A larger $\omega_k$ means imposing a higher cost on energy consumption at user $k$ due to the limited battery power. Note that, if $\omega_k=1$ for all $k\in\mathcal{K}$, the average weighted sum energy consumption $E( {{\mathbf{o}},{\mathbf{t}}_{\texttt{u}}^{\mathrm{dat}}, {\mathbf{t}}_{\texttt{d}}^{\mathrm{dat}}, {\mathbf{t}}_{\texttt{d}}^{\mathrm{sfw}},  \mathbf{Q}} )$ represents the average total energy consumption (in Joule). Furthermore, the {three summation terms}  in (\ref{eqTDMAtotalEngery}) {correspond} to the energy consumptions for software downloading, tasks being offloaded and tasks being computed locally, respectively.

\section{Problem Formulation and Transformation}\label{subsectionTDMAProbForum}
In this section, we first formulate an energy minimization problem. {Then, we} transform it into a {convex mixed integer nonlinear programming (MINLP)} problem.

\subsection{Problem Formulation}
In this part, we {formulate} the energy minimization problem.  {First, we define} a feasible joint caching, offloading and time allocation policy.
\begin{Definition}[Feasible Joint Policy]\label{defTDMA} We consider a joint caching, offloading and time allocation policy $(\mathbf{c}, \mathbf{O}, \mathbf{T}_{\texttt{u}}^{\mathrm{dat}}, \mathbf{T}_{\texttt{d}}^{\mathrm{dat}}, \mathbf{T}_{\texttt{d}}^{\mathrm{sfw}})$, where the caching action $\mathbf{c}$ does not change with the system state $\mathbf{Q}$, and the offloading and time allocation $(\mathbf{O}, \mathbf{T}_{\texttt{u}}^{\mathrm{dat}}, \mathbf{T}_{\texttt{d}}^{\mathrm{dat}}, \mathbf{T}_{\texttt{d}}^{\mathrm{sfw}})$ is a vector mapping (i.e., function) from the system state $\mathbf{Q}$ to {the offloading and time allocation actions} $(\mathbf{o}, \mathbf{t}_{\texttt{u}}^{\mathrm{dat}}, \mathbf{t}_{\texttt{d}}^{\mathrm{dat}}, \mathbf{t }_{\texttt{d}}^{\mathrm{sfw}})$, i.e., $\mathbf{o}=\mathbf{O}(\mathbf{Q})$, $\mathbf{t}_{\texttt{u}}^{\mathrm{dat}} =\mathbf{T}_{\texttt{u}}^{\mathrm{dat}}(\mathbf{Q})$, $\mathbf{t}_{\texttt{d}}^{\mathrm{dat}} = \mathbf{T}_{\texttt{d}}^{\mathrm{dat}} (\mathbf{Q})$ and $\mathbf{t}_{\texttt{d}}^{\mathrm{sfw}} = \mathbf{T}_{\texttt{d}}^{\mathrm{sfw}} (\mathbf{Q})$. Here, $ \mathbf{O} \triangleq (O_{k} )_{k\in\mathcal{K}}$, $ \mathbf{T}_{\texttt{u}}^{\mathrm{dat}} \triangleq (T_{\texttt{u}, k}^{\mathrm{dat}} )_{k\in\mathcal{K}}$, $ \mathbf{T}_{\texttt{d}}^{\mathrm{dat}} \triangleq (T_{\texttt{d}, k}^{\mathrm{dat}} )_{k\in\mathcal{K}}$ and $ \mathbf{T}_{\texttt{d}}^{\mathrm{sfw}} \triangleq (T_{\texttt{d}, n}^{\mathrm{sfw}} )_{n\in\mathcal{X}}$. We call  a policy $(\mathbf{c}, \mathbf{O}, \mathbf{T}_{\texttt{u}}^{\mathrm{dat}}, \mathbf{T}_{\texttt{d}}^{\mathrm{dat}}, \mathbf{T}_{\texttt{d}}^{\mathrm{sfw}})$ feasible, if the caching action $\mathbf{c}$ satisfies (\ref{eqconstrcn1}) and (\ref{eqconstrcn2}), and the offloading and time allocation actions $(\mathbf{o}, \mathbf{t}_{\texttt{u}}^{\mathrm{dat}}, \mathbf{t}_{\texttt{d}}^{\mathrm{dat}}, \mathbf{t}_{\texttt{d}}^{\mathrm{sfw}})$ at each $\mathbf{Q}$  together with $\mathbf{c}$ satisfy (\ref{eqconstrok}), (\ref{eqconstrSumTimeTDMA}), (\ref{eqconstrSumTimeTDMAlocaluser}), (\ref{eqconstrtuk}), (\ref{eqconstrtdk}) and (\ref{eqconstrtdksfw}).
\end{Definition}

\begin{Remark}[Interpretation of Definition \ref{defTDMA}] \label{remarkInterpre} {\cite{MECCachingCui2017}}
Caching generally follows a much larger timescale (e.g., hours or days) and should reflect statistics of the MEC system. On the contrary, the computation offloading and the time allocation actions evolve at a much shorter timescale (e.g., miliseconds) and should utilize instantaneous information of the MEC system. As a result, we assume in Definition~1 that the caching action only depends on the probability mass functions $p_{\mathbf{Q}}(\mathbf{q})$, $\mathbf{q}\in \mathcal{Q}$, and does not change with the $\mathbf{Q}$, while the computation offloading and time allocation actions are adaptive to $\mathbf{Q}$.
\end{Remark}

Under a feasible joint policy $(\mathbf{c}, \mathbf{O}, \mathbf{T}_{\texttt{u}}^{\mathrm{dat}}, \mathbf{T}_{\texttt{d}}^{\mathrm{dat}}, \mathbf{T}_{\texttt{d}}^{\mathrm{sfw}})$, the average weighted sum energy consumption is given by
\begin{align}
&\overline{E}( {{\mathbf{O}},{\mathbf{T}}_{\texttt{u}}^{\mathrm{dat}}, {\mathbf{T}}_{\texttt{d}}^{\mathrm{dat}}, {\mathbf{T}}_{\texttt{d}}^{\mathrm{sfw}}} ) \nonumber\\
&= \mathbb{E} \left[E( {{\mathbf{O}(\mathbf{Q})}, {\mathbf{T}}_{\texttt{u}}^{\mathrm{dat}}(\mathbf{Q}), {\mathbf{T}}_{\texttt{d}}^{\mathrm{dat}}(\mathbf{Q}), {\mathbf{T}}_{\texttt{d}}^{\mathrm{sfw}}(\mathbf{Q}), \mathbf{Q}} ) \right],\label{eqTDMAaverageSumEnergy}
\end{align}where the expectation $\mathbb{E}$ is taken over the random system state $\mathbf{Q}\in \mathcal{Q}$ and $E(\cdot)$ is given by~(\ref{eqTDMAtotalEngery}).
In this paper, we would like to minimize {the average weighted sum energy consumption in} (\ref{eqTDMAaverageSumEnergy}). Specifically, we have the following problem.\footnote{Although Problem \ref{eqTDMAPrimalProblem} is for time interval $[0,D]$, a solution of Problem \ref{eqTDMAPrimalProblem} can be applied to a practical MEC system over a long time during which {the statistics of the random system state does not change.}}

\begin{Problem}[Energy Minimization]\label{eqTDMAPrimalProblem}
{\begingroup\makeatletter\def\f@size{9.3}\check@mathfonts
\def\maketag@@@#1{\hbox{\m@th\normalsize\normalfont#1}}\setlength{\arraycolsep}{0.0em}
\begin{eqnarray}
&&\overline{E}^* \triangleq \min_{\substack{ {\mathbf{c}, {\mathbf{O}},
{\mathbf{T}}_{\texttt{u}}^{\mathrm{dat}}, {\mathbf{T}}_{\texttt{d}}^{\mathrm{dat}}, {\mathbf{T}}_{\texttt{d}}^{\mathrm{sfw}}}}} \overline{E}( {{\mathbf{O}}, {\mathbf{T}}_{\texttt{u}}^{\mathrm{dat}}, {\mathbf{T}}_{\texttt{d}}^{\mathrm{dat}}, {\mathbf{T}}_{\texttt{d}}^{\mathrm{sfw}}} )\nonumber\\
\mathrm{s.t.}\;&&
 (\ref{eqconstrcn1}),(\ref{eqconstrcn2}),\nonumber\\
&&  O_k(\mathbf{Q}) \in \{0, 1\},\label{eqconstrOkQstateprimal}\\
&&  0 \le T_{\texttt{u},k}^{\mathrm{dat}}(\mathbf{Q}) \le  O_k(\mathbf{Q})D, \label{eqconstrTuk} \\
&&  0 \le T_{\texttt{d},k}^{\mathrm{dat}}(\mathbf{Q}) \le  O_k(\mathbf{Q})D, \label{eqconstrTdk} \\
&&  0 \le T_{\texttt{d},n}^{\mathrm{sfw}}(\mathbf{Q}) \le  \max\nolimits_{k\in\mathcal{K}_n(\mathbf{X})}(1-O_k(\mathbf{Q}))D, \label{eqconstrTdn} \\
&& \sum _{n\in\mathcal{X}} \Big(\mathbbm{1}\left(K_n(\mathbf{X})\ge 1\right)(1-c_n)T_{\mathrm{B,\texttt{d} },n}^{\mathrm{sfw}} \nonumber\\
&& \qquad+ \left(\max\nolimits_{k\in\mathcal{K}_n(\mathbf{X})} (1-O_k(\mathbf{Q}))\right) T_{\texttt{d},n}^{\mathrm{sfw}}\Big) \label{eqconstrTotalTimeAllusersQstateprimal}\\
  &&+ \sum_{k\in\mathcal{K}}O_k(\mathbf{Q}) \left(T_{\texttt{u},k}^{\mathrm{dat}}(\mathbf{Q})  + {L_{\texttt{e},k}^{\mathrm{dat}}} / {F_{\mathrm{{sn}}}} + T_{\texttt{d},k}^{\mathrm{dat}}(\mathbf{Q}) \right)\le D,\nonumber\\
&& \sum _{n\in\mathcal{X}} \Big(\mathbbm{1}\left(K_n(\mathbf{X})\ge 1\right)(1-c_n)T_{\mathrm{B,\texttt{d} },n}^{\mathrm{sfw}} \nonumber\\
&& \qquad+ \left(\max\nolimits_{k\in\mathcal{K}_n(\mathbf{X})} (1-O_k(\mathbf{Q}))\right) T_{\texttt{d},n}^{\mathrm{sfw}}\Big) \label{eqconstrTotalTimeAllusersQstateprimal2}\\
&&\quad  + (1-O_k(\mathbf{Q}))  {L_{\texttt{e},k} ^{\mathrm{dat}}}/{F_k} \le D, \nonumber
\end{eqnarray}\setlength{\arraycolsep}{5pt}\endgroup}
\end{Problem}where $k\in\mathcal{K}$, $n\in\mathcal{X}$, $\mathbf{Q}\in\mathcal{Q}$, and $\overline{E}(\cdot)$ is given by  (\ref{eqTDMAaverageSumEnergy}). Let $({\mathbf{c}^{*}, {\mathbf{O}^{*}}, {\mathbf{{T}}}_{\texttt{u}}^{\mathrm{dat}^{*}}, {\mathbf{{T}}}_{\texttt{d}}^{\mathrm{dat}^{*}}, {\mathbf{{T}}}_{\texttt{d}}^{\mathrm{sfw}^{*}}})$ denote an optimal solution of Problem~\ref{eqTDMAPrimalProblem}.

Problem \ref{eqTDMAPrimalProblem} is a challenging two-timescale {non-convex} MINLP problem and is NP-hard in general \cite{BURER201297}.

\subsection{Problem Transformation}

In this part, using several optimization techniques, we transform Problem \ref{eqTDMAPrimalProblem} to an equivalent {convex} MINLP problem.  Let $(\mathbf{Y}, \mathbf{\hat{T}}_{\texttt{u}}^{\mathrm{dat}}, \mathbf{\hat{T}}_{\texttt{d}}^{\mathrm{dat}}, \mathbf{\hat{T}}_{\texttt{d}}^{\mathrm{sfw}})$ denote a vector mapping (i.e., function) from the system state $\mathbf{Q}$ to $(\mathbf{y}, \mathbf{\hat{t}}_{\texttt{u}}^{\mathrm{dat}}, \mathbf{\hat{t}}_{\texttt{d}}^{\mathrm{dat}}, \mathbf{\hat{t}}_{\texttt{d}}^{\mathrm{sfw}})$, i.e., $\mathbf{y}=\mathbf{Y}(\mathbf{Q})$, $\mathbf{\hat{t}}_{\texttt{u}}^{\mathrm{dat}} =\mathbf{\hat{T}}_{\texttt{u}}^{\mathrm{dat}}(\mathbf{Q})$, $\mathbf{\hat{t}}_{\texttt{d}}^{\mathrm{dat}} = \mathbf{\hat{T}}_{\texttt{d}}^{\mathrm{dat}} (\mathbf{Q})$ and $\mathbf{\hat{t}}_{\texttt{d}}^{\mathrm{sfw}} = \mathbf{\hat{T}}_{\texttt{d}}^{\mathrm{sfw}} (\mathbf{Q})$, where $\mathbf{y} \triangleq (y_n)_{n\in\mathcal{X}}$, $\mathbf{\hat{t}}_{\texttt{u}}^{\mathrm{dat}} \triangleq (\hat{t}_{\texttt{u},k}^{\mathrm{dat}} )_{k\in\mathcal{K}}$, $\mathbf{\hat{t}}_{\texttt{d}}^{\mathrm{dat}} \triangleq (\hat{t}_{\texttt{d},k}^{\mathrm{dat}} )_{k\in\mathcal{K}}$, $\mathbf{\hat{t}}_{\texttt{d}}^{\mathrm{sfw}} \triangleq (\hat{t}_{\texttt{d},n}^{\mathrm{sfw}}  )_{n\in\mathcal{X}}$, $ \mathbf{Y} \triangleq (Y_n )_{n\in\mathcal{X}}$, $ \mathbf{\hat{T}}_{\texttt{u}}^{\mathrm{dat}} \triangleq (\hat{T}_{\texttt{u}, k}^{\mathrm{dat}} )_{k\in\mathcal{K}}$, $ \mathbf{\hat{T}}_{\texttt{d}}^{\mathrm{dat}} \triangleq (\hat{T}_{\texttt{d}, k}^{\mathrm{dat}} )_{k\in\mathcal{K}}$ and $ \mathbf{\hat{T}}_{\texttt{d}}^{\mathrm{sfw}} \triangleq (\hat{T}_{\texttt{d}, n}^{\mathrm{sfw}} )_{n\in\mathcal{X}}$. In addition, define \begin{align}\label{eqTDMAtotalEngeryrew2obj2}
&\overline{e} ( {{\mathbf{O}}, {\mathbf{Y}}, {\mathbf{\hat{T}}}_{\texttt{u}}^{\mathrm{dat}}, {\mathbf{\hat{T}}}_{\texttt{d}}^{\mathrm{dat}}, {\mathbf{\hat{T}}}_{\texttt{d}}^{\mathrm{sfw}}}  ) \\
&\triangleq  \mathbb{E} [e  (  {\mathbf{O}(\mathbf{Q})},  {\mathbf{Y}}(\mathbf{Q}), {\mathbf{\hat{T}}}_{\texttt{u}}^{\mathrm{dat}}(\mathbf{Q}), {\mathbf{\hat{T}}}_{\texttt{d}}^{\mathrm{dat}}(\mathbf{Q}),  {\mathbf{\hat{T}}}_{\texttt{d}}^{\mathrm{sfw}}(\mathbf{Q}),  \mathbf{Q}   ) ],\nonumber
\end{align}where $e  (\cdot)$ is given by (\ref{eqTDMAtotalEngeryrew}) with $e_{\texttt{u},k}^{\mathrm{dat}} ( o_k, \hat{t}_{\texttt{u},k}^{\mathrm{dat}}, L_{\texttt{u},k}^{\mathrm{dat}}, H_k ) \triangleq {\hat{t}_{\texttt{u},k}^{\mathrm{dat}}}/{H_k} g\left( {o_kL_{\texttt{u},k}^{\mathrm{dat}}} / {\hat{t}_{\texttt{u},k}^{\mathrm{dat}}}\right)$, $
e_{\texttt{d},k}^{\mathrm{dat}} ( o_k, \hat{t}_{\texttt{d},k}^{\mathrm{dat}}, L_{\texttt{d},k}^{\mathrm{dat}}, H_k ) \triangleq {\hat{t}_{\texttt{d},k}^{\mathrm{dat}}}/{H_k} g\left( {o_kL_{\texttt{d},k}^{\mathrm{dat}}} / {\hat{t}_{\texttt{d},k}^{\mathrm{dat}}}\right)$ and $
e_{\texttt{d},n}^{\mathrm{sfw}} ( y_n, \hat{t}_{\texttt{d},n}^{\mathrm{sfw}}, \mathbf{X,H})  \triangleq  {\hat{t}_{\texttt{d},n}^{\mathrm{sfw}}} / {{H_{\texttt{d},n}(\mathbf{X,H})}} g\left(  { y_n l_{\texttt{d},n}^{\mathrm{sfw}}} / { \hat{t}_{\texttt{d},n}^{\mathrm{sfw}} } \right)$. Then, we introduce the following problem.
\begin{figure*}[!t]
\begin{align}
e(  {\mathbf{o}},  {\mathbf{y}}, {\mathbf{\hat{t}}}_{\texttt{u}}^{\mathrm{dat}}, {\mathbf{\hat{t}}}_{\texttt{d}}^{\mathrm{dat}},  {\mathbf{\hat{t}}}_{\texttt{d}}^{\mathrm{sfw}},  \mathbf{Q}   ) & \triangleq   \sum _{n \in {\cal X}} e_{\texttt{d},n}^{\mathrm{sfw}}(y_n, \hat{t}_{\texttt{d},n}^{\mathrm{sfw}},\mathbf{X,H}) + \sum_{k \in {\cal K}}  \Big( {\omega_k}e_{\texttt{u},k}^{\mathrm{dat}}({o_k}, {t_{\texttt{u},k}^{\mathrm{dat}}, L_{\texttt{u},k}^{\mathrm{dat}}, {H_k}} )\nonumber \\
& + {o_k}\mu_{\mathrm{sn}} L_{\texttt{e},k}^{\mathrm{dat}} F_{\mathrm{sn}}^2 + e_{\texttt{d},k}^{\mathrm{dat}}({o_k}, \hat{t}_{\texttt{d},k}^{\mathrm{dat}} , L_{\texttt{d},k}^{\mathrm{dat}}, H_k ) \Big)   + \sum_{k \in {\cal K}} (1 - {o_k}){\omega_k}\mu_k L_{\texttt{e},k}^{\mathrm{dat}} F_k^2.\label{eqTDMAtotalEngeryrew}
\end{align}
\noindent\rule[0.0\baselineskip]{\textwidth}{0.1pt}
\end{figure*}

\begin{Problem}[Equivalent Problem of Problem \ref{eqTDMAPrimalProblem}]\label{eqTDMAPrimalProblem2}
{\begingroup\makeatletter\def\f@size{9.0}\check@mathfonts
\def\maketag@@@#1{\hbox{\m@th\normalsize\normalfont#1}}\setlength{\arraycolsep}{0.0em}
\begin{align}
& \min_{\substack{{\mathbf{c}, {\mathbf{O}}, {\mathbf{Y}},
{\mathbf{\hat{T}}}_{\texttt{u}}^{\mathrm{dat}}, {\mathbf{\hat{T}}}_{\texttt{d}}^{\mathrm{dat}}, {\mathbf{\hat{T}}}_{\texttt{d}}^{\mathrm{sfw}}}}} \overline{e} ( {{\mathbf{O}}, {\mathbf{Y}}, {\mathbf{\hat{T}}}_{\texttt{u}}^{\mathrm{dat}}, {\mathbf{\hat{T}}}_{\texttt{d}}^{\mathrm{dat}}, {\mathbf{\hat{T}}}_{\texttt{d}}^{\mathrm{sfw}}}  )\nonumber\\
\mathrm{s.t.}&
\; (\ref{eqconstrcn1}), (\ref{eqconstrcn2}), (\ref{eqconstrOkQstateprimal}),\nonumber\\
& \begin{cases}\label{eqconstrynmaxQstate}
     1-O_i(\mathbf{Q}) \le Y_n(\mathbf{Q}) \le 1, \; i\in\mathcal{K}_n(\mathbf{X}), &\mbox{if}\; K_n(\mathbf{X})\ge1,\\
     Y_n(\mathbf{Q}) = 0,&\mbox{if}\; K_n(\mathbf{X}) = 0,
    \end{cases}\\
&   0\le {{\hat{T}}}_{\texttt{u},k}^{\mathrm{dat}}(\mathbf{Q}) \le O_k(\mathbf{Q})D,   \label{eqconstrhattukQstate}\\
&  0 \le {{\hat{T}}}_{\texttt{d},k}^{\mathrm{dat}}(\mathbf{Q}) \le O_k(\mathbf{Q})D, \label{eqconstrhattdkQstate}\\
&  0 \le {{\hat{T}}}_{\texttt{d},n}^{\mathrm{sfw}}(\mathbf{Q}) \le Y_n(\mathbf{Q}) D, \label{eqconstrhattdnQstate}\\
& \sum _{n\in\mathcal{X}} \left( \mathbbm{1}\left(K_n(\mathbf{X})\ge 1\right)(1-c_n)T_{\mathrm{B,\texttt{d} },n}^{\mathrm{sfw}} + \hat{T}_{\texttt{d},n}^{\mathrm{sfw}}(\mathbf{Q}) \right) \nonumber\\
&\;  + \sum_{k\in\mathcal{K}} \left(\hat{T}_{\texttt{u},k}^{\mathrm{dat}}(\mathbf{Q}) + O_k(\mathbf{Q}) {L_{\texttt{e},k}^{\mathrm{dat}}} / {F_{\mathrm{{sn}}}} + \hat{T}_{\texttt{d},k}^{\mathrm{dat}}(\mathbf{Q}) \right)\le D,\label{eqconstrsumtasksTimeQstate}\\
& \sum _{n\in\mathcal{X}} \left( \mathbbm{1}\left(K_n(\mathbf{X})\ge 1\right)(1-c_n)T_{\mathrm{B,\texttt{d} },n}^{\mathrm{sfw}} + \hat{T}_{\texttt{d},n}^{\mathrm{sfw}}(\mathbf{Q}) \right) \nonumber\\
&\;     + (1-O_k(\mathbf{Q}))  {L_{\texttt{e},k} ^{\mathrm{dat}}} / {F_k} \le D,\label{eqconstrhatsumTimeTDMAQstate}
\end{align}\endgroup\setlength{\arraycolsep}{5pt}}where $k\in\mathcal{K}$, $n\in\mathcal{X}$ and $\mathbf{Q}\in\mathcal{Q}$. Let $(\mathbf{c}^{\star}, {\mathbf{O}^{\star}}, {\mathbf{Y}^{\star}},  {\mathbf{\hat{T}}}_{\texttt{u}}^{\mathrm{dat}^{\star}}, \\ {\mathbf{\hat{T}}}_{\texttt{d}}^{\mathrm{dat}^{\star}}, {\mathbf{\hat{T}}}_{\texttt{d}}^{\mathrm{sfw}^{\star}})$ denote an optimal solution of Problem~\ref{eqTDMAPrimalProblem2}.
\end{Problem}

Note that the constraints in (\ref{eqconstrhattukQstate}), (\ref{eqconstrhattdkQstate}), (\ref{eqconstrsumtasksTimeQstate}) and (\ref{eqconstrhatsumTimeTDMAQstate}) correspond to the constraints in (\ref{eqconstrTuk}), (\ref{eqconstrTdk}), (\ref{eqconstrTotalTimeAllusersQstateprimal}) and (\ref{eqconstrTotalTimeAllusersQstateprimal2}), and the constraints in (\ref{eqconstrynmaxQstate}) and (\ref{eqconstrhattdnQstate}) correspond to the constraints in (\ref{eqconstrTdn}). Since the objective and constraint functions are convex, Problem~\ref{eqTDMAPrimalProblem2} is a two-timescale {convex} MINLP problem and is generally still NP-hard. The relationship between Problem \ref{eqTDMAPrimalProblem} and Problem \ref{eqTDMAPrimalProblem2} is shown below.

\begin{Lemma}[Equivalence between Problems \ref{eqTDMAPrimalProblem} and  \ref{eqTDMAPrimalProblem2}]\label{lemmaequivalent}
Problems~\ref{eqTDMAPrimalProblem} and \ref{eqTDMAPrimalProblem2} are equivalent, i.e., $c_n^* = c_n^{\star}$, $n\in\mathcal{X}$, $O_k^* (\mathbf{Q}) = O_k^{\star}(\mathbf{Q})$, $T_{\texttt{u},k}^{\mathrm{dat}^*}(\mathbf{Q}) = \hat{T}_{\texttt{u},k}^{\mathrm{dat}^{\star}}(\mathbf{Q})$, $T_{\texttt{d},k}^{\mathrm{dat}^*}(\mathbf{Q}) = \hat{T}_{\texttt{d},k}^{\mathrm{dat}^{\star}}(\mathbf{Q})$, $k\in\mathcal{K}$, $\mathbf{Q}\in\mathcal{Q}$, and $T_{\texttt{d},n}^{\mathrm{sfw}^*}(\mathbf{Q}) = \hat{T}_{\texttt{d},n}^{\mathrm{sfw}^{\star}}(\mathbf{Q})$, $n\in\mathcal{X}$, $\mathbf{Q}\in\mathcal{Q}$.
\end{Lemma}

\indent\indent \emph{Proof:} See Appendix \ref{prooflemmaequivalent}. \hfill $\blacksquare$

Note that, although both {Problems~\ref{eqTDMAPrimalProblem} and \ref{eqTDMAPrimalProblem2}} are NP-hard, Problem~\ref{eqTDMAPrimalProblem2} (convex MINLP) is more tractable than Problem~\ref{eqTDMAPrimalProblem} (non-convex MINLP). In the following two sections, based on Problem \ref{eqTDMAPrimalProblem2}, we develop two low-complexity algorithms to obtain a feasible solution and a stationary point of Problem~\ref{eqTDMAPrimalProblem2}, which have promising {performance} and can be treated as suboptimal solutions of Problem \ref{eqTDMAPrimalProblem2}. Specifically, the first algorithm is based on continuous relaxation and the second one bases on penalty convex-concave procedure {(Penalty-CCP)}.

\section{A Low-complexity Algorithm Based on Continuous Relaxation}
In this section, we develop a low-complexity algorithm to obtain a suboptimal solution of Problem~\ref{eqTDMAPrimalProblem2} based on continuous relaxation. First, we obtain an optimal solution of the continuous relaxation of Problem~\ref{eqTDMAPrimalProblem2} using {consensus} alternating direction method of multipliers (ADMM). Then, we construct a suboptimal solution of Problem~\ref{eqTDMAPrimalProblem2}.

\subsection{Optimal Solution of Relaxed Problem}

By relaxing the discrete constraints in (\ref{eqconstrcn1}) and (\ref{eqconstrOkQstateprimal}) to
{\setlength{\arraycolsep}{0.0em}
\begin{eqnarray}
&&0 \le O_k(\mathbf{Q}) \le 1,\quad k\in\mathcal{K},\quad \mathbf{Q}\in\mathcal{Q},\label{eqconstrOkQstate}\\
&&0 \le c_n  \le 1,\quad n\in\mathcal{X},\label{eqconstrcn1new1}
\end{eqnarray}\setlength{\arraycolsep}{5pt}}we obtain the following continuous relaxation of Problem~\ref{eqTDMAPrimalProblem2}.
\begin{Problem}[Continuous Relaxation of Problem~\ref{eqTDMAPrimalProblem2}] \label{eqTDMAPrimalProblem2CR}
{\setlength{\arraycolsep}{0.0em}
\begin{eqnarray}
&&  \overline{e}^* \triangleq \min_{\substack{{\mathbf{c}, {\mathbf{O}}, {\mathbf{Y}},
{\mathbf{\hat{T}}}_{\texttt{u}}^{\mathrm{dat}}, {\mathbf{\hat{T}}}_{\texttt{d}}^{\mathrm{dat}}, {\mathbf{\hat{T}}}_{\texttt{d}}^{\mathrm{sfw}}}}} \overline{e} ( {{\mathbf{O}}, {\mathbf{Y}}, {\mathbf{\hat{T}}}_{\texttt{u}}^{\mathrm{dat}}, {\mathbf{\hat{T}}}_{\texttt{d}}^{\mathrm{dat}}, {\mathbf{\hat{T}}}_{\texttt{d}}^{\mathrm{sfw}}})\nonumber\\
\mathrm{s.t.}\quad&&
(\ref{eqconstrcn2}), (\ref{eqconstrynmaxQstate}),(\ref{eqconstrhattukQstate}), (\ref{eqconstrhattdkQstate}), (\ref{eqconstrhattdnQstate}), (\ref{eqconstrsumtasksTimeQstate}), (\ref{eqconstrhatsumTimeTDMAQstate}) , (\ref{eqconstrOkQstate}), (\ref{eqconstrcn1new1}).\nonumber
\end{eqnarray}\setlength{\arraycolsep}{5pt}}Let  $(\mathbf{c}^{\ddag}, {\mathbf{O}}^{\ddag},   {\mathbf{Y}}^{\ddag},
{\mathbf{\hat{T}}}_{\texttt{u}}^{\mathrm{dat}^{\ddag}}, {\mathbf{\hat{T}}}_{\texttt{d}}^{\mathrm{dat}^{\ddag}}, {\mathbf{\hat{T}}}_{\texttt{d}}^{\mathrm{sfw}^{\ddag}})$ denote an optimal solution of Problem~\ref{eqTDMAPrimalProblem2CR}.
\end{Problem}

It is easy to verify that Problem~\ref{eqTDMAPrimalProblem2CR} is  convex, and hence we can obtain an optimal solution of Problem~\ref{eqTDMAPrimalProblem2CR} using an interior point method. However, this method has high computational complexity. Specifically, the computational complexity of each iteration of an interior point method used for solving Problem~\ref{eqTDMAPrimalProblem2CR} is given by {$\mathcal{O}\left((K+N)^3N^{3K}\Delta^{3K}\right)$, where $\Delta = \left|\mathcal{L}^{\mathrm{dat}}_{\texttt{u}}\right| \times \left|\mathcal{L}^{\mathrm{dat}}_{\texttt{e}}\right| \times \left|\mathcal{L}^{\mathrm{dat}}_{\texttt{d}}\right| \times \left|\mathcal{H}\right|$ \cite{convexoptimization}.} In the following, we will develop a fast algorithm based on ADMM, which is more efficient than an interior point method in solving Problem~\ref{eqTDMAPrimalProblem2CR}. { Note that, ADMM is a simple but powerful first-order method for solving convex optimization problems with a large number of variables and constraints \cite{admmSboyd, LargCVXSHI2015}. A significant advantage of ADMM is that it allows us to decompose the original problem into a series of subproblems, each of which only contains a small number of variables and constraints. These subproblems are separated from each other and can be effectively solved in a distributed manner. It has been shown in \cite{admmSboyd} that ADMM can quickly converge to an optimal solution of the original problem with modest accuracy.} 

In order to enable the application of ADMM, we need to transform Problem~\ref{eqTDMAPrimalProblem2CR} into an ADMM form that is separable. We notice that the objective function and the constraint functions in (\ref{eqconstrynmaxQstate})--(\ref{eqconstrOkQstate}) are separable with respective to (w.r.t.) system state $\mathbf{Q}\in\mathcal{Q}$, while the caching constraint functions in  (\ref{eqconstrcn2}) and (\ref{eqconstrcn1new1}) are not. In order to make Problem~\ref{eqTDMAPrimalProblem2CR} separable, we first introduce new variables $\mathbf{\hat{C}}(\mathbf{Q})$ and the corresponding constraints on $\mathbf{\hat{C}}(\mathbf{Q})$, given by (\ref{eqconstrhatcn3ADMMnew})--(\ref{eqconstrhatcn1ADMM}),
\begin{figure*}[!t]
{\begingroup\makeatletter\def\f@size{9.5}\check@mathfonts
\def\maketag@@@#1{\hbox{\m@th\normalsize\normalfont#1}}
\begin{eqnarray}
&&\sum _{n\in\mathcal{X}} \left( \mathbbm{1}\left(K_n(\mathbf{X})\ge 1\right)(1-\hat{C}_n(\mathbf{Q}))T_{\mathrm{B,\texttt{d} },n}^{\mathrm{sfw}} + \hat{T}_{\texttt{d},n}^{\mathrm{sfw}}(\mathbf{Q}) \right) + \sum\limits_{k\in\mathcal{K}}  \left(\hat{T}_{\texttt{u},k}^{\mathrm{dat}}(\mathbf{Q}) + O_k(\mathbf{Q}) {L_{\texttt{e},k}^{\mathrm{dat}}} / {F_{\mathrm{{sn}}}} + \hat{T}_{\texttt{d},k}^{\mathrm{dat}}(\mathbf{Q}) \right)\le D,  \label{eqconstrhatcn3ADMMnew}  \\
&&\sum\limits_{n\in\mathcal{X}}  \left( \mathbbm{1}\left(K_n(\mathbf{X})\ge 1\right)(1-\hat{C}_n(\mathbf{Q}))T_{\mathrm{B,\texttt{d} },n}^{\mathrm{sfw}} + \hat{T}_{\texttt{d},n}^{\mathrm{sfw}}(\mathbf{Q}) \right) + (1-O_k(\mathbf{Q}))  {L_{\texttt{e},k} ^{\mathrm{dat}}}/{F_k} \le D,   \label{eqconstrhatcn3ADMM}\\
&&\sum\limits_{n\in\mathcal{X}}\hat{C}_n(\mathbf{Q}) l_{\texttt{d},n}^{\mathrm{sfw}}\le C, \label{eqconstrsumhatcn1ADMM}\\
&&0 \le \hat{C}_n(\mathbf{Q}) \le 1. \label{eqconstrhatcn1ADMM}
\end{eqnarray}\endgroup}
\noindent\rule[1\baselineskip]{\textwidth}{0.1pt}
\end{figure*}
where $k\in\mathcal{K}$, $n\in\mathcal{X}$, $\mathbf{Q}\in\mathcal{Q}$ and $\mathbf{\hat{C}}\triangleq(\hat{C}_n)_{n\in\mathcal{X}}$ is a vector mapping of $\mathbf{Q}$.  Then, define
{\begingroup\makeatletter\def\f@size{9.5}\check@mathfonts
\def\maketag@@@#1{\hbox{\m@th\normalsize\normalfont#1}}
\begin{align}
&v( {\mathbf{O}(\mathbf{Q})}, \mathbf{Y}(\mathbf{Q}), {\mathbf{\hat{T}}}_{\texttt{u}}^{\mathrm{dat}}(\mathbf{Q}), {\mathbf{\hat{T}}}_{\texttt{d}}^{\mathrm{dat}}(\mathbf{Q}), {\mathbf{\hat{T}}}_{\texttt{d}}^{\mathrm{sfw}}(\mathbf{Q}) ) \triangleq \label{eqdefv} \\
&  p_{\mathbf{Q}}(\mathbf{q})  e ( {{\mathbf{O}(\mathbf{Q})},\mathbf{Y}(\mathbf{Q}),
{\mathbf{\hat{T}}}_{\texttt{u}}^{\mathrm{dat}}(\mathbf{Q}), {\mathbf{\hat{T}}}_{\texttt{d}}^{\mathrm{dat}}(\mathbf{Q}), {\mathbf{\hat{T}}}_{\texttt{d}}^{\mathrm{sfw}}(\mathbf{Q}), \mathbf{Q}} ) \nonumber\\
&  + \mathbb{I}_{\mathcal{F}_1(\mathbf{Q})}\left[(\mathbf{\hat{C}}(\mathbf{Q}), {\mathbf{O}(\mathbf{Q})}, \mathbf{Y}(\mathbf{Q}), {\mathbf{\hat{T}}}_{\texttt{u}}^{\mathrm{dat}}(\mathbf{Q}), {\mathbf{\hat{T}}}_{\texttt{d}}^{\mathrm{dat}}(\mathbf{Q}), {\mathbf{\hat{T}}}_{\texttt{d}}^{\mathrm{sfw}}(\mathbf{Q}) ) \right],\nonumber
\end{align}\endgroup}where $e (\cdot)$ is given by (\ref{eqTDMAtotalEngeryrew}),  $\mathbb{I}_{\mathcal{F}_1(\mathbf{Q})}[s]$ is defined as
\begin{eqnarray*}
\mathbb{I}_{\mathcal{F}_1(\mathbf{Q})}[s]\triangleq
\begin{cases}
0, &\mbox{if  } s\in \mathcal{F}_1(\mathbf{Q}), \\
+\infty,   &\mbox{otherwise},
\end{cases}
\end{eqnarray*}and $\mathcal{F}_1(\mathbf{Q}) \triangleq  \{(\mathbf{\hat{C}}(\mathbf{Q}), {\mathbf{O}}(\mathbf{Q}), {\mathbf{Y}}(\mathbf{Q}),
{\mathbf{\hat{T}}}_{\texttt{u}}^{\mathrm{dat}}(\mathbf{Q}), {\mathbf{\hat{T}}}_{\texttt{d}}^{\mathrm{dat}}(\mathbf{Q}), \\ {\mathbf{\hat{T}}}_{\texttt{d}}^{\mathrm{sfw}}(\mathbf{Q}))|(\ref{eqconstrynmaxQstate}), (\ref{eqconstrhattukQstate}), (\ref{eqconstrhattdkQstate}), (\ref{eqconstrhattdnQstate}), (\ref{eqconstrOkQstate}), (\ref{eqconstrhatcn3ADMMnew}),  (\ref{eqconstrhatcn3ADMM}),(\ref{eqconstrsumhatcn1ADMM}), (\ref{eqconstrhatcn1ADMM})\}$.  Consider the following problem in ADMM form that is separable.
\begin{Problem}[Equivalent Problem of Problem \ref{eqTDMAPrimalProblem2CR}]\label{probADMMform}
{\begingroup\makeatletter\def\f@size{9.5}\check@mathfonts
\def\maketag@@@#1{\hbox{\m@th\normalsize\normalfont#1}}
\begin{align}
\min_{\substack{{ \mathbf{c}, \mathbf{\hat{C}}, {\mathbf{O}}, {\mathbf{Y}},}\\ {
{\mathbf{\hat{T}}}_{\texttt{u}}^{\mathrm{dat}}, {\mathbf{\hat{T}}}_{\texttt{d}}^{\mathrm{dat}}, {\mathbf{\hat{T}}}_{\texttt{d}}^{\mathrm{sfw}}}}}& \sum_{\mathbf{Q}\in \mathcal{Q}}  v( {\mathbf{O}(\mathbf{Q})}, \mathbf{Y}(\mathbf{Q}),
{\mathbf{\hat{T}}}_{\texttt{u}}^{\mathrm{dat}}(\mathbf{Q}), {\mathbf{\hat{T}}}_{\texttt{d}}^{\mathrm{dat}}(\mathbf{Q}), {\mathbf{\hat{T}}}_{\texttt{d}}^{\mathrm{sfw}}(\mathbf{Q}) )\nonumber\\
& \; \mathrm{s.t.}  \; \hat{C}_n(\mathbf{Q})=c_n,\quad n\in\mathcal{X},\quad \mathbf{Q}\in\mathcal{Q}.\label{eqconstrhatcnandcADMM}
\end{align}\endgroup}
\end{Problem}

{Note that, since the variable $\mathbf{c}$ accounts for all system states and $( \hat{C}_n(\mathbf{Q}), {\mathbf{O}(\mathbf{Q})}, \mathbf{Y}(\mathbf{Q}),
{\mathbf{\hat{T}}}_{\texttt{u}}^{\mathrm{dat}}(\mathbf{Q}), {\mathbf{\hat{T}}}_{\texttt{d}}^{\mathrm{dat}}(\mathbf{Q}), {\mathbf{\hat{T}}}_{\texttt{d}}^{\mathrm{sfw}}(\mathbf{Q}) )$ is only related to a specific system state $\mathbf{Q}$, we refer to the former and the latter as ``global variable'' and ``local variable'' \cite{MECCachingWang2017}, respectively.} The constraints in (\ref{eqconstrhatcnandcADMM}) is used to guarantee the consistency of the local variables $\hat{C}_n(\mathbf{Q})$ for all $\mathbf{Q}\in\mathcal{Q}$ with each other, as well as with the global variable $\mathbf{c}$, so it is called a consensus constraints \cite{admmSboyd}. In addition, the objective function of Problem~\ref{probADMMform} is separable w.r.t. the system state $\mathbf{Q}$. By exploiting this separable structure, we can apply ADMM to solve Problem~\ref{probADMMform}. Before continuing, we first show the equivalence between Problems \ref{eqTDMAPrimalProblem2CR} and \ref{probADMMform} as follows.

\begin{Lemma}[Equivalence between Problem \ref{eqTDMAPrimalProblem2CR} and Problem~\ref{probADMMform}]\label{lemmaequivalentADMM}
Problem \ref{eqTDMAPrimalProblem2CR} and Problem~\ref{probADMMform} are equivalent.
\end{Lemma}

\indent\indent \emph{Proof:} See Appendix \ref{prooflemmaequivalentADMM}.\hfill $\blacksquare$

{
\begin{algorithm}[!t]
\caption{Solving Problem \ref{eqTDMAPrimalProblem2CR} via ADMM}\label{algADMM}
\begin{algorithmic}[1]  \small
\STATE{\textbf{input:} $B$, $C$, $D$, $K$, $N$, $R$, $F_{\mathrm{sn}}$, $\mu_{\mathrm{sn}}$, $l_{\texttt{d},n}^{\mathrm{sfw}}$, $n\in\mathcal{X}$, $F_k$, $\mu_k$, $\omega_k$, $k\in\mathcal{K}$, $n_0$, $\epsilon$ and $\rho$}
\STATE{\textbf{output:} $(\mathbf{c}^{\ddag}, {\mathbf{O}}^{\ddag}, {\mathbf{Y}}^{\ddag},
{\mathbf{\hat{T}}}_{\texttt{u}}^{\mathrm{dat}^{\ddag}}, {\mathbf{\hat{T}}}_{\texttt{d}}^{\mathrm{dat}^{\ddag}}, {\mathbf{\hat{T}}}_{\texttt{d}}^{\mathrm{sfw}^{\ddag}})$}
\STATE{\textbf{initialization:} choose any $c_{0,n}\in[0,1]$, $\theta_{0,n}(\mathbf{Q})>0$, $n\in\mathcal{X}$, $\mathbf{Q}\in\mathcal{Q}$,  and set $t:= 0$}
\REPEAT
\STATE {obtain $(\mathbf{\hat{C}}_{t+1}(\mathbf{Q}), {\mathbf{O}_{t+1}}(\mathbf{Q}), \mathbf{Y}_{t+1}(\mathbf{Q}), {\mathbf{\hat{T}}}_{\texttt{u},{t+1}}^{\mathrm{dat}}(\mathbf{Q}), {\mathbf{\hat{T}}}_{\texttt{d},{t+1}}^{\mathrm{dat}}(\mathbf{Q})$, $ {\mathbf{\hat{T}}}_{\texttt{d},{t+1}}^{\mathrm{sfw}}(\mathbf{Q}) )$ by solving the problem in (\ref{eqTDMAPrimalProblemADMMupdateslocalvar}) for all $\mathbf{Q}\in\mathcal{Q}$, via an interior point method }
\STATE { compute $c_{t+1,n}$, $n\in\mathcal{X}$ according to (\ref{eqcalglobalvar})}
\STATE { compute ${\theta}_{t+1,n}(\mathbf{Q})$, $n\in\mathcal{X}$, $\mathbf{Q}\in\mathcal{Q}$ according to (\ref{eqADMMThetaupdate})}
\STATE {set $t:= t+1$}
\UNTIL {{$\|\mathbf{\hat{C}}_{t+1}(\mathbf{Q})-\mathbf{c}_{t+1}\|_2 \le \epsilon$ and $\|\left(\mathbf{c}_{t+1}-\mathbf{c}{_{t}}\right)\rho \|_2 \le \epsilon$}}
\STATE{set $(\mathbf{c}^{\ddag}, {\mathbf{O}}^{\ddag}, {\mathbf{Y}}^{\ddag},
{\mathbf{\hat{T}}}_{\texttt{u}}^{\mathrm{dat}^{\ddag}}, {\mathbf{\hat{T}}}_{\texttt{d}}^{\mathrm{dat}^{\ddag}}, {\mathbf{\hat{T}}}_{\texttt{d}}^{\mathrm{sfw}^{\ddag}}) := (\mathbf{c}_t, {\mathbf{O}_t}, {\mathbf{Y}_t},
{\mathbf{\hat{T}}}_{\texttt{u},t}^{\mathrm{dat}}, {\mathbf{\hat{T}}}_{\texttt{d},t}^{\mathrm{dat}}, {\mathbf{\hat{T}}}_{\texttt{d},t}^{\mathrm{sfw}})$}
\end{algorithmic}
\end{algorithm}
}

Based on Lemma~\ref{lemmaequivalentADMM}, instead of solving Problem~\ref{eqTDMAPrimalProblem2CR}, we now solve Problem~\ref{lemmaequivalentADMM} using ADMM \cite{admmSboyd}. Specifically, we choose any $c_{0,n}$ and $\theta_{0,n}(\mathbf{Q})$, $n\in\mathcal{X}$, $\mathbf{Q}\in\mathcal{Q}$, and at iteration $t+1$, compute (\ref{eqTDMAPrimalProblemADMMupdateslocalvar}), (\ref{eqcalglobalvar}) and (\ref{eqADMMThetaupdate}), respectively.
\begin{figure*}[!t]
{\begingroup\makeatletter\def\f@size{9.5}\check@mathfonts
\def\maketag@@@#1{\hbox{\m@th\normalsize\normalfont#1}}
\begin{align}\label{eqTDMAPrimalProblemADMMupdateslocalvar}
\left( \begin{array}{l}
{{{\bf{\hat C}}}_{t + 1}}({\bf{Q}}),{{\bf{O}}_{t + 1}}({\bf{Q}}),{{\bf{Y}}_{t + 1}}({\bf{Q}}),\\
{\bf{\hat T}}_{u,t + 1}^{{\rm{dat}}}({\bf{Q}}),{\bf{\hat T}}_{d,t + 1}^{{\rm{dat}}}({\bf{Q}}),{\bf{\hat T}}_{d,t + 1}^{{\rm{sfw}}}({\bf{Q}})
\end{array} \right)
& \triangleq \arg\min_{\substack{ {(\mathbf{\hat{C}}(\mathbf{Q}), {\mathbf{O}}(\mathbf{Q}),\mathbf{Y}(\mathbf{Q}),}\\  { {\mathbf{\hat{T}}}_{\texttt{u}}^{\mathrm{dat}}(\mathbf{Q}), {\mathbf{\hat{T}}}_{\texttt{d}}^{\mathrm{dat}}(\mathbf{Q}),} \\ { {\mathbf{\hat{T}}}_{\texttt{d}}^{\mathrm{sfw}}(\mathbf{Q})) \in\mathcal{F}_1(\mathbf{Q}) }}} \hspace{-6mm}  f_1 ( {{\mathbf{\hat{C}}(\mathbf{Q})},{\mathbf{O}(\mathbf{Q})},\mathbf{Y}(\mathbf{Q}),
{\mathbf{\hat{T}}}_{\texttt{u}}^{\mathrm{dat}}(\mathbf{Q}), {\mathbf{\hat{T}}}_{\texttt{d}}^{\mathrm{dat}}(\mathbf{Q}), {\mathbf{\hat{T}}}_{\texttt{d}}^{\mathrm{sfw}}(\mathbf{Q})} ) , \\
c_{t+1,n}&=\frac{1}{\left|\mathcal{Q}\right|} \sum_{\mathbf{Q}\in\mathcal{Q}}\left({\hat{C}}_{t+1,n}(\mathbf{Q}) + ({1}/{\rho}){\theta}_{t,n}(\mathbf{Q})\right),\quad n\in\mathcal{X},\label{eqcalglobalvar}\\
{\theta}_{t+1,n}(\mathbf{Q}) &= {\theta}_{t,n}(\mathbf{Q}) + \rho ({\hat{C}}_{t+1, n}(\mathbf{Q})-{c}_{t+1, n}),\quad n\in\mathcal{X},\quad\mathbf{Q}\in\mathcal{Q}.\label{eqADMMThetaupdate}
\end{align}\endgroup}
\noindent\rule[1.00\baselineskip]{\textwidth}{0.1pt}
\end{figure*}
Here,
\begin{align*}
& f_1 ( {{\mathbf{\hat{C}}(\mathbf{Q})},{\mathbf{O}(\mathbf{Q})},\mathbf{Y}(\mathbf{Q}),
{\mathbf{\hat{T}}}_{\texttt{u}}^{\mathrm{dat}}(\mathbf{Q}), {\mathbf{\hat{T}}}_{\texttt{d}}^{\mathrm{dat}}(\mathbf{Q}), {\mathbf{\hat{T}}}_{\texttt{d}}^{\mathrm{sfw}}(\mathbf{Q})} )\nonumber\\
& \triangleq p_{\mathbf{Q}}(\mathbf{q})  e ( {{\mathbf{O}(\mathbf{Q})},\mathbf{Y}(\mathbf{Q}),
{\mathbf{\hat{T}}}_{\texttt{u}}^{\mathrm{dat}}(\mathbf{Q}), {\mathbf{\hat{T}}}_{\texttt{d}}^{\mathrm{dat}}(\mathbf{Q}), {\mathbf{\hat{T}}}_{\texttt{d}}^{\mathrm{sfw}}(\mathbf{Q}), \mathbf{Q}} )   \\
&\quad\quad  +  \displaystyle \sum_{n\in\mathcal{X}} \theta_{t,n}(\mathbf{Q}) ( {\hat{C}}_n(\mathbf{Q})-c_{t,n}) + \frac{\rho}{2}  \sum_{n\in\mathcal{X}} ( {\hat{C}}_n(\mathbf{Q})-c_{t,n})^2,
\end{align*}where $\rho>0$ is a penalty parameter, and {the update equations in} {(\ref{eqADMMThetaupdate}) are to obtain the  Lagrange multipliers corresponding to the constraints in (\ref{eqconstrhatcnandcADMM})}. Note that the problem in (\ref{eqTDMAPrimalProblemADMMupdateslocalvar}) is a convex problem, and hence can be solved by an interior point method.
The details of ADMM are summarized in Algorithm~\ref{algADMM}.  According to \cite{admmSboyd}, we have the following lemma.
\begin{Lemma}[Convergence of Algorithm \ref{algADMM}]\label{lemmaConvergenceADMM}
As $t\to\infty$, {we have $\sum\limits_{\mathbf{Q}\in \mathcal{Q}}  v( {\mathbf{O}_t}(\mathbf{Q}), {\mathbf{Y}_t}(\mathbf{Q}),
{\mathbf{\hat{T}}}_{\texttt{u},t}^{\mathrm{dat}}(\mathbf{Q}), {\mathbf{\hat{T}}}_{\texttt{d},t}^{\mathrm{dat}}(\mathbf{Q}), {\mathbf{\hat{T}}}_{\texttt{d},t}^{\mathrm{sfw}}(\mathbf{Q})) \to \overline{e}^*$.}
\end{Lemma}

\indent\indent \emph{Proof:} See Appendix \ref{prooflemmaConvergenceADMM}.\hfill $\blacksquare$

We now analyze the computational complexity of Algorithm~\ref{algADMM}. Specifically, the computational complexity of each iteration of an interior point method used for solving the problem in (\ref{eqTDMAPrimalProblemADMMupdateslocalvar}) for all $\mathbf{Q}\in\mathcal{Q}$ is given by $\mathcal{O}((K+N)^3N^K\Delta^K)$. The computational complexities for computing $c_{t+1,n}$, $n\in\mathcal{X}$, and $\theta_{t+1,n}(\mathbf{Q})$, $n\in\mathcal{X}$, $\mathbf{Q}\in\mathcal{Q}$, are given by $\mathcal{O}(N^{K+1}\Delta^K)$. Although the number of iterations of Algorithm~\ref{algADMM} cannot be analytically characterized, numerical results show that Algorithm~\ref{algADMM} terminates in a few iterations. {Therefore, solving  Problem \ref{eqTDMAPrimalProblem2CR} using Algorithm~\ref{algADMM} is of much lower computational complexity than using an interior point method.}

\subsection{Binary Variable Recovery}

As $(\mathbf{c}^{\ddag}, {\mathbf{O}}^{\ddag})$ are usually  not  binary,  the optimal solution $(\mathbf{c}^{\ddag}, {\mathbf{O}}^{\ddag}, {\mathbf{Y}}^{\ddag}, {\mathbf{\hat{T}}}_{\texttt{u}}^{\mathrm{dat}^{\ddag}}, {\mathbf{\hat{T}}}_{\texttt{d}}^{\mathrm{dat}^{\ddag}}, {\mathbf{\hat{T}}}_{\texttt{d}}^{\mathrm{sfw}^{\ddag}})$ of Problem~\ref{eqTDMAPrimalProblem2CR} may not be in the feasible set of Problem~\ref{eqTDMAPrimalProblem2}.  In this part, based on $(\mathbf{c}^{\ddag}, {\mathbf{O}}^{\ddag}, {\mathbf{Y}}^{\ddag}, {\mathbf{\hat{T}}}_{\texttt{u}}^{\mathrm{dat}^{\ddag}},  {\mathbf{\hat{T}}}_{\texttt{d}}^{\mathrm{dat}^{\ddag}}, {\mathbf{\hat{T}}}_{\texttt{d}}^{\mathrm{sfw}^{\ddag}})$, we construct a suboptimal solution, denoted by $(\mathbf{c}^{\dag}, {\mathbf{O}}^{\dag}, {\mathbf{Y}}^{\dag}, {\mathbf{\hat{T}}}_{\texttt{u}}^{\mathrm{dat}^{\dag}}, {\mathbf{\hat{T}}}_{\texttt{d}}^{\mathrm{dat}^{\dag}}, {\mathbf{\hat{T}}}_{\texttt{d}}^{\mathrm{sfw}^{\dag}})$, of Problem~\ref{eqTDMAPrimalProblem2} with promising performance, which will be seen in Section VI.

First, we determine $\mathbf{c}^{\dag}$. Both the cache size constraints in (\ref{eqconstrcn1}) and (\ref{eqconstrcn2}) and the deadline constraints in (\ref{eqconstrTotalTimeAllusersQstateprimal}) and (\ref{eqconstrTotalTimeAllusersQstateprimal2}) are affected by $\mathbf{c}^{\dag}$ but in different manners. {Caching software $n$ consumes storage resource of $l_{\texttt{d},n}^{\mathrm{sfw}}$ and saves time duration of {$\Pr\left[\mathcal{K}_n(\mathbf{X})\ge 1\right]T_{\mathrm{B,\texttt{d} },n}^{\mathrm{sfw}}$} for software fetching {on average. Define $z_n\triangleq \Pr\left[\mathcal{K}_n(\mathbf{X})\ge 1\right] {T_{\mathrm{B,\texttt{d} },n}^{\mathrm{sfw}}}/{l_{\texttt{d},n}^{\mathrm{sfw}}}$, $n\in\mathcal{X}$.} Jointly considering the storage cost and time cost for software fetching, we construct $\mathbf{c}^{\dag}$ as follows:
\begin{equation}
  c^{\dag}_{(i)}=
  \begin{cases}\label{eqdeterc}
    1, & \mbox{if } i \le \overline{i}, \\
    0, & \mbox{otherwise},
  \end{cases}
\end{equation}where {$z_{(1)} \ge z_{(2)} \ge \cdots \ge z_{(N)}$ are $z_1$, $z_2$, $\cdots$, $z_N$} sorted in descending order and $\overline{i} \triangleq \max\{j|\sum_{i=1}^{j}l_{\texttt{d},(i)}^{\mathrm{sfw}} \le C\}$. Note that the constructed $\mathbf{c}^{\dag}$ satisfies the constraints in (\ref{eqconstrcn1}) and (\ref{eqconstrcn2}).}

Then, we determine $({\mathbf{O}}^{\dag}, {\mathbf{Y}}^{\dag})$. By (\ref{eqconstrOkQstateprimal}) and (\ref{eqconstrynmaxQstate}), we know that any feasible solution $(\mathbf{c}, {\mathbf{O}}, {\mathbf{Y}},  {\mathbf{\hat{T}}}_{\texttt{u}}^{\mathrm{dat}}, {\mathbf{\hat{T}}}_{\texttt{d}}^{\mathrm{dat}}, {\mathbf{\hat{T}}}_{\texttt{d}}^{\mathrm{sfw}})$ of Problem~\ref{eqTDMAPrimalProblem2} {satisfies: (i) $O_k(\mathbf{Q})\ge Y_n(\mathbf{Q})$ if and only if $O_k(\mathbf{Q})=1$, for all $k\in \mathcal{K}_n(\mathbf{X})$, $n\in\mathcal{X}$, $\mathbf{Q}\in\mathcal{Q}$; (ii) $Y_n(\mathbf{Q}) = \max_{k\in\mathcal{K}_n(\mathbf{X})}(1-O_k(\mathbf{Q}))$, for all $n\in\mathcal{X}$, $\mathbf{Q}\in\mathcal{Q}$}. Thus, based on $({\mathbf{O}}^{\ddag}, {\mathbf{Y}}^{\ddag})$, we construct $({\mathbf{O}}^{\dag}, {\mathbf{Y}}^{\dag})$ as follows:
\begin{eqnarray}
&&\begin{cases}\label{eqdeterOk}
  O_k^{\dag}(\mathbf{Q})=1, & \mbox{if } k\in\mathcal{K}_n^{\texttt{off}}(\mathbf{X}) , \\
  O_k^{\dag}(\mathbf{Q})=0, & \mbox{if } k\in \mathcal{K}_n(\mathbf{X}) \setminus \mathcal{K}_n^{\texttt{off}}(\mathbf{X}) , \\
\end{cases} \\
&&Y_n^{\dag}(\mathbf{Q}) = \max_{k\in\mathcal{K}_n(\mathbf{X})}(1-O_k^{\dag}(\mathbf{Q})),  \label{eqdeterYn}
\end{eqnarray}where  $\mathcal{K}_n^{\texttt{off}}(\mathbf{X}) \triangleq  \left\{k\left.|\right. { O_k^{\ddag}(\mathbf{Q}) \ge Y_n^{\ddag}(\mathbf{Q}),k\in \mathcal{K}_n(\mathbf{X})}\right\}$, $n\in\mathcal{X}$, $\mathbf{Q}\in\mathcal{Q}$. {Note that the constructed $({\mathbf{O}}^{\dag}, {\mathbf{Y}}^{\dag})$ satisfies the constraints in (\ref{eqconstrOkQstateprimal}) and (\ref{eqconstrynmaxQstate}).}

Finally, based on the obtained $(\mathbf{c}^{\dag}, {\mathbf{O}}^{\dag}, {\mathbf{Y}}^{\dag})$, we determine $({\mathbf{\hat{T}}}_{\texttt{u}}^{\mathrm{dat}^{\dag}}, {\mathbf{\hat{T}}}_{\texttt{d}}^{\mathrm{dat}^{\dag}}, {\mathbf{\hat{T}}}_{\texttt{d}}^{\mathrm{sfw}^{\dag}})$ by solving Problem~\ref{eqTDMAPrimalProblem2} with $\mathbf{c}=\mathbf{c}^{\dag}$, $\mathbf{O}=\mathbf{O}^{\dag}$ and $\mathbf{Y}=\mathbf{Y}^{\dag}$. Using decomposition, it is equivalent to solve the following problem for each $\mathbf{Q}\in\mathcal{Q}$:
{\begingroup\makeatletter\def\f@size{9.5}\check@mathfonts
\def\maketag@@@#1{\hbox{\m@th\normalsize\normalfont#1}}
\begin{align}
& ({{\mathbf{\hat{T}}}_{\texttt{u}}^{\mathrm{dat}}(\mathbf{Q}), {\mathbf{\hat{T}}}_{\texttt{d}}^{\mathrm{dat}}(\mathbf{Q}), {\mathbf{\hat{T}}}_{\texttt{d}}^{\mathrm{sfw}}(\mathbf{Q})})\triangleq \nonumber \\
&\arg\min_{{\mathbf{\hat{T}}}_{\texttt{u}}^{\mathrm{dat}}(\mathbf{Q}), {\mathbf{\hat{T}}}_{\texttt{d}}^{\mathrm{dat}}(\mathbf{Q}), {\mathbf{\hat{T}}}_{\texttt{d}}^{\mathrm{sfw}}(\mathbf{Q})} \hat{e} ( {\mathbf{\hat{T}}}_{\texttt{u}}^{\mathrm{dat}}(\mathbf{Q}), {\mathbf{\hat{T}}}_{\texttt{d}}^{\mathrm{dat}}(\mathbf{Q}), {\mathbf{\hat{T}}}_{\texttt{d}}^{\mathrm{sfw}}(\mathbf{Q})  ) \nonumber \\
\mathrm{s.t.}\;
&  0 \le \hat{T}_{\texttt{u},k}^{\mathrm{dat}}(\mathbf{Q}) \le  O_k^{\dag}(\mathbf{Q})D,  \nonumber\\
&  0 \le \hat{T}_{\texttt{d},k}^{\mathrm{dat}}(\mathbf{Q}) \le  O_k^{\dag}(\mathbf{Q})D,  \label{probreallocatesubprob}\\
&  0 \le \hat{T}_{\texttt{d},n}^{\mathrm{sfw}}(\mathbf{Q}) \le  Y_n^{\dag}(\mathbf{Q}))D,  \nonumber\\
& \sum _{n\in\mathcal{X}} \left( \mathbbm{1}\left[K_n(\mathbf{X})\ge 1\right](1-c_n^{\dag})T_{\mathrm{B,\texttt{d} },n}^{\mathrm{sfw}} + \hat{T}_{\texttt{d},n}^{\mathrm{sfw}}(\mathbf{Q}) \right) \nonumber\\
&\qquad  + \sum_{k\in\mathcal{K}} \left(\hat{T}_{\texttt{u},k}^{\mathrm{dat}}(\mathbf{Q}) + O_k^{\dag}(\mathbf{Q}) {L_{\texttt{e},k}^{\mathrm{dat}}} / {F_{\mathrm{{sn}}}} + \hat{T}_{\texttt{d},k}^{\mathrm{dat}}(\mathbf{Q}) \right)\le D,  \nonumber\\
& \sum _{n\in\mathcal{X}} \left( \mathbbm{1}\left[K_n(\mathbf{X})\ge 1\right](1-c_n^{\dag})T_{\mathrm{B,\texttt{d} },n}^{\mathrm{sfw}} + \hat{T}_{\texttt{d},n}^{\mathrm{sfw}}(\mathbf{Q}) \right) \nonumber\\
&\qquad + (1-O_k^{\dag}(\mathbf{Q})) {L_{\texttt{e},k} ^{\mathrm{dat}}}/{F_k} \le D, \nonumber
\end{align}\endgroup}where $k\in\mathcal{K}$, $n\in\mathcal{X}$ and $\hat{e} ( {\mathbf{\hat{T}}}_{\texttt{u}}^{\mathrm{dat}}(\mathbf{Q}), {\mathbf{\hat{T}}}_{\texttt{d}}^{\mathrm{dat}}(\mathbf{Q}), {\mathbf{\hat{T}}}_{\texttt{d}}^{\mathrm{sfw}}(\mathbf{Q})  ) \triangleq  \bar{e} ( {\mathbf{O}}^{\dag}(\mathbf{Q}), {\mathbf{Y}^{\dag}}(\mathbf{Q}), {\mathbf{\hat{T}}}_{\texttt{u}}^{\mathrm{dat}}(\mathbf{Q}), {\mathbf{\hat{T}}}_{\texttt{d}}^{\mathrm{dat}}(\mathbf{Q}), {\mathbf{\hat{T}}}_{\texttt{d}}^{\mathrm{sfw}}(\mathbf{Q})  ) $ with $\bar{e}(\cdot)$ being given by (\ref{eqTDMAtotalEngeryrew2obj2}). {Note that $({\mathbf{O}}^{\dag}, {\mathbf{Y}}^{\dag},{\mathbf{\hat{T}}}_{\texttt{u}}^{\mathrm{dat}^{\dag}}, {\mathbf{\hat{T}}}_{\texttt{d}}^{\mathrm{dat}^{\dag}}, {\mathbf{\hat{T}}}_{\texttt{d}}^{\mathrm{sfw}^{\dag}})$ satisfies the remanning constraints, i.e., (\ref{eqconstrhattukQstate})--(\ref{eqconstrhatsumTimeTDMAQstate}).}

The above details are summarized in Algorithm~\ref{algrecovery}. The following lemma provides a sufficient condition for Algorithm~\ref{algrecovery} to find a feasible  solution of Problem~\ref{eqTDMAPrimalProblem2}.

\begin{algorithm}[t]
\caption{Feasible Solution of Problem~\ref{eqTDMAPrimalProblem2} Based on Continuous Relaxation }\label{algrecovery}
\begin{algorithmic}[1]   \small
\STATE{\textbf{input:} $B$, $C$, $D$, $K$, $N$, $R$, $F_{\mathrm{sn}}$, $\mu_{\mathrm{sn}}$, $l_{\texttt{d},n}^{\mathrm{sfw}}$, $n\in\mathcal{X}$, $F_k$, $\mu_k$, $\omega_k$, $k\in\mathcal{K}$, $n_0$, $\epsilon$ and $\rho$}
\STATE {\textbf{output:} $(\mathbf{c}^{\dag}, {\mathbf{O}}^{\dag},{\mathbf{T}}_{\texttt{u}}^{\mathrm{dat}^{\dag}}, {\mathbf{T}}_{\texttt{d}}^{\mathrm{dat}^{\dag}}, {\mathbf{T}}_{\texttt{d}}^{\mathrm{sfw}^{\dag}})$}
\STATE {obtain $(\mathbf{c}^{\ddag}, {\mathbf{O}}^{\ddag}, {\mathbf{Y}}^{\ddag},
{\mathbf{\hat{T}}}_{\texttt{u}}^{\mathrm{dat}^{\ddag}}, {\mathbf{\hat{T}}}_{\texttt{d}}^{\mathrm{dat}^{\ddag}}, {\mathbf{\hat{T}}}_{\texttt{d}}^{\mathrm{sfw}^{\ddag}})$ by solving Problem~\ref{eqTDMAPrimalProblem2CR} using Algorithm~\ref{algADMM}}
\STATE{compute $(\mathbf{c}^{\dag}, {\mathbf{O}}^{\dag}, {\mathbf{Y}}^{\dag})$ according to (\ref{eqdeterc}), (\ref{eqdeterOk}) and (\ref{eqdeterYn})}
\STATE{compute $({\mathbf{T}}_{\texttt{u}}^{\mathrm{dat}^{\dag}}, {\mathbf{T}}_{\texttt{d}}^{\mathrm{dat}^{\dag}}, {\mathbf{T}}_{\texttt{d}}^{\mathrm{sfw}^{\dag}})$ by solving the problem  in (\ref{probreallocatesubprob}) using an interior point method}

\end{algorithmic}
\end{algorithm}

\begin{Lemma}[Applicability of Algorithm~\ref{algrecovery}]\label{lemmafeasisuffnec}
If $(\mathbf{c}^{\dag}, {\mathbf{O}}^{\dag}, {\mathbf{Y}}^{\dag})$ satisfies
{\begingroup\makeatletter\def\f@size{8}\check@mathfonts
\def\maketag@@@#1{\hbox{\m@th\normalsize\normalfont#1}}
\begin{align}
& \sum _{n\in\mathcal{X}} \mathbbm{1}\left(K_n(\mathbf{X})\ge 1\right)(1-c_n^{\dag})T_{\mathrm{B,\texttt{d} },n}^{\mathrm{sfw}}  + \sum_{k\in\mathcal{K}}O_k^{\dag}(\mathbf{Q}) \frac{L_{\texttt{e},k}^{\mathrm{dat}}} {F_{\mathrm{{sn}}}}\le D, \label{eqconstrSNC1}\\
& \sum _{n\in\mathcal{X}}  \mathbbm{1}\left(K_n(\mathbf{X})\ge 1\right)(1-c_n^{\dag})T_{\mathrm{B,\texttt{d} },n}^{\mathrm{sfw}}   + (1-O_k^{\dag}(\mathbf{Q})) \frac{L_{\texttt{e},k} ^{\mathrm{dat}}}{F_k} \le D, \label{eqconstrSNC2}
\end{align}\endgroup}where $k\in\mathcal{K}$, $\mathbf{Q}\in\mathcal{Q}$, then $(\mathbf{c}^{\dag}, {\mathbf{O}}^{\dag}, {\mathbf{Y}}^{\dag}, {\mathbf{\hat{T}}}_{\texttt{u}}^{\mathrm{dat}^{\dag}},{\mathbf{\hat{T}}}_{\texttt{d}}^{\mathrm{dat}^{\dag}}, {\mathbf{\hat{T}}}_{\texttt{d}}^{\mathrm{sfw}^{\dag}})$ is a feasible solution of Problem~\ref{eqTDMAPrimalProblem2}.
\end{Lemma}

\indent\indent {\textit{Proof:}} The fact that $(\mathbf{c},{\mathbf{O}},{\mathbf{Y}})$ satisfies the constraints in (\ref{eqconstrcn1}), (\ref{eqconstrcn2}), (\ref{eqconstrOkQstateprimal}), (\ref{eqconstrynmaxQstate}), (\ref{eqconstrSNC1}) and (\ref{eqconstrSNC2}) implies that $(\mathbf{c}, {\mathbf{O}},{\mathbf{Y}},{\mathbf{0}},{\mathbf{0}}, {\mathbf{0}})$ (i.e., $\mathbf{\hat{T}} _{\texttt{u}}^{\mathrm{dat}}=\mathbf{0}$, $\mathbf{\hat{T}} _{\texttt{u}}^{\mathrm{dat}}=\mathbf{0}$ and $\mathbf{\hat{T}} _{\texttt{d}}^{\mathrm{sfw}}=\mathbf{0}$) satisfies the constraints of Problem~\ref{eqTDMAPrimalProblem2}. That is, $(\mathbf{c}, {\mathbf{O}},{\mathbf{Y}},{\mathbf{0}},{\mathbf{0}}, {\mathbf{0}})$ is a feasible solution of Problem~\ref{eqTDMAPrimalProblem2}, which completes the proof.~\hfill~$\blacksquare$

Later in Section VI, we shall see that the suboptimal solution $({\mathbf{O}}^{\dag}, {\mathbf{Y}}^{\dag},{\mathbf{\hat{T}}}_{\texttt{u}}^{\mathrm{dat}^{\dag}}, {\mathbf{\hat{T}}}_{\texttt{d}}^{\mathrm{dat}^{\dag}}, {\mathbf{\hat{T}}}_{\texttt{d}}^{\mathrm{sfw}^{\dag}})$ obtained by Algorithm~\ref{algrecovery} offers a promising performance.

\section{A Low-complexity Algorithm Based on Penalty-CCP}
In this section, we develop a low-complexity algorithm to obtain a suboptimal solution of Problem~\ref{eqTDMAPrimalProblem2} with a promising performance based on Penalty-CCP. {First, we equivalently convert the discrete constraints in (\ref{eqconstrcn1}) and (\ref{eqconstrOkQstateprimal}) to those in (\ref{eqconstrOkQstate}), (\ref{eqconstrcn1new1}) and
{\setlength{\arraycolsep}{0.0em}
\begin{eqnarray}
&& F(\mathbf{c}, \mathbf{O}(\mathbf{Q})) \le 0,\quad \mathbf{Q}\in\mathcal{Q},\label{eqconstrOkQstate3}
\end{eqnarray}\setlength{\arraycolsep}{5pt}}where
$F(\mathbf{c}, \mathbf{O}(\mathbf{Q})) \triangleq \sum_{k\in\mathcal{K}}O_k(\mathbf{Q})\left(1-O_k(\mathbf{Q})\right) +\sum_{n\in\mathcal{X}}c_n \left(1-c_n\right)$. Then, Problem~\ref{eqTDMAPrimalProblem2} can be equivalently transformed to the following DC problem.}

\begin{Problem}[Equivalent DC Problem of Problem~\ref{eqTDMAPrimalProblem2}] \label{eqTDMAPrimalProblem2DC}
{\setlength{\arraycolsep}{0.0em}
\begin{eqnarray}
&&\min_{\substack{{\mathbf{c}, {\mathbf{O}}, {\mathbf{Y}},
{\mathbf{\hat{T}}}_{\texttt{u}}^{\mathrm{dat}}, {\mathbf{\hat{T}}}_{\texttt{d}}^{\mathrm{dat}}, {\mathbf{\hat{T}}}_{\texttt{d}}^{\mathrm{sfw}}}}} \overline{e} ( {{\mathbf{O}}, {\mathbf{Y}}, {\mathbf{\hat{T}}}_{\texttt{u}}^{\mathrm{dat}}, {\mathbf{\hat{T}}}_{\texttt{d}}^{\mathrm{dat}}, {\mathbf{\hat{T}}}_{\texttt{d}}^{\mathrm{sfw}}})\nonumber\\
\mathrm{s.t.}\;&&
(\ref{eqconstrcn2}), (\ref{eqconstrynmaxQstate}),(\ref{eqconstrhattukQstate}), (\ref{eqconstrhattdkQstate}), (\ref{eqconstrhattdnQstate}), (\ref{eqconstrsumtasksTimeQstate}), (\ref{eqconstrhatsumTimeTDMAQstate}) , (\ref{eqconstrOkQstate}), (\ref{eqconstrcn1new1}), (\ref{eqconstrOkQstate3}).\nonumber
\end{eqnarray}\setlength{\arraycolsep}{5pt}}
\end{Problem}

Note that the constraint function in (\ref{eqconstrOkQstate3}) is concave and the objective function and the other constraint functions are convex. Thus, Problem~\ref{eqTDMAPrimalProblem2DC} is a DC problem (which is non-convex). A classical goal of solving a non-convex problem is to obtain a stationary point. In the following, we adopt the Penalty-CCP method in \cite{Lipp2016} to obtain a stationary point of Problem~\ref{eqTDMAPrimalProblem2DC}. Specifically, at iteration $j+1$, we solve the following convex approximation problem, which is obtained via linearizing the concave portion in $F(\mathbf{c}, \mathbf{O}(\mathbf{Q}))$ at the solution obtained at iteration $j$, and relaxing the constraint in (\ref{eqconstrOkQstate3}) by introducing a slack variable $s$.
\begin{Problem}[Convex Approximation Problem at Iteration $j+1$] \label{eqsubproblem1DC}
{\setlength{\arraycolsep}{0.0em}
\begin{eqnarray}
{\overline{e}^{(j+1)}\triangleq} &&\min_{\substack{{s\ge 0, \mathbf{c}, \mathbf{O}, \mathbf{Y}},\\ {
{\mathbf{\hat{T}}}_{\texttt{u}}^{\mathrm{dat}}, {\mathbf{\hat{T}}}_{\texttt{d}}^{\mathrm{dat}}, {\mathbf{\hat{T}}}_{\texttt{d}}^{\mathrm{sfw}}}}} \overline{e} ( {{\mathbf{O}}, {\mathbf{Y}}, {\mathbf{\hat{T}}}_{\texttt{u}}^{\mathrm{dat}}, {\mathbf{\hat{T}}}_{\texttt{d}}^{\mathrm{dat}}, {\mathbf{\hat{T}}}_{\texttt{d}}^{\mathrm{sfw}}})  + \tau^{(j)}s  \nonumber\\
\mathrm{s.t.}\quad&&
(\ref{eqconstrcn2}), (\ref{eqconstrynmaxQstate}),(\ref{eqconstrhattukQstate}), (\ref{eqconstrhattdkQstate}), (\ref{eqconstrhattdnQstate}), (\ref{eqconstrsumtasksTimeQstate}), (\ref{eqconstrhatsumTimeTDMAQstate}), (\ref{eqconstrOkQstate}), (\ref{eqconstrcn1new1}),  \nonumber\\
&& \hat{F}(\mathbf{c}, \mathbf{O}(\mathbf{Q}); \mathbf{c}^{(j)}, \mathbf{O}^{(j)}(\mathbf{Q})) \le s, \quad \mathbf{Q}\in \mathcal{Q}, \label{eqconstrcn1new2newlinear}
\end{eqnarray}\setlength{\arraycolsep}{5pt}}where {$\overline{e}^{(j+1)}$ denotes the optimal value,} $\tau^{(j+1)}=\min\{\nu\tau^{(j)}, \tau_{\max}\}$ for some $\nu>1$ and $\tau_{\max}>0$, and $\hat{F}(\mathbf{c}, \mathbf{O}(\mathbf{Q}); \mathbf{c}^{(j)}, \mathbf{O}^{(j)}(\mathbf{Q}))
\triangleq  \sum_{k\in\mathcal{K}}\left((O_k^{(j)}(\mathbf{Q}))^2  + O_k(\mathbf{Q})( 1 - 2O_k^{(j)}(\mathbf{Q}))\right)+\sum_{n\in\mathcal{X}}\left((c_n^{(j)})^2+c_n(1-2c_n^{(j)})\right)$. Let $ (s^{(j+1)}, \mathbf{c}^{(j+1)}, \mathbf{O}^{(j+1)}, \mathbf{Y}^{(j+1)},
{\mathbf{\hat{T}}}_{\texttt{u}}^{\mathrm{dat}^{(j+1)}}, {\mathbf{\hat{T}}}_{\texttt{d}}^{\mathrm{dat}^{(j+1)}}, {\mathbf{\hat{T}}}_{\texttt{d}}^{\mathrm{sfw}^{(j+1)}} )$ denote an optimal solution at iteration $j+1$.
\end{Problem}

\begin{algorithm*}[!t]
\caption{Solving Problem \ref{eqsubproblem1DC} via ADMM}\label{algADMMCCP}
\begin{algorithmic}[1]   \small
\STATE{\textbf{input:} $B$, $C$, $D$, $K$, $N$, $R$, $F_{\mathrm{sn}}$, $\mu_{\mathrm{sn}}$, $l_{\texttt{d},n}^{\mathrm{sfw}}$, $n\in\mathcal{X}$, $F_k$, $\mu_k$, $\omega_k$, $k\in\mathcal{K}$, $n_0$, $\epsilon$ and $\rho$}
\STATE{\textbf{output:} $(s^{(j+1)}, \mathbf{c}^{(j+1)}, \mathbf{O}^{(j+1)}, \mathbf{Y}^{(j+1)},
{\mathbf{\hat{T}}}_{\texttt{u}}^{\mathrm{dat}^{(j+1)}}, {\mathbf{\hat{T}}}_{\texttt{d}}^{\mathrm{dat}^{(j+1)}}, {\mathbf{\hat{T}}}_{\texttt{d}}^{\mathrm{sfw}^{(j+1)}})$}
\STATE{\textbf{initialization:} choose any $s_0>0$, $c_{0,n}\in[0,1]$, $\theta_{0,n}(\mathbf{Q})>0$, $\xi_{0}(\mathbf{Q})>0$, $n\in\mathcal{X}$, $\mathbf{Q}\in\mathcal{Q}$, and set $t:= 0$}
\REPEAT
\STATE {obtain $({{\hat{S}}_{t+1}(\mathbf{Q}), \mathbf{\hat{C}}_{t+1}(\mathbf{Q}), \mathbf{O}_{t+1}(\mathbf{Q}), \mathbf{Y}_{t+1}(\mathbf{Q}), {\mathbf{\hat{T}}}_{\texttt{u},{t+1}}^{\mathrm{dat}}(\mathbf{Q}), {\mathbf{\hat{T}}}_{\texttt{d},{t+1}}^{\mathrm{dat}}(\mathbf{Q}), {\mathbf{\hat{T}}}_{\texttt{d},{t+1}}^{\mathrm{sfw}}(\mathbf{Q}) } )$ by solving the problem in (\ref{eqprobCCPSub}) for all $\mathbf{Q}\in \mathcal{Q}$ via an interior point method }
\STATE {compute $c_{t+1,n}$, ${\theta}_{t+1,n}(\mathbf{Q})$, $n\in\mathcal{X}$, $\mathbf{Q}\in\mathcal{Q}$, $s_{t+1}$ and ${\xi}_{t+1}(\mathbf{Q})$, $\mathbf{Q}\in\mathcal{Q}$ according to (\ref{eqcalglobalvar}), (\ref{eqADMMThetaupdate}), (\ref{eqcandsupdateCCP}) and (\ref{eqthetaandxiupdateCCP})}
\STATE {set $t:= t+1$}
\UNTIL {{$\|(\mathbf{\hat{C}}_{t+1}(\mathbf{Q})-\mathbf{c}_{t+1},  {\hat{S}}_{t+1}(\mathbf{Q})-s_{t+1})\|_2\le\epsilon$ and $\|\left(\mathbf{c}_{t+1}-\mathbf{c}{_{t}}, s_{t+1}(\mathbf{Q})-s_t\right)\rho\|_2\le\epsilon$}}
\STATE{set $ (s^{(j+1)}, \mathbf{c}^{(j+1)}, \mathbf{O}^{(j+1)}, \mathbf{Y}^{(j+1)},
{\mathbf{\hat{T}}}_{\texttt{u}}^{\mathrm{dat}^{(j+1)}}, {\mathbf{\hat{T}}}_{\texttt{d}}^{\mathrm{dat}^{(j+1)}}, {\mathbf{\hat{T}}}_{\texttt{d}}^{\mathrm{sfw}^{(j+1)}})$ \\ $ := (s_{t+1}, \mathbf{c}_{t+1,n}, {\hat{S}}_{t+1}(\mathbf{Q}), \mathbf{\hat{C}}_{t+1}(\mathbf{Q}), \mathbf{O}_{t+1}(\mathbf{Q}), \mathbf{Y}_{t+1}(\mathbf{Q}), {\mathbf{\hat{T}}}_{\texttt{u},{t+1}}^{\mathrm{dat}}(\mathbf{Q}), {\mathbf{\hat{T}}}_{\texttt{d},{t+1}}^{\mathrm{dat}}(\mathbf{Q}), {\mathbf{\hat{T}}}_{\texttt{d},{t+1}}^{\mathrm{sfw}}(\mathbf{Q})  )$}
\end{algorithmic}
\end{algorithm*}

\begin{algorithm*}[t]
\caption{  Stationary Point of Problem~\ref{eqTDMAPrimalProblem2} Based on Penalty-CCP}\label{algCCPAndADMM}
\begin{algorithmic}[1]     \small
\STATE{\textbf{input:} $B$, $C$, $D$, $K$, $N$, $R$, $F_{\mathrm{sn}}$, $\mu_{\mathrm{sn}}$, $l_{\texttt{d},n}^{\mathrm{sfw}}$, $n\in\mathcal{X}$, $F_k$, $\mu_k$, $\omega_k$, $k\in\mathcal{K}$, $n_0$, $\epsilon$, $\rho$, $\tau^{(0)}$, $\tau_{\max}$ and $\nu $}
\STATE{ \textbf{output:} $(\mathbf{c}^{(j+1)}, \mathbf{O}^{(j+1)}, \mathbf{Y}^{(j+1)},
{\mathbf{\hat{T}}}_{\texttt{u}}^{\mathrm{dat}^{(j+1)}}, {\mathbf{\hat{T}}}_{\texttt{d}}^{\mathrm{dat}^{(j+1)}}, {\mathbf{\hat{T}}}_{\texttt{d}}^{\mathrm{sfw}^{(j+1)}})$}
\STATE{\textbf{initialization:} choose any $(s^{(0)}, \mathbf{c}^{(0)}, \mathbf{O}^{(0)}, \mathbf{Y}^{(0)},
{\mathbf{\hat{T}}}_{\texttt{u}}^{\mathrm{dat}^{(0)}}, {\mathbf{\hat{T}}}_{\texttt{d}}^{\mathrm{dat}^{(0)}}, {\mathbf{\hat{T}}}_{\texttt{d}}^{\mathrm{sfw}^{(0)}})$ and set $j:= 0$.}
\REPEAT
\STATE {obtain $(s^{(j+1)}, \mathbf{c}^{(j+1)}, \mathbf{O}^{(j+1)}, \mathbf{Y}^{(j+1)},
{\mathbf{\hat{T}}}_{\texttt{u}}^{\mathrm{dat}^{(j+1)}}, {\mathbf{\hat{T}}}_{\texttt{d}}^{\mathrm{dat}^{(j+1)}}, {\mathbf{\hat{T}}}_{\texttt{d}}^{\mathrm{sfw}^{(j+1)}})$ by using Algorithm~\ref{algADMMCCP}}
\STATE {set $\tau^{(j+1)}:=\min\{\nu\tau^{(j)}, \tau_{\max}\}$}
\STATE {set $j:= j+1$}
\UNTIL {$\|(s^{(j)}, \mathbf{c}^{(j)}, \mathbf{O}^{(j)}, \mathbf{Y}^{(j)},
{\mathbf{\hat{T}}}_{\texttt{u}}^{\mathrm{dat}^{(j)}}, {\mathbf{\hat{T}}}_{\texttt{d}}^{\mathrm{dat}^{(j)}}, {\mathbf{\hat{T}}}_{\texttt{d}}^{\mathrm{sfw}^{(j)}})$ \\ \qquad $- (s^{(j-1)}, \mathbf{c}^{(j-1)}, \mathbf{O}^{(j-1)}, \mathbf{Y}^{(j-1)},
{\mathbf{\hat{T}}}_{\texttt{u}}^{\mathrm{dat}^{(j-1)}}, {\mathbf{\hat{T}}}_{\texttt{d}}^{\mathrm{dat}^{(j-1)}}, {\mathbf{\hat{T}}}_{\texttt{d}}^{\mathrm{sfw}^{(j-1)}})\|_2\le\epsilon$}
\end{algorithmic}
\end{algorithm*}

Since the objective function is convex and all constraint functions are linear, Problem~\ref{eqsubproblem1DC} is convex, and can be solved by an interior point method. Note that, the computational complexity of each iteration of an interior point method used for solving Problem~\ref{eqsubproblem1DC} is given by {$\mathcal{O}\left((K+N)^3N^{3K}\Delta^{3K}\right)$ \cite{convexoptimization}.} Similarly, to reduce computation time in practice, we propose a fast algorithm to solve Problem~\ref{eqsubproblem1DC} using ADMM \cite{admmSboyd}. We first introduce new variables $\hat{S}(\mathbf{Q})$ and new constraints
\begin{align}
  &\hat{S}(\mathbf{Q})=s, \quad \mathbf{Q}\in\mathcal{Q},\label{eqconsCCP}\\
  &\hat{F}(\mathbf{c}, \mathbf{O}(\mathbf{Q}); \mathbf{c}^{(j)}, \mathbf{O}^{(j)}(\mathbf{Q})) \le \hat{S}(\mathbf{Q}), \quad \mathbf{Q}\in \mathcal{Q}, \label{eqconstrcn1new2newlinearhatS}
\end{align}and define a new set as $\mathcal{F}_2(\mathbf{Q}) \triangleq  \{(\hat{S}(\mathbf{Q}), \mathbf{\hat{C}}(\mathbf{Q}), {\mathbf{O}}(\mathbf{Q}), {\mathbf{Y}}(\mathbf{Q}), {\mathbf{\hat{T}}}_{\texttt{u}}^{\mathrm{dat}}(\mathbf{Q}), {\mathbf{\hat{T}}}_{\texttt{d}}^{\mathrm{dat}}(\mathbf{Q}), {\mathbf{\hat{T}}}_{\texttt{d}}^{\mathrm{sfw}}(\mathbf{Q}))|\\ (\ref{eqconstrynmaxQstate}), (\ref{eqconstrhattukQstate}), (\ref{eqconstrhattdkQstate}), (\ref{eqconstrhattdnQstate}), (\ref{eqconstrOkQstate}), (\ref{eqconstrhatcn3ADMMnew}),   (\ref{eqconstrhatcn3ADMM}), (\ref{eqconstrsumhatcn1ADMM}), (\ref{eqconstrhatcn1ADMM}), (\ref{eqconstrcn1new2newlinearhatS})\}$,~where the constraints in (\ref{eqconsCCP}) are referred to as consensus constraints and $\hat{S}$ is a mapping of $\mathbf{Q}$. {Here, the global variables and the local variables are identified as $(\mathbf{c},s)$ and $(\hat{S}(\mathbf{Q}), \mathbf{\hat{C}}(\mathbf{Q}), {\mathbf{O}(\mathbf{Q})}, {\mathbf{Y}(\mathbf{Q})},
{\mathbf{\hat{T}}}_{\texttt{u}}^{\mathrm{dat}}(\mathbf{Q}), {\mathbf{\hat{T}}}_{\texttt{d}}^{\mathrm{dat}}(\mathbf{Q}), {\mathbf{\hat{T}}}_{\texttt{d}}^{\mathrm{sfw}}(\mathbf{Q}))$ for all $\mathbf{Q}\in\mathcal{Q}$, respectively.} Then, we choose any $s_0$, $c_{0,n}$, $\theta_{0,n}(\mathbf{Q})$ and $\xi_{0}(\mathbf{Q})$, $n\in\mathcal{X}$, $\mathbf{Q}\in\mathcal{Q}$, and at iteration $t+1$, for each $\mathbf{Q} \in \mathcal{Q}$, obtain $({\hat{S}}_{t+1}(\mathbf{Q}), \mathbf{\hat{C}}_{t+1}(\mathbf{Q}), \mathbf{O}_{t+1}(\mathbf{Q}), \mathbf{Y}_{t+1}(\mathbf{Q}), {\mathbf{\hat{T}}}_{\texttt{u},{t+1}}^{\mathrm{dat}}(\mathbf{Q}), \\ {\mathbf{\hat{T}}}_{\texttt{d},{t+1}}^{\mathrm{dat}}(\mathbf{Q}), {\mathbf{\hat{T}}}_{\texttt{d},{t+1}}^{\mathrm{sfw}}(\mathbf{Q}) )$ by solving the problem in  (\ref{eqprobCCPSub})
\begin{figure*}[!t]
{\begingroup\makeatletter\def\f@size{9.5}\check@mathfonts
\def\maketag@@@#1{\hbox{\m@th\normalsize\normalfont#1}}
\begin{align}\label{eqprobCCPSub}
& ({{\hat{S}}_{t+1}(\mathbf{Q}), \mathbf{\hat{C}}_{t+1}(\mathbf{Q}), \mathbf{O}_{t+1}(\mathbf{Q}), \mathbf{Y}_{t+1}(\mathbf{Q}), {\mathbf{\hat{T}}}_{\texttt{u},{t+1}}^{\mathrm{dat}}(\mathbf{Q}), {\mathbf{\hat{T}}}_{\texttt{d},{t+1}}^{\mathrm{dat}}(\mathbf{Q}), {\mathbf{\hat{T}}}_{\texttt{d},{t+1}}^{\mathrm{sfw}}(\mathbf{Q}) }  )  \triangleq \nonumber\\
    &\arg\min\limits_{\substack{ {({\hat{S}}(\mathbf{Q}), \mathbf{\hat{C}}(\mathbf{Q}), {\mathbf{O}}(\mathbf{Q}),}\\  { \mathbf{Y}(\mathbf{Q}),{\mathbf{\hat{T}}}_{\texttt{u}}^{\mathrm{dat}}(\mathbf{Q}),  {\mathbf{\hat{T}}}_{\texttt{d}}^{\mathrm{dat}}(\mathbf{Q}), {\mathbf{\hat{T}}}_{\texttt{d}}^{\mathrm{sfw}}(\mathbf{Q}) ) \in\mathcal{F}_2(\mathbf{Q})}}}  f_2 ( {\hat{S}}(\mathbf{Q}), \mathbf{\hat{C}}(\mathbf{Q}), {\mathbf{O}(\mathbf{Q})},\mathbf{Y}(\mathbf{Q}),
    {\mathbf{\hat{T}}}_{\texttt{u}}^{\mathrm{dat}}(\mathbf{Q}), {\mathbf{\hat{T}}}_{\texttt{d}}^{\mathrm{dat}}(\mathbf{Q}), {\mathbf{\hat{T}}}_{\texttt{d}}^{\mathrm{sfw}}(\mathbf{Q})  )
\end{align}\endgroup}
 \noindent\rule[0.8\baselineskip]{\textwidth}{0.1pt}
\end{figure*}
and compute $c_{t+1,n}$, ${\theta}_{t+1,n}(\mathbf{Q})$, $s_{t+1}$ and ${\xi}_{t+1}(\mathbf{Q})$ by using (\ref{eqcalglobalvar}), (\ref{eqADMMThetaupdate}) and
\begin{align}
&s_{t+1}=\frac{1}{\left|\mathcal{Q}\right|} \sum_{\mathbf{Q}\in\mathcal{Q}}\left({\hat{S}}_{t+1}(\mathbf{Q})+(1/\rho)\xi_{t}(\mathbf{Q})\right),\label{eqcandsupdateCCP}\\
&{\xi}_{t+1}(\mathbf{Q}) = {\xi}_{t}(\mathbf{Q}) + \rho ({\hat{S}}_{t+1}(\mathbf{Q})-{s}_{t+1}),\quad\mathbf{Q}\in\mathcal{Q}.\label{eqthetaandxiupdateCCP}
\end{align}Here,
\begin{align*}
& f_2 ( {\hat{S}}(\mathbf{Q}), \mathbf{\hat{C}}(\mathbf{Q}), {\mathbf{O}(\mathbf{Q})},\mathbf{Y}(\mathbf{Q}),
    {\mathbf{\hat{T}}}_{\texttt{u}}^{\mathrm{dat}}(\mathbf{Q}), {\mathbf{\hat{T}}}_{\texttt{d}}^{\mathrm{dat}}(\mathbf{Q}), {\mathbf{\hat{T}}}_{\texttt{d}}^{\mathrm{sfw}}(\mathbf{Q})  ) \nonumber\\
     &\triangleq p_{\mathbf{Q}}(\mathbf{q})  e ( {{\mathbf{O}(\mathbf{Q})},\mathbf{Y}(\mathbf{Q}),
    {\mathbf{\hat{T}}}_{\texttt{u}}^{\mathrm{dat}}(\mathbf{Q}), {\mathbf{\hat{T}}}_{\texttt{d}}^{\mathrm{dat}}(\mathbf{Q}), {\mathbf{\hat{T}}}_{\texttt{d}}^{\mathrm{sfw}}(\mathbf{Q}), \mathbf{Q}} )  \\
    &\quad\quad  +  \displaystyle \sum_{n\in\mathcal{X}} \left(\theta_{t,n}(\mathbf{Q}) ( {\hat{C}}_n(\mathbf{Q})-c_{t,n}) + \frac{\rho}{2}  ( {\hat{C}}_n(\mathbf{Q})-c_{t,n})^2\right)\\
    &\quad\quad  +  \displaystyle  \xi_{t}(\mathbf{Q}) ( \hat{S}(\mathbf{Q})-s_t) + \frac{\rho}{2}  ( {\hat{S}}(\mathbf{Q})-s_t)^2 + \tau^{(j+1)} \hat{S}(\mathbf{Q})  ,
\end{align*}and the update equations in (\ref{eqthetaandxiupdateCCP}) are to obtain the Lagrange multipliers corresponding to the consensus constraints in (\ref{eqconsCCP}). Note that the problem in (\ref{eqprobCCPSub}) is a convex problem, and hence can be solved by an interior point method. The details of ADMM are summarized in Algorithm~\ref{algADMMCCP}. Note that, the computational complexity of each iteration of an interior point method used for solving the problem in (\ref{eqprobCCPSub}) for all $\mathbf{Q}\in\mathcal{Q}$ is given by $\mathcal{O}((K+N)^3N^K\Delta^K)$ and the computational complexities for computing $c_{t+1,n}$, ${\theta}_{t+1,n}(\mathbf{Q})$, $n\in\mathcal{X}$, $\mathbf{Q}\in\mathcal{Q}$, $s_{t+1}$ and ${\xi}_{t+1}(\mathbf{Q})$, $\mathbf{Q}\in\mathcal{Q}$ are given by $\mathcal{O}(N^{K+1}\Delta^K)$. Although the number of iterations of Algorithm~\ref{algADMMCCP} cannot be analytically characterized, numerical results show that Algorithm~\ref{algADMMCCP} terminates in a few iterations. Therefore, solving Problem \ref{eqsubproblem1DC} using Algorithm~\ref{algADMMCCP} is of much lower computational complexity than using an interior point~method.

The details of Penalty-CCP are summarized in Algorithm~\ref{algCCPAndADMM}.  It is known that the sequence $\{(s^{(j)}, \mathbf{c}^{(j)}, \mathbf{O}^{(j)}, \mathbf{Y}^{(j)}, {\mathbf{\hat{T}}}_{\texttt{u}}^{\mathrm{dat}^{(j)}}, \\ {\mathbf{\hat{T}}}_{\texttt{d}}^{\mathrm{dat}^{(j)}}, {\mathbf{\hat{T}}}_{\texttt{d}}^{\mathrm{sfw}^{(j)}})\}$ generated by Algorithm \ref{algCCPAndADMM} is convergent, and the limit point of this sequence is a stationary point of Problem~\ref{eqTDMAPrimalProblem2DC} \cite{Lipp2016}. We can run Algorithm \ref{algCCPAndADMM} multiple times, each with a random initial point {satisfying the constraints in (\ref{eqconstrcn2}), (\ref{eqconstrynmaxQstate})--(\ref{eqconstrcn1new1})}, and select the stationary point with the minimum weighted sum energy consumption among those with zero penalty, denoted by $(s^{\dag}, \mathbf{c}^{\dag}, {\mathbf{O}^{\dag}}, {\mathbf{Y}^{\dag}}, {\mathbf{\hat{T}}}_{\texttt{u}}^{\mathrm{dat}^{\dag}}, {\mathbf{\hat{T}}}_{\texttt{d}}^{\mathrm{dat}^{\dag}}, {\mathbf{\hat{T}}}_{\texttt{d}}^{\mathrm{sfw}^{\dag}})$, with slight abuse of notation. {By the equivalence between Problem \ref{eqTDMAPrimalProblem2} and Problem~\ref{eqTDMAPrimalProblem2DC}, $(\mathbf{c}^{\dag}, {\mathbf{O}^{\dag}}, {\mathbf{Y}^{\dag}}, {\mathbf{\hat{T}}}_{\texttt{u}}^{\mathrm{dat}^{\dag}}, {\mathbf{\hat{T}}}_{\texttt{d}}^{\mathrm{dat}^{\dag}}, {\mathbf{\hat{T}}}_{\texttt{d}}^{\mathrm{sfw}^{\dag}})$ is a stationary point of Problem \ref{eqTDMAPrimalProblem2}. Later in Section VI,  we shall see that this stationary point has promising performance.}

\begin{figure*}[!t]
    \centering
         \subfloat[Convergence of Algorithm~\ref{algADMM}.]{\includegraphics[width=0.33\textwidth]{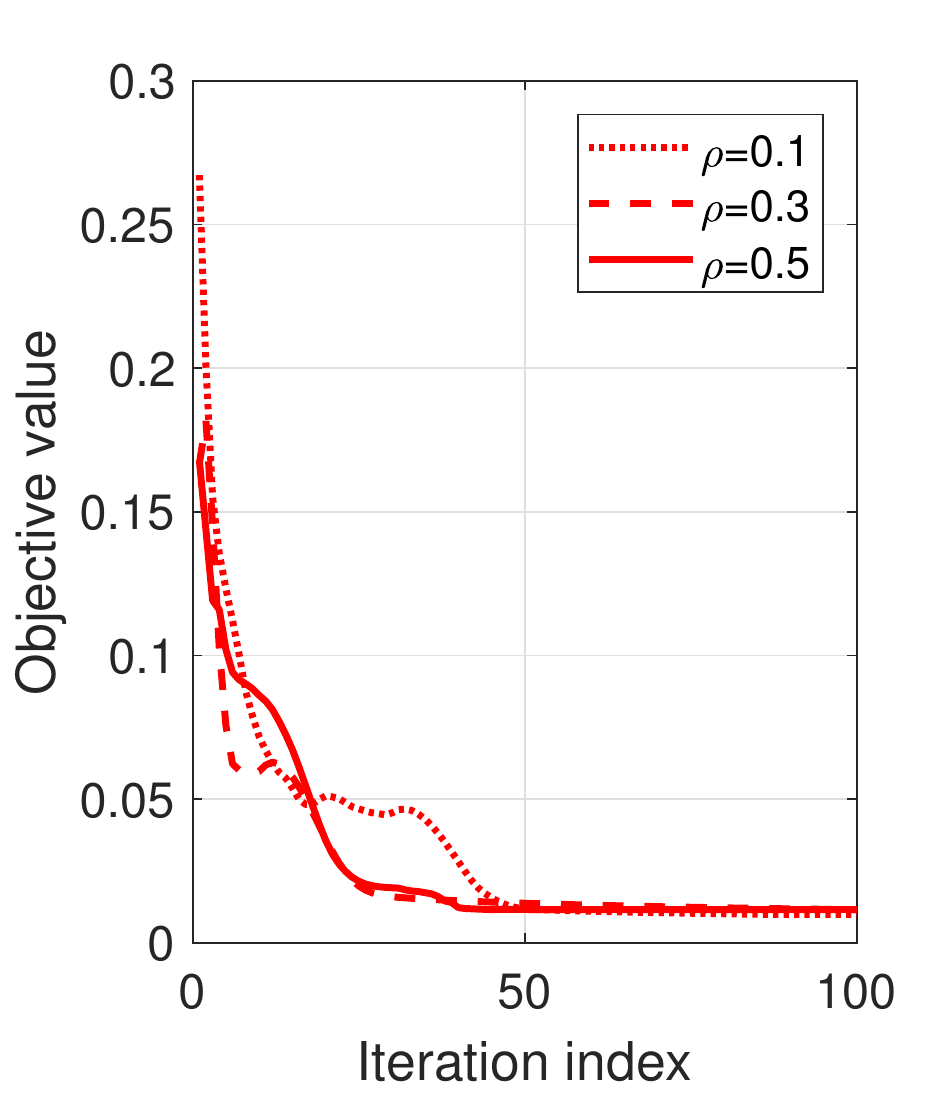}}
         \subfloat[Convergence of Algorithm~\ref{algADMMCCP}.]{\includegraphics[width=0.33\textwidth]{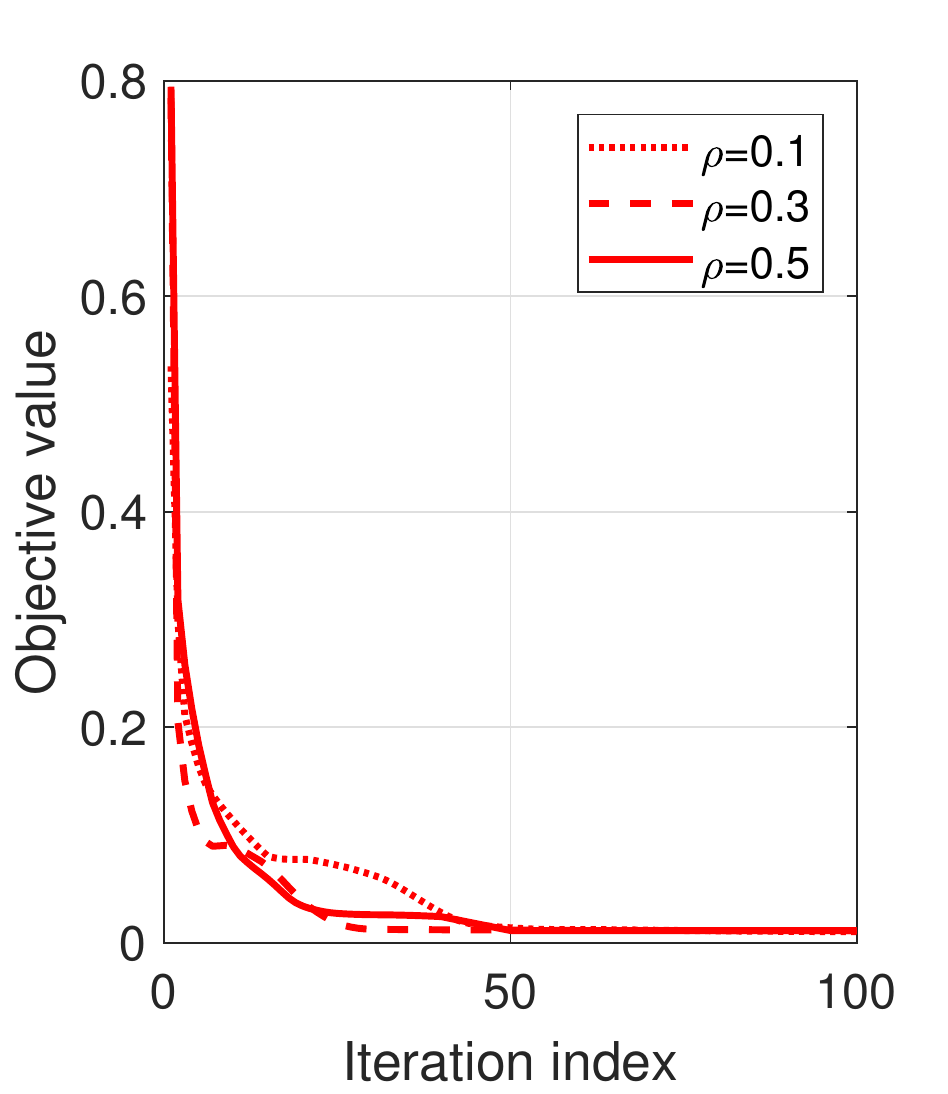}}
         \subfloat[Convergence of Algorithm~\ref{algCCPAndADMM}.]{\includegraphics[width=0.33\textwidth]{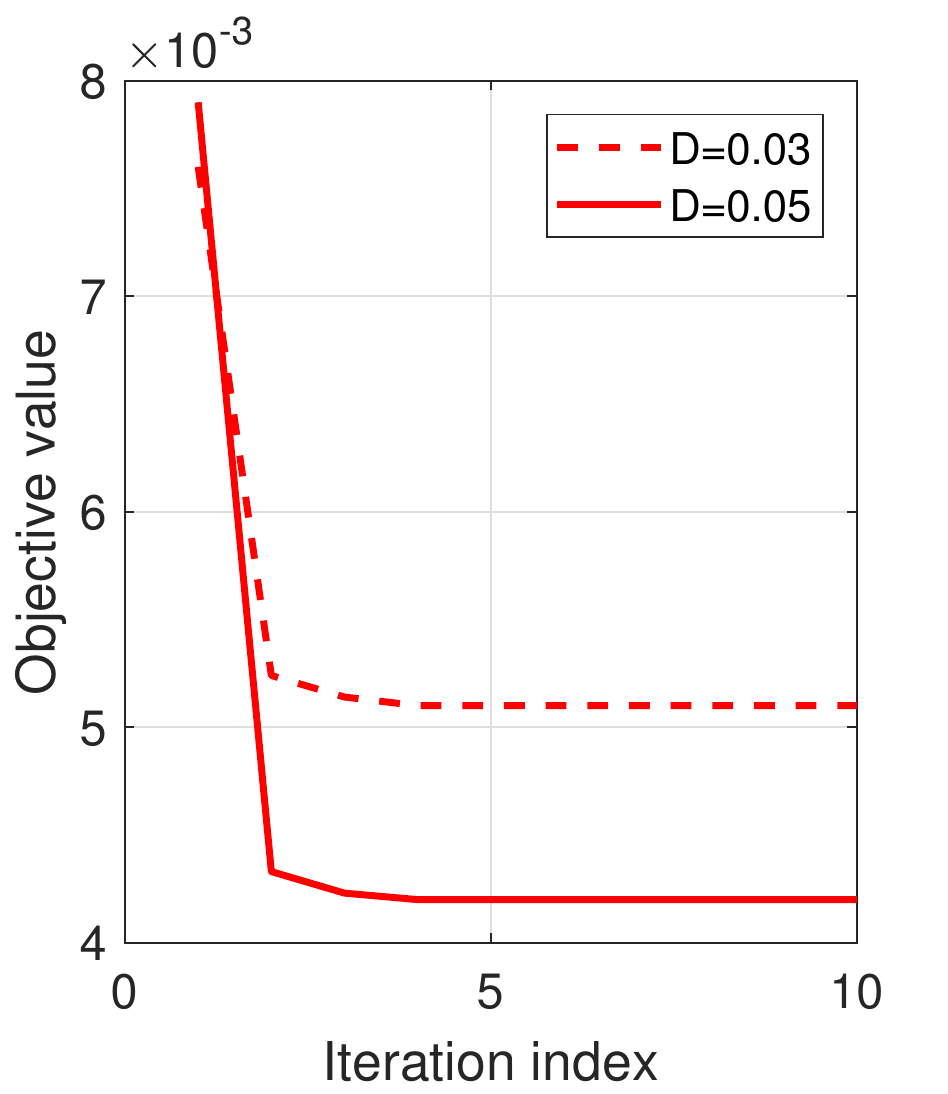}}
        \caption{(a) Objective value of Problem \ref{eqTDMAPrimalProblem2CR} obtained by Algorithm~\ref{algADMM} versus iteration index at $D=0.03$. (b) Objective value of Problem \ref{eqsubproblem1DC} obtained by Algorithm~\ref{algADMMCCP} versus iteration index at $\tau^{(j+1)}=0$ and $D=0.03$. (c) Objective value of Problem \ref{eqTDMAPrimalProblem2} obtained by Algorithm~\ref{algCCPAndADMM} versus iteration index.  }\label{figsdpasymptoticZero}
\end{figure*}

\section{Numerical Results}\label{secSimResult}

\begin{table}
\centering\small
\caption{ Simulation Parameters.}\label{simpara}
\begin{tabular}{p{1.2cm} p{1.5cm} | p{1.2cm} p{1.8cm}}
  \hline\hline
  Parameter                             &   Value   & Parameter                             &   Value                                          \\ \hline
  $F_{\mathrm{sn}}$                     &   6 GHz   & $l_{\texttt{d},n}^{\mathrm{sfw}}$     &   $4n\times 10^5$ bits \\
  $F_k$                                 &   0.7 GHz & $\mu_{\mathrm{sn}}$                   &   $10^{-29}$      \\
  $\mu_k$                               &   $5\times10^{-27}$   & $B$                                   &   20 MHz         \\
  $n_0$                                 &   $10^{-9}$ W         & $R$                                   &   $10^8$ bps      \\
  $C$                                   & $1\times10^6$ bits    & $\epsilon$                            & $10^{-3}$ \\
  $\rho$                                & 0.1                   & $\tau^{(0)}$                          & $10^{-3}$\\
  $\nu$                                 & 2                     & $\tau^{\max}$                         & 1    \\
  \hline\hline
\end{tabular}
\end{table}

In this section,  we first show the convergence and complexity of the proposed solutions, and then compare the average total energy consumption of the proposed solutions with those of some baseline schemes. The proposed solutions and baseline schemes are implemented using MATLAB. Unless otherwise stated, the main simulation environment settings are summarized in Table~\ref{simpara} \cite{MECCachingCui2017, MECYouTWC}.
We consider the same weight factor for all users and thus the average weighted sum energy consumption becomes the average total energy consumption (in Joule).  Assume that $X_k$, $k\in\mathcal{K}$ follows the same Zipf distribution, i.e., $p_{X_k}(n) = \frac{n^{-\gamma}}{\sum_{i\in\mathcal{X}}i^{-\gamma}}$ for all $k\in\mathcal{K}$, $n\in\mathcal{X}$, where $\gamma=0.8$ is the Zipf exponent.
Set $\mathcal{L}_{\texttt{u}}^{\mathrm{dat}}=\{1\times 10^5, 9\times 10^5\}$, $p_{L_{\texttt{u},k}^{\mathrm{dat}}} (1\times 10^5) = 0.4$, $p_{L_{\texttt{u},k}^{\mathrm{dat}}} (9\times 10^5) = 0.6$, $\mathcal{L}_{\texttt{e}}^{\mathrm{dat}}=\{1\times 10^6, 2\times 10^6\}$, $p_{L_{\texttt{e},k}^{\mathrm{dat}}} (1\times 10^6) = 0.9$, $p_{L_{\texttt{e},k}^{\mathrm{dat}}} (2\times 10^6) = 0.1$, $\mathcal{L}_{\texttt{d}}^{\mathrm{dat}}=\{1\times 10^4, 3\times 10^4\}$, $p_{L_{\texttt{d},k}^{\mathrm{dat}}} (1\times 10^4) = 0.9$, $p_{L_{\texttt{d},k}^{\mathrm{dat}}} (2 \times 10^4) = 0.1$, $\mathcal{H}=\{1\times 10^{-7},1\times10^{-8}\}$, $p_{H_k}(1\times 10^{-7})=0.65$ and  $p_{H_k}(1\times10^{-8})=0.35$, for all $k\in\mathcal{K}$. For ease of illustration, we assume that all random variables are independent.

\begin{figure*}[!t]
    \centering
         \subfloat[Complexity of Algorithm \ref{algADMM}.]{\includegraphics[width=0.33\textwidth]{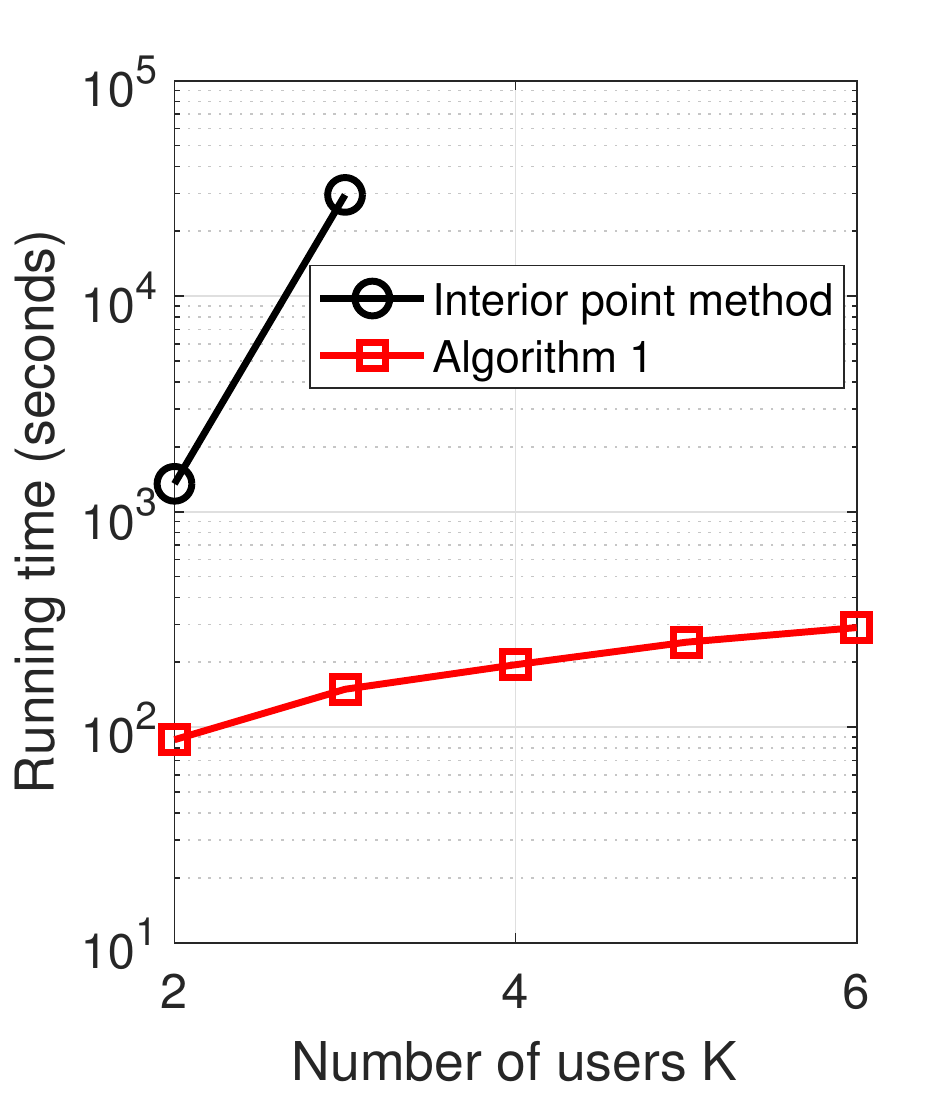}}
         \subfloat[Complexity of Algorithm \ref{algADMMCCP}.]{\includegraphics[width=0.33\textwidth]{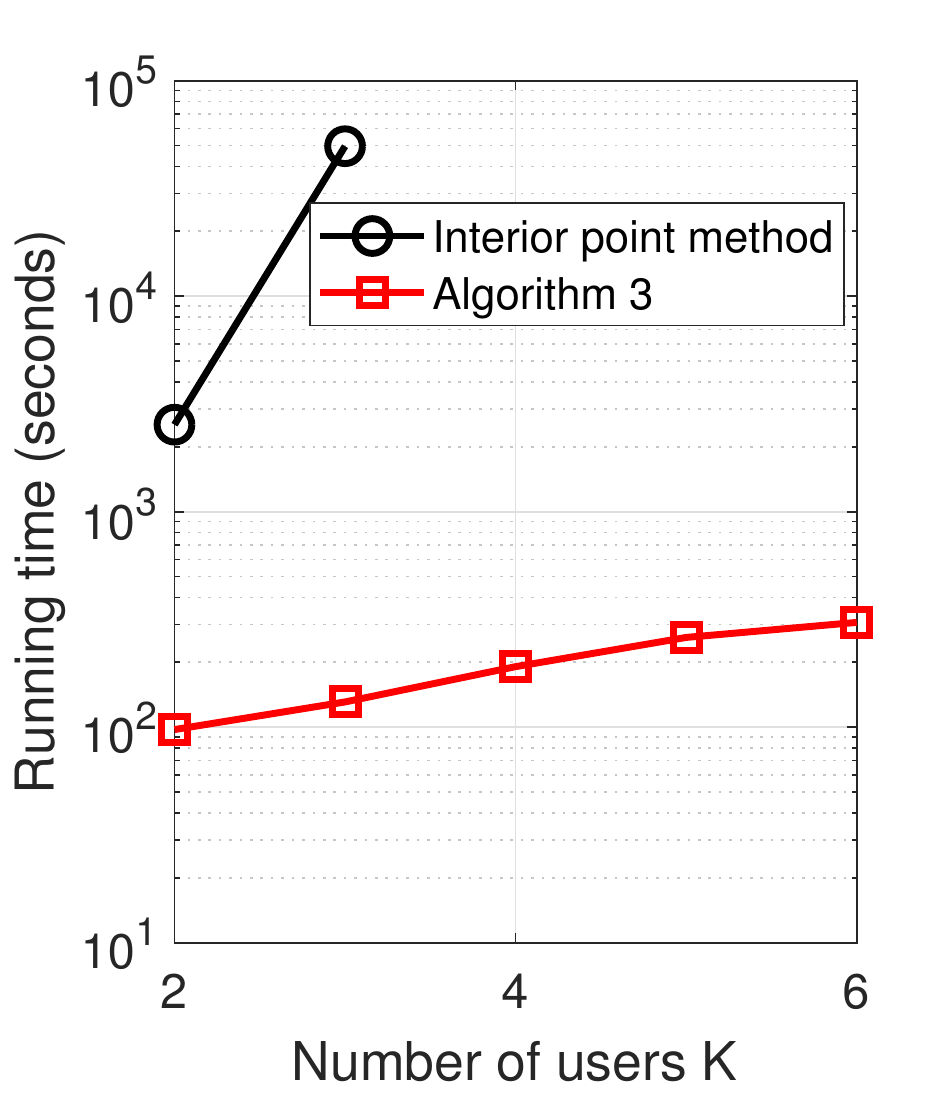}}
         \subfloat[Complexity of Algorithms~\ref{algrecovery} and \ref{algCCPAndADMM}.]{\includegraphics[width=0.33\textwidth]{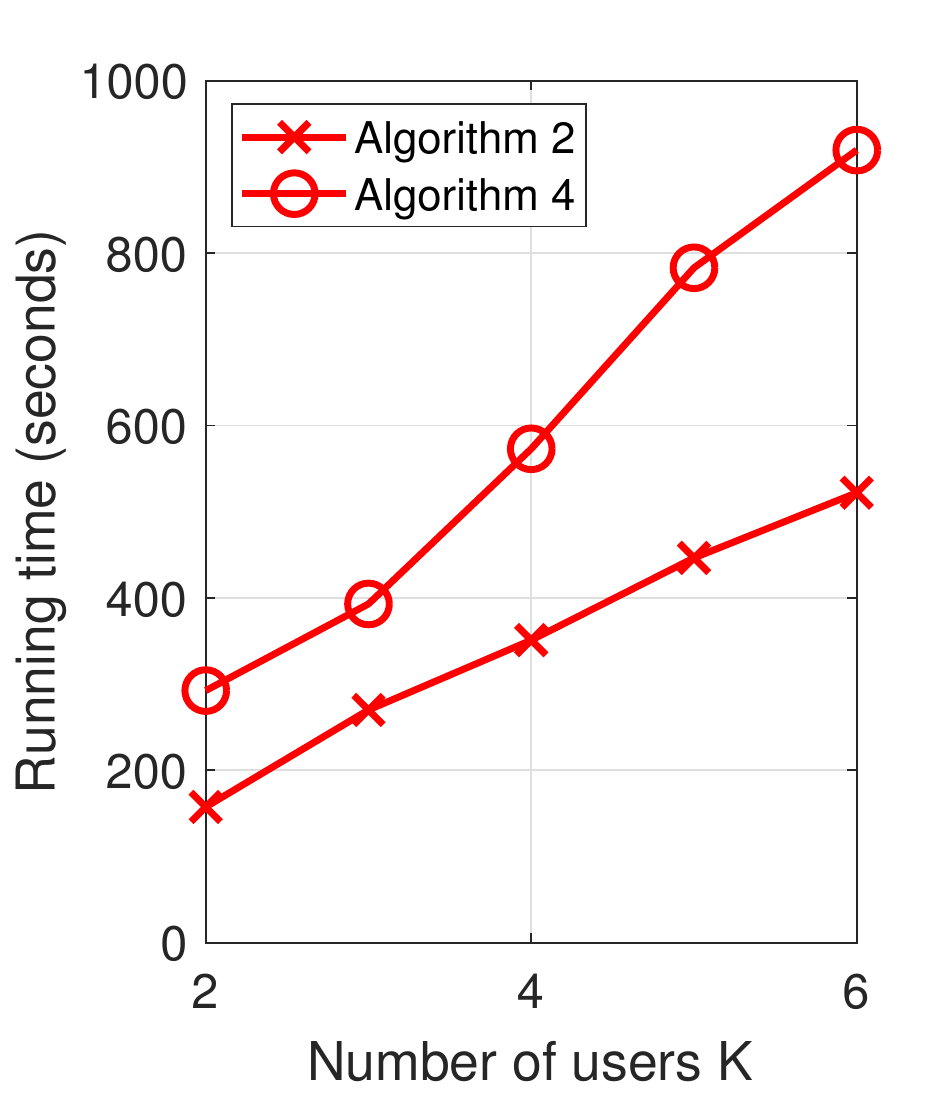}}
        \caption{(a) Running time of Algorithm \ref{algADMM} versus number of users $K$. (b) Running time of Algorithm~\ref{algADMMCCP} versus number of users $K$. (c) Running times of Algorithms~\ref{algrecovery} and \ref{algCCPAndADMM} versus number of users $K$.}\label{figcomputingtime}
\end{figure*}

\subsection{Convergence and Complexity of Proposed Solutions}
In this subsection, we show the convergence and complexity of the proposed solutions.  Fig.~\ref{figsdpasymptoticZero} shows the convergence of the iterative algorithms, i.e., Algorithm~\ref{algADMM}, Algorithm~\ref{algADMMCCP} and  Algorithm~\ref{algCCPAndADMM}, where $K=2$ and $N=4$. From Fig. \ref{figsdpasymptoticZero}, we can see that Algorithm~\ref{algADMM}, Algorithm~\ref{algADMMCCP} and Algorithm \ref{algCCPAndADMM} converge very fast. In addition, for each of Algorithm~\ref{algADMM} and Algorithm~\ref{algADMMCCP}, the iterative progresses at different $\rho$  converge to the same objective value eventually.   Fig.~\ref{figcomputingtime} shows the {computational} complexities of the proposed {algorithms}, where $N=4$ and  $D=0.03$.  From Fig.~\ref{figcomputingtime}(a) and Fig.~\ref{figcomputingtime}(b), we can {see that} the computational complexity of an interior point method is much higher than {those of the two ADMM algorithms, demonstrating the advantages of the proposed solutions in terms of computational complexity reduction. In addition, from Fig.~\ref{figcomputingtime}(c), we can see that the computational complexity of Algorithm~\ref{algrecovery} is much lower than that of Algorithm~\ref{algCCPAndADMM}.}

\begin{figure*}[!t]
    \centering
        \subfloat[$N=5$, $\gamma=0.8$, $C=5\times 10^6$, $D=0.1$.]{\includegraphics[width=0.5\textwidth]{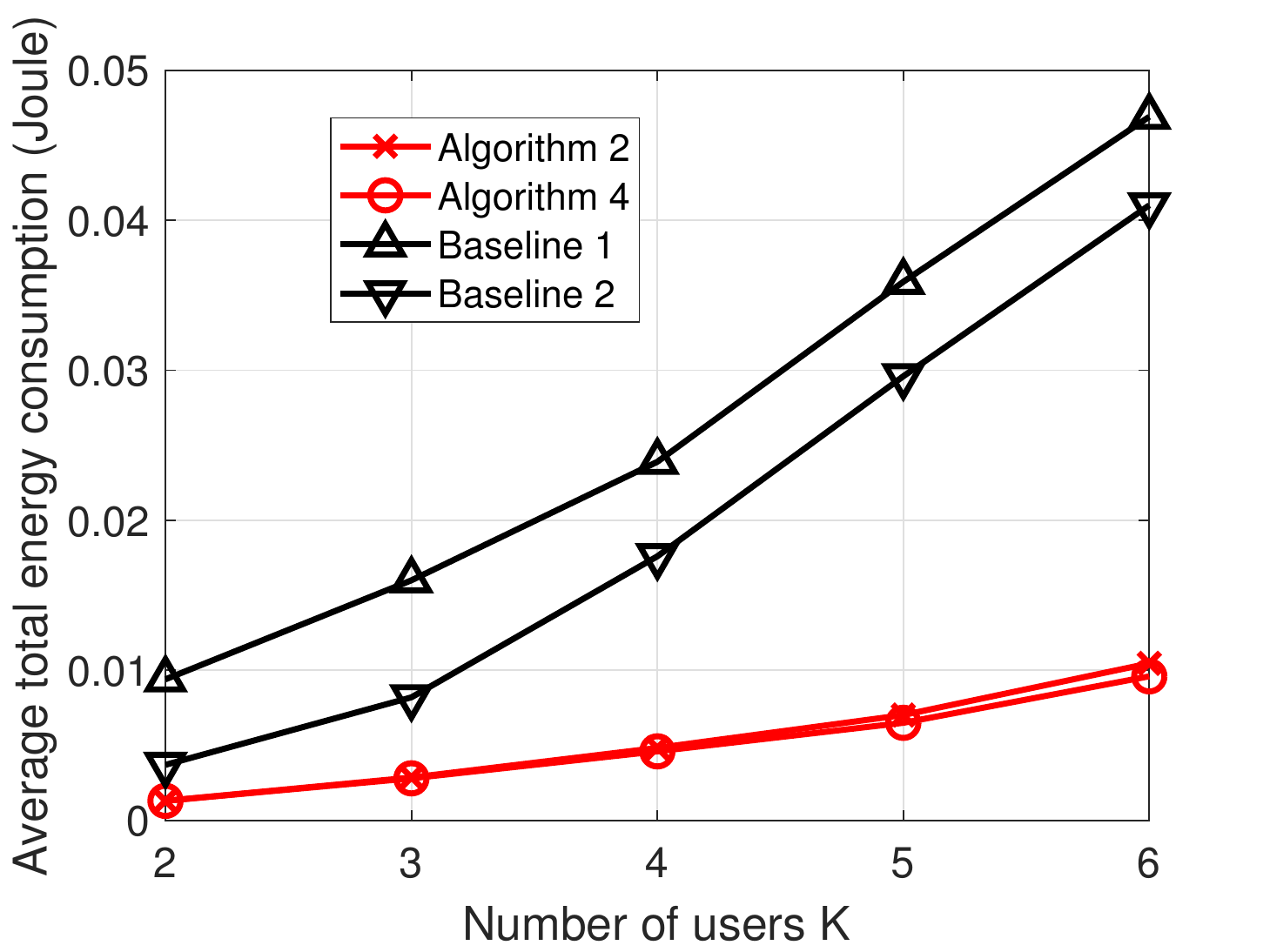}}
        \subfloat[$K=2$, $\gamma=0.6$, $C=1\times 10^6$, $D=0.08$.]{\includegraphics[width=0.5\textwidth]{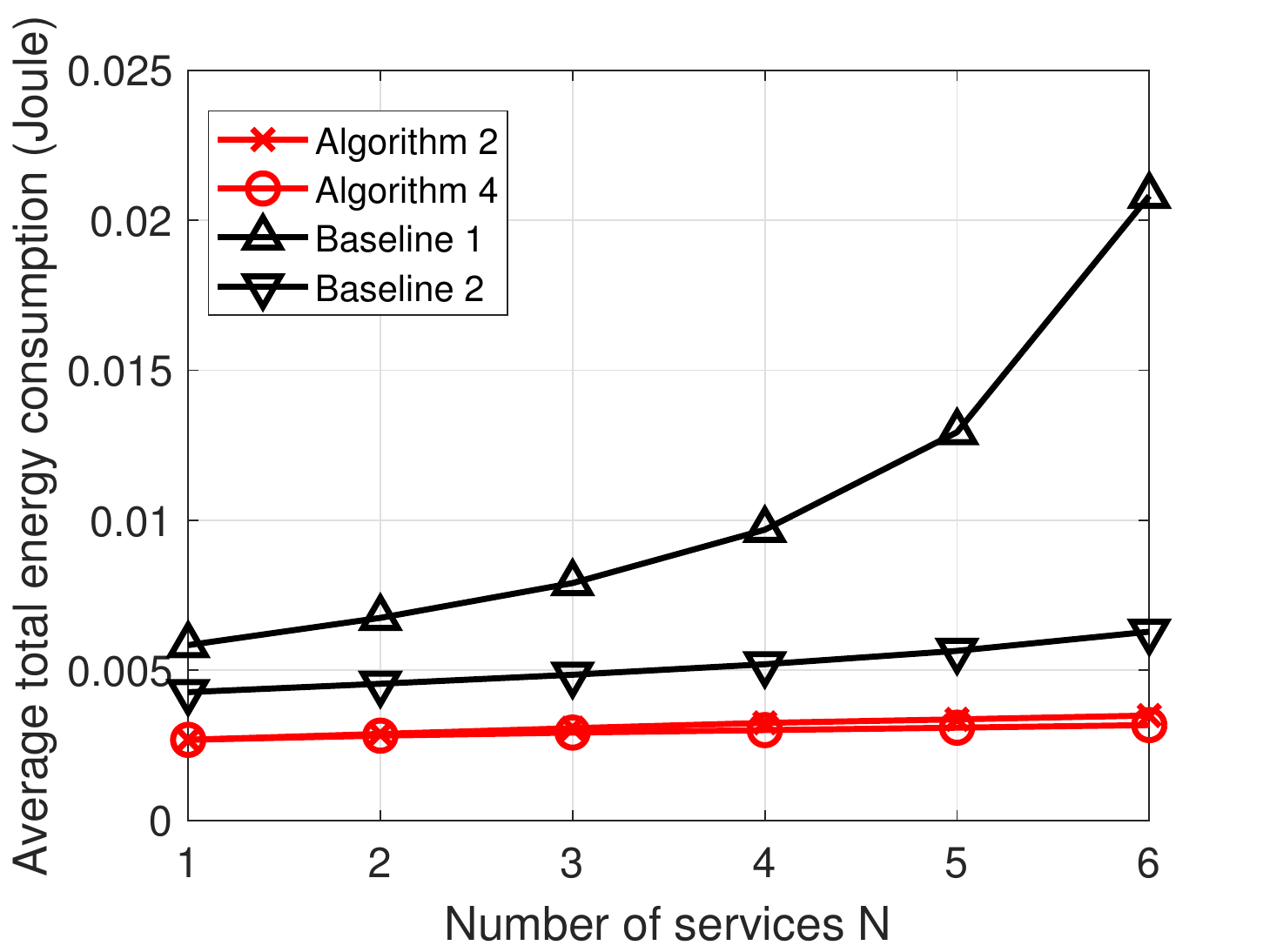}}\\
        \subfloat[$K=2$, $N=4$, $C=2\times 10^6$, $D=0.05$. ]{\includegraphics[width=0.5\textwidth]{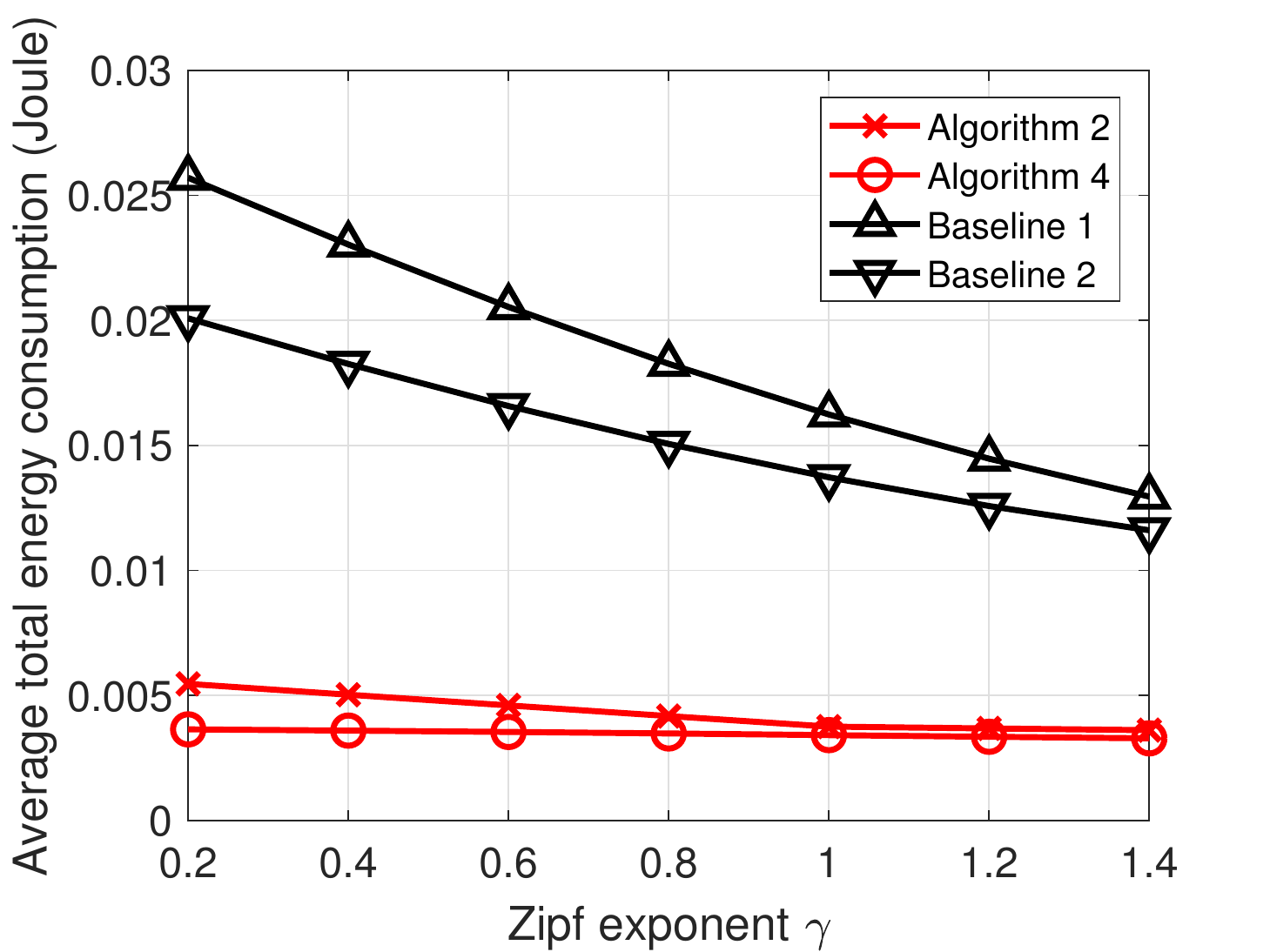}}
        \subfloat[$K=2$, $N=4$, $\gamma=0.8$, $D=0.05$.]{\includegraphics[width=0.5\textwidth]{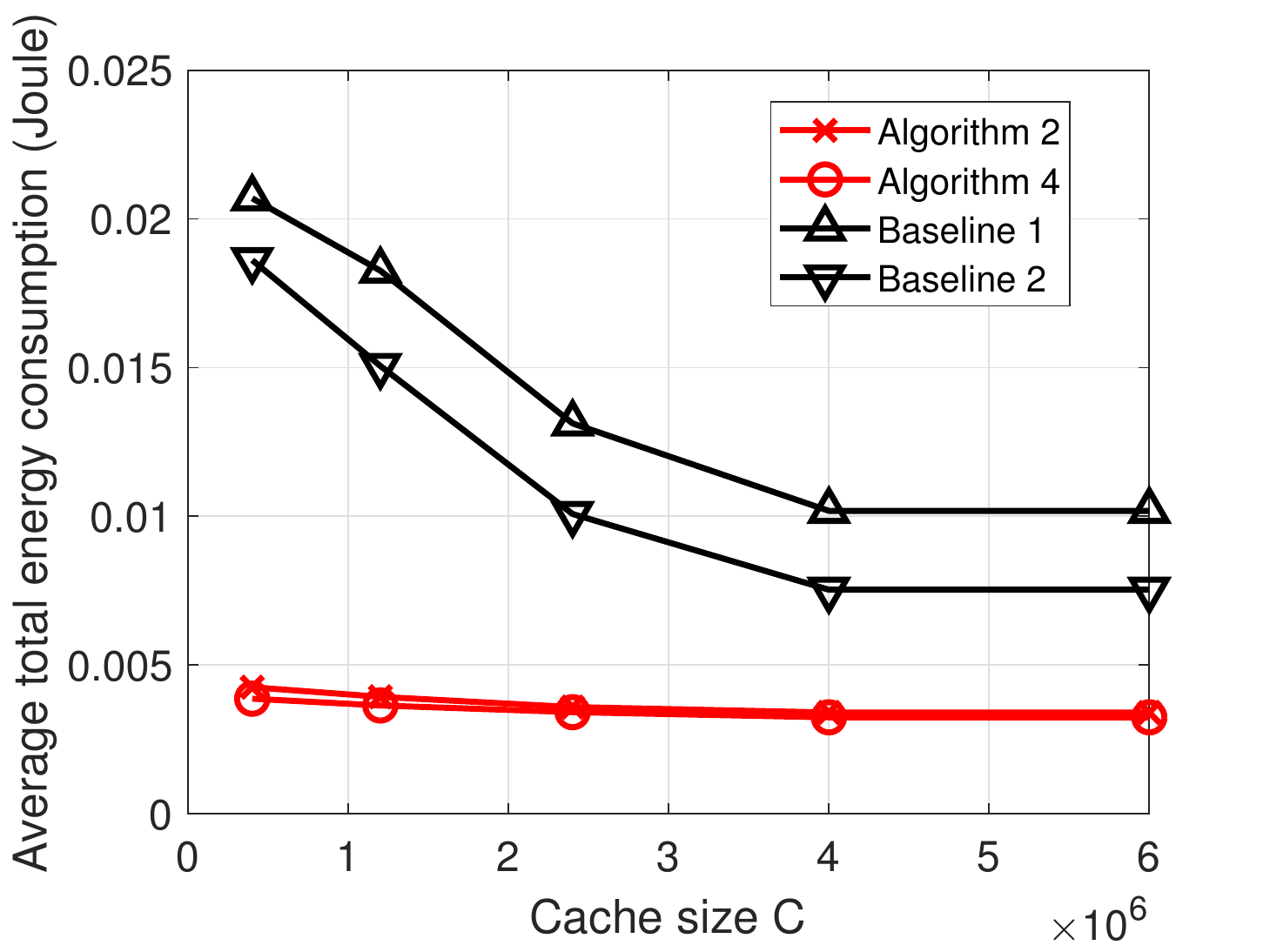}}
      \caption{(a) Effects of the number of users $K$. (b) Effects of the number of services $N$. (c) Effects of the Zipf exponent $\gamma$. (d)  Effects of the cache size $C$.}\label{figtdmaEffects}
\end{figure*}

\subsection{Performance Comparison between Proposed Solutions and Baselines}
In this subsection, we compare the  average total energy consumption of our proposed solutions with two baseline schemes, which adopt the most popular caching scheme. {Specifically, Baseline~1 executes all tasks locally at users and adopts} equal multicasting time allocation, i.e., $\hat T_{\texttt{u},k}^{{\rm{dat}}}({\bf{Q}}) = \hat T_{\texttt{d},k}^{{\rm{dat}}}({\bf{Q}}) = 0$ and $\hat T_{\texttt{d},n}^{{\rm{sfw}}}({\bf{Q}}) = \frac{1}{|\mathcal{X}'|}\left( {D - \sum_{n \in \mathcal{X}'} {(1 - {c_n})T_{{\rm{B}},\texttt{d},n}^{{\rm{sfw}}}}  - \max \left( { {{L_{\texttt{e},k}^{{\rm{dat}}}}}/{{{F_k}}}} \right)} \right)$.~Baseline~2 executes all tasks remotely at the serving node {and adopts} equal uploading and downloading time allocation, i.e., $\hat{T}_{\texttt{u}, k}^{\mathrm{dat}}(\mathbf{Q})=\hat{T}_{\texttt{d}, k}^{\mathrm{dat}}(\mathbf{Q})=\frac{1}{2 K}\left(D-\sum_{n \in\mathcal{X}'}\left(1-c_{n}\right) T_{\mathrm{B}, \texttt{d}, n}^{\mathrm{sfw}}-\sum_{k \in \mathcal{K}}  {L_{\texttt{e}, k}^{\mathrm{dat}}}/{F_{\mathrm{sn}}}\right)$ and $\hat T_{\texttt{d},n}^{{\rm{sfw}}}({\bf{Q}})=0$. {Here,} $\mathcal{X}'\triangleq\left\{ {i \in {\cal X}:{K_i}({\bf{X}}) \ge 1} \right\}$.

Fig. \ref{figtdmaEffects} illustrates the average total energy consumption versus the number of users $K$, number of services $N$, Zipf exponent $\gamma$ and cache size $C$, respectively. From Fig. \ref{figtdmaEffects}, we can see that Algorithm~\ref{algrecovery} and Algorithm~\ref{algCCPAndADMM} outperform the two baselines, demonstrating the advantage of the proposed Algorithm~\ref{algrecovery} and Algorithm~\ref{algCCPAndADMM} in efficiently utilizing the storage, computation and {communications} resources. In addition, from Fig. \ref{figtdmaEffects}, we also see that Algorithm~\ref{algCCPAndADMM} outperforms Algorithm~\ref{algrecovery}.\footnote{Note that, although Algorithm~\ref{algCCPAndADMM} outperforms Algorithm~\ref{algrecovery}, it has a larger computational complexity than Algorithm~\ref{algrecovery}, as shown in Fig.~\ref{figcomputingtime}(c). However, unlike Algorithm~\ref{algrecovery}, Algorithm~\ref{algCCPAndADMM} can be theoretically guaranteed to converge to a stationary point, which generally has a promising performance \cite{Lipp2016}. Thus Algorithm~\ref{algCCPAndADMM} is suitable for a scenario where the performance of the algorithm needs to be theoretically guaranteed.} Specifically, when $K$ or $N$ increases, the average total energy consumption of each scheme increases, due to the increase of the load for fetching and multicasting software and {the increase of the load for} executing tasks at the wireless edge. When $\gamma$ increases, the average total energy consumption of each scheme decreases, due to the decrease of the load {for} multicasting software. When $C$ increases, the average total energy consumptions of Algorithm~\ref{algrecovery}, Algorithm~\ref{algCCPAndADMM} and Baseline~2 decrease. This can be explained as follows. When $C$ increases, the time duration for the serving node to fetch software over the backhaul link decreases, {due to the decrease of the load for fetching software}. {Given} the deadline constraints, the time duration for multicasting software, transmitting the task input data or computation results can be increased, leading to the decrease of the {transmission} energy consumption.

\section{Conclusions}

In this paper, we have proposed a joint caching, computation and communications mechanism which involves software fetching, caching and multicasting, as well as task input data uploading, task executing and computation result downloading, and have mathematically characterized it. We optimized the joint caching, offloading and time allocation policy to minimize the weighted sum energy consumption subject to the caching and deadline constraints. The problem is a challenging two-timescale MINLP problem and is NP-hard in general. {We proposed two low-complexity suboptimal solutions, based on continuous relaxation and penalty CCP, respectively. Numerical results show  that the proposed solutions outperform existing schemes, and the suboptimal solution based on penalty CCP outperforms the one based on continuous relaxation at the cost of computational complexity~increase.}

\appendices
\section{Proof of Lemma \ref{lemmaequivalent}}\label{prooflemmaequivalent}

First, by a change of variables
\begin{eqnarray}
Y_n(\mathbf{Q}) = \max_{k\in\mathcal{K}_n(\mathbf{X})}(1-O_k(\mathbf{Q})),\; n\in\mathcal{X},\; \mathbf{Q}\in\mathcal{Q},\label{eqYnMax}
\end{eqnarray}we can replace $\max _{k\in\mathcal{K}_n(\mathbf{X})}(1-O_k(\mathbf{Q}))$ in (\ref{eqTDMAaverageSumEnergy}), (\ref{eqconstrTotalTimeAllusersQstateprimal}) and (\ref{eqconstrTotalTimeAllusersQstateprimal2}) with $Y_n(\mathbf{Q})$, and thus can equivalently transform the objective function $\overline{E}( {{\mathbf{O}},{\mathbf{T}}_{\texttt{u}}^{\mathrm{dat}}, {\mathbf{T}}_{\texttt{d}}^{\mathrm{dat}}, {\mathbf{T}}_{\texttt{d}}^{\mathrm{sfw}}} ) $ in (\ref{eqTDMAaverageSumEnergy}) to $\overline{e} ( {{\mathbf{O}}, {\mathbf{Y}}, {\mathbf{\hat{T}}}_{\texttt{u}}^{\mathrm{dat}}, {\mathbf{\hat{T}}}_{\texttt{d}}^{\mathrm{dat}}, {\mathbf{\hat{T}}}_{\texttt{d}}^{\mathrm{sfw}}}  )$ in (\ref{eqTDMAtotalEngeryrew2obj2}). In addition, by noting that $\overline{e} ( {{\mathbf{O}}, {\mathbf{Y}}, {\mathbf{\hat{T}}}_{\texttt{u}}^{\mathrm{dat}}, {\mathbf{\hat{T}}}_{\texttt{d}}^{\mathrm{dat}}, {\mathbf{\hat{T}}}_{\texttt{d}}^{\mathrm{sfw}}}  )$ increases with $Y_n(\mathbf{Q})$, we can further equivalently transform (\ref{eqYnMax}) to (\ref{eqconstrynmaxQstate}). Next, by a change of variables $\hat{T}_{\texttt{u},k}^{\mathrm{dat}}(\mathbf{Q}) \triangleq {O_k(\mathbf{Q})}T_{\texttt{u},k}^{\mathrm{dat}}(\mathbf{Q})$, $\hat{T}_{\texttt{d},k}^{\mathrm{dat}}(\mathbf{Q}) \triangleq {O_k(\mathbf{Q})}T_{\texttt{d},k}^{\mathrm{dat}}(\mathbf{Q})$, {$k\in\mathcal{K}$, $\mathbf{Q}\in\mathcal{Q}$,} and $\hat{T}_{\texttt{d},n}^{\mathrm{sfw}}(\mathbf{Q}) \triangleq {Y_n(\mathbf{Q})}T_{\texttt{d},n}^{\mathrm{sfw}}(\mathbf{Q})$, {$n\in\mathcal{X}$, $\mathbf{Q}\in\mathcal{Q}$,} we can equivalently convert (\ref{eqconstrTuk}), (\ref{eqconstrTdk}), (\ref{eqconstrTdn}), (\ref{eqconstrTotalTimeAllusersQstateprimal}) and (\ref{eqconstrTotalTimeAllusersQstateprimal2}) to (\ref{eqconstrhattukQstate}), (\ref{eqconstrhattdkQstate}), (\ref{eqconstrhattdnQstate}), (\ref{eqconstrsumtasksTimeQstate}) and (\ref{eqconstrhatsumTimeTDMAQstate}), respectively. Therefore, we can equivalently transform Problem \ref{eqTDMAPrimalProblem} to Problem~\ref{eqTDMAPrimalProblem2}, which completes the proof.

\section{Proof of Lemma \ref{lemmaequivalentADMM}}\label{prooflemmaequivalentADMM}

{By introducing the consensus constraints in (\ref{eqconstrhatcnandcADMM}), Problem~\ref{eqTDMAPrimalProblem2CR} can be equivalently converted} to the following problem:
{\setlength{\arraycolsep}{0.0em}
\begin{eqnarray}
&&\min_{\substack{{\mathbf{c},\mathbf{\hat{C}}, {\mathbf{O}}, {\mathbf{Y}},
{\mathbf{\hat{T}}}_{\texttt{u}}^{\mathrm{dat}}, {\mathbf{\hat{T}}}_{\texttt{d}}^{\mathrm{dat}}, {\mathbf{\hat{T}}}_{\texttt{d}}^{\mathrm{sfw}}}}}
\overline{e} ( {{\mathbf{O}}, {\mathbf{Y}}, {\mathbf{\hat{T}}}_{\texttt{u}}^{\mathrm{dat}}, {\mathbf{\hat{T}}}_{\texttt{d}}^{\mathrm{dat}}, {\mathbf{\hat{T}}}_{\texttt{d}}^{\mathrm{sfw}}})    \label{eqTDMAPrimalProblemADMM}\\
\mathrm{s.t.} \;&& (\ref{eqconstrynmaxQstate}), (\ref{eqconstrhattukQstate}), (\ref{eqconstrhattdkQstate}), (\ref{eqconstrhattdnQstate}), (\ref{eqconstrOkQstate}), (\ref{eqconstrhatcnandcADMM}), (\ref{eqconstrhatcn3ADMMnew}),  (\ref{eqconstrhatcn3ADMM}), (\ref{eqconstrsumhatcn1ADMM}), (\ref{eqconstrhatcn1ADMM})\nonumber.
\end{eqnarray}\setlength{\arraycolsep}{5pt}}Then, based on the definition of $ v(\cdot)$ in (\ref{eqdefv}), the problem in (\ref{eqTDMAPrimalProblemADMM}) can be equivalently rewritten as Problem~\ref{probADMMform}.  Therefore, we complete the proof of Lemma \ref{lemmaequivalentADMM}.

\section{Proof of Lemma \ref{lemmaConvergenceADMM}}\label{prooflemmaConvergenceADMM}

{Following ADMM, we first form the augmented Lagrangian function \cite{admmSboyd} of Problem~\ref{probADMMform}}, which is given by (\ref{eqTDMAPrimalProblemADMM2LarFunc}),
\begin{figure*}[!t]
{\begingroup\makeatletter\def\f@size{9.5}\check@mathfonts
\def\maketag@@@#1{\hbox{\m@th\normalsize\normalfont#1}}
\begin{eqnarray}\label{eqTDMAPrimalProblemADMM2LarFunc}
&& L_{\rho} (\mathbf{c}, \mathbf{\hat{C}}, {\mathbf{O}}, {\mathbf{Y}},
{\mathbf{\hat{T}}}_{\texttt{u}}^{\mathrm{dat}}, {\mathbf{\hat{T}}}_{\texttt{d}}^{\mathrm{dat}}, {\mathbf{\hat{T}}}_{\texttt{d}}^{\mathrm{sfw}}, \boldsymbol{\theta}) \nonumber\\
&&  = \sum_{\mathbf{Q}\in \mathcal{Q}} \left( v( {\mathbf{O}(\mathbf{Q})}, \mathbf{Y}(\mathbf{Q}),
{\mathbf{\hat{T}}}_{\texttt{u}}^{\mathrm{dat}}(\mathbf{Q}), {\mathbf{\hat{T}}}_{\texttt{d}}^{\mathrm{dat}}(\mathbf{Q}), {\mathbf{\hat{T}}}_{\texttt{d}}^{\mathrm{sfw}}(\mathbf{Q}) ) +  \sum_{n\in\mathcal{X}} \theta_n(\mathbf{Q}) \left( {\hat{C}}_n(\mathbf{Q})-c_n\right) + \frac{\rho}{2}   \sum_{n\in\mathcal{X}} \left( {\hat{C}}_n(\mathbf{Q})-c_n\right)^2\right).
\end{eqnarray}
 \endgroup}
\noindent\rule[0.25\baselineskip]{\textwidth}{0.1pt}
\end{figure*}where $\theta_n(\mathbf{Q})$, $n\in\mathcal{X}$, $\mathbf{Q}\in\mathcal{Q}$ {are the Lagrange multipliers corresponding to the constraints in (\ref{eqconstrhatcnandcADMM}), and $\boldsymbol{\theta} \triangleq (\theta_n)_{n\in\mathcal{X}}$ denotes the vector mapping of $\mathbf{Q}$. At iteration $t+1$, we execute} three steps of ADMM, i.e., (\ref{eqADMMlocalupdate}), (\ref{eqADMMglobalupdate}) and (\ref{eqADMMThetaupdate}).
\begin{figure*}[!t]
{\begingroup\makeatletter\def\f@size{9.5}\check@mathfonts
\def\maketag@@@#1{\hbox{\m@th\normalsize\normalfont#1}}
\begin{align}
\left(\mathbf{\hat{C}}_{t+1},  {\mathbf{O}}_{t+1}, \mathbf{Y}_{t+1}, {\mathbf{\hat{T}}}_{\texttt{u},t+1}^{\mathrm{dat}},   {\mathbf{\hat{T}}}_{\texttt{d},t+1}^{\mathrm{dat}}, {\mathbf{\hat{T}}}_{\texttt{d},t+1}^{\mathrm{sfw}}\right)  &=  \arg\min_{ { {  \mathbf{\hat{C}}, {\mathbf{O}},  {\mathbf{Y}},}   {
{\mathbf{\hat{T}}}_{\texttt{u}}^{\mathrm{dat}}, {\mathbf{\hat{T}}}_{\texttt{d}}^{\mathrm{dat}}, {\mathbf{\hat{T}}}_{\texttt{d}}^{\mathrm{sfw}}} }  } L_{\rho} (\mathbf{c}_{t}, \mathbf{\hat{C}}, {\mathbf{O}}, {\mathbf{Y}}, {\mathbf{\hat{T}}}_{\texttt{u}}^{\mathrm{dat}}, {\mathbf{\hat{T}}}_{\texttt{d}}^{\mathrm{dat}}, {\mathbf{\hat{T}}}_{\texttt{d}}^{\mathrm{sfw}}, \boldsymbol{\theta}_{t}) , \label{eqADMMlocalupdate}\\
\mathbf{c}_{t+1} &= \arg\min_{\mathbf{c}} L_{\rho} (\mathbf{c},\mathbf{\hat{C}}_{t+1},  {\mathbf{O}}_{t+1}, \mathbf{Y}_{t+1}, {\mathbf{\hat{T}}}_{\texttt{u},t+1}^{\mathrm{dat}},   {\mathbf{\hat{T}}}_{\texttt{d},t+1}^{\mathrm{dat}}, {\mathbf{\hat{T}}}_{\texttt{d},t+1}^{\mathrm{sfw}}, \boldsymbol{\theta}_{t}). \label{eqADMMglobalupdate}
\end{align} \endgroup}
\noindent\rule[0.25\baselineskip]{\textwidth}{0.1pt}
\end{figure*}
{Then, we show that (\ref{eqADMMlocalupdate}) is equivalent to (\ref{eqTDMAPrimalProblemADMMupdateslocalvar}).
Substituting (\ref{eqTDMAPrimalProblemADMM2LarFunc}) into (\ref{eqADMMlocalupdate}) and dropping the constant terms in the objective function of the problem in (\ref{eqADMMlocalupdate}) which do not affect the solution, the problem in (\ref{eqADMMlocalupdate}) can be decomposed into $\left|\mathcal{Q}\right|$ subproblems, one for each $\mathbf{Q}\in\mathcal{Q}$, given by (\ref{eqTDMAPrimalProblemADMMupdateslocalvar}). Next, we show that (\ref{eqADMMglobalupdate}) is equivalent to (\ref{eqcalglobalvar}). Specifically, substituting (\ref{eqTDMAPrimalProblemADMM2LarFunc}) into (\ref{eqADMMglobalupdate}) and dropping the constant terms in the objective function of the problem in (\ref{eqADMMglobalupdate}) which do not affect the solution, the problem in (\ref{eqADMMglobalupdate}) becomes the following unconstrained quadratic problem:
\begin{align}
\mathbf{c}_{t+1} & = \arg\min_{\mathbf{c}} \sum_{\mathbf{Q}\in \mathcal{Q}}  \sum_{n\in\mathcal{X}} \theta_{t,n}(\mathbf{Q}) \left( {\hat{C}}_{t+1,n}(\mathbf{Q})-c_n\right) \nonumber\\
& + \frac{\rho}{2} \sum_{\mathbf{Q}\in \mathcal{Q}}  \sum_{n\in\mathcal{X}} \left( {\hat{C}}_{t+1,n}(\mathbf{Q})-c_n\right)^2  \label{eqADMMglobalupdatetmp}.
\end{align}It is clear that the problem in (\ref{eqADMMglobalupdatetmp}) is strictly convex. By setting the gradient of the objective function of the problem in (\ref{eqADMMglobalupdatetmp}) to be zero, we can obtain a closed-form  solution, given by (\ref{eqcalglobalvar}).

It can be easily verified that the objective function of Problem~\ref{probADMMform} is convex, closed and proper. Problem~\ref{probADMMform}  and its dual problem have optimal solutions and there is no duality gap, which implies that the unaugmented Lagrangian $L_{0} (\mathbf{c}, \mathbf{\hat{C}}, {\mathbf{O}}, {\mathbf{Y}}, {\mathbf{\hat{T}}}_{\texttt{u}}^{\mathrm{dat}}, {\mathbf{\hat{T}}}_{\texttt{d}}^{\mathrm{dat}}, {\mathbf{\hat{T}}}_{\texttt{d}}^{\mathrm{sfw}}, \boldsymbol{\theta})$ has a saddle point. Hence, the two assumptions required by the convergence analysis in \cite[Chapter 3.2]{admmSboyd} are satisfied. As a result, we know that the objective function of the iterates of Algorithm~\ref{algADMM} approaches the optimal value of Problem~\ref{probADMMform}, as $k\to\infty$.
Therefore, we complete the proof of Lemma \ref{lemmaConvergenceADMM}.}

\bibliographystyle{IEEEtran}


\end{document}